\documentclass[universe,review,accept,pdftex,moreauthors]{mdpi} 

\usepackage{natbib}
\usepackage{aas_macros}

\usepackage{yfonts}
\usepackage{xcolor}
\newcommand{\vidak}[1]{\textcolor{cyan}{#1}}

\newcommand{\seli}[1]{\textcolor{brown}{#1}}
\newcommand{\martin}[1]{\textcolor{red}{#1}}
\newcommand{\petra}[1]{\textcolor{violet}{#1}}

\newcommand{\kovari}[1]{\textcolor{purple}{#1}}

\renewcommand{\vidak}[1]{\textcolor{black}{#1}}

\renewcommand{\seli}[1]{\textcolor{black}{#1}}
\renewcommand{\martin}[1]{\textcolor{black}{#1}}
\renewcommand{\petra}[1]{\textcolor{black}{#1}}

\renewcommand{\kovari}[1]{\textcolor{black}{#1}}
\firstpage{1} 
\pubvolume{1}
\issuenum{1}
\articlenumber{0}
\pubyear{2024}
\copyrightyear{2024}
\datereceived{ } 
\daterevised{ } 
\dateaccepted{ } 
\datepublished{ } 
\hreflink{https://doi.org/} 

\Title{
Stellar flares, superflares and coronal mass ejections -- entering the Big data era}

\TitleCitation{Stellar flares, superflares and coronal mass ejections -- entering the Big data era}

\Author{Kriszti\'{a}n Vida $^{1}$\orcidA{}, 
Zsolt K\H{o}v\'{a}ri $^{1}$\orcidC{}, 
Martin Leitzinger $^{2}$\orcidE{},
Petra Odert $^{2}$\orcidF{},
Katalin Ol\'{a}h $^{1}$\orcidG{}, 
B\'{a}lint Seli $^{1,3}$\orcidH{}, 
Levente Kriskovics $^{1}$\orcidD{},  
Robert Greimel $^{4}$\orcidB{}, 
and
Anna G\"{o}rgei $^{1,3}$\orcidI{}}

\AuthorNames{Kriszti\'{a}n Vida, Zsolt K\H{o}v\'{a}ri, Martin Leitzinger, Petra Odert,  Katalin Ol\'{a}h, B\'{a}lint Seli, Levente Kriskovics, Robert Greimel and Anna G\"{o}rgei}

\AuthorCitation{Vida, K.; K\H{o}v\'{a}ri, Zs.; Leitzinger, M. et al.}

\address{%
$^{1}$ \quad Konkoly Observatory, HUN-REN Research Centre for Astronomy and Earth Sciences, Budapest, Hungary; vida.krisztian@csfk.org\\
$^{2}$ \quad Department for Astrophysics and Geophysics (AGP), Institute of Physics, University of Graz, Graz, Austria\\
$^{3}$ \quad E\"otv\"os University, Department of Astronomy, Budapest, Hungary\\
$^{4}$ \quad RG Science, Graz, Austria
}

\abstract{Flares, sometimes accompanied by coronal mass ejections (CMEs), are the result of sudden changes in the magnetic field of stars with high energy release through magnetic reconnection, which can be observed across a wide range of the electromagnetic spectrum from radio waves to the optical range to X-rays. In our \vidak{observational} review, we attempt to collect some fundamental new results, which can largely be linked to the Big data era that has arrived due to the expansion of space photometric observations of the last two decades. We list the different types of stars showing flare activity, their observation strategies, and discuss how their main stellar properties relate to the characteristics of the flares (or even CMEs) they emit. Our goal is to focus, without claiming to be complete, on those results that may in one way or another challenge the "standard" flare model based on the solar paradigm.
}


\begin{document}

\section{Introduction}\label{intro}

The effect of stellar activity on the \vidak{surrounding extended astrospheres, i.e., the AU scale region}, where planets may also orbit, has always been an exciting topic. We already know that stellar flares can greatly affect their environment \citep[e.g.,][]{Vida2017ApJ...841..124V}, just as solar activity and solar flares affect the Earth's astroclimate. The famous Carrington event, the most intense, documented geomagnetic storm peaking between 1--2 September 1859 during Solar Cycle 10 \citep{2003JGRA..108.1268T} resulted in intense auroral lights and reportedly caused sparks and in a few cases, even fires in telegraph stations. According to \citet{Carrington1859MNRAS..20...13C}, the event was associated with a highly energetic solar white-light flare. Nowadays, a geomagnetic storm of this magnitude could even lead to total chaos in the operation of satellite-supported electronic devices and systems. For this reason alone, it is important to understand what leads to these kinds of processes on the surface of the Sun.
The mechanism behind flaring is thought to be magnetic reconnection, which is closely related to magnetic activity including sunspots. The source of the magnetic activity is dynamo operation, which can be related to the interaction between the Sun's convective envelope and its rotation. At the same time, we can also observe similar activity features (spots, flares, etc.) on other stars, the formation of which, we believe, requires similar conditions. By observing flare stars with different properties (temperature, rotation period, size, age, etc.) and comparing them with the Sun, we can get a more comprehensive picture not only of the underlying physics but also of the solar flares themselves, together with associated phenomena such as the coronal mass ejections (CMEs) which, for example, caused the intense geomagnetic storm during the Carrington-event.
Therefore, in this review paper, we attempt to summarize all the knowledge we have gathered in the past few years on observing different types of flare stars.

\vidak{Flares are typically characterized by their peak flux or their energy. In the case of the Sun, the classification (A, B, C, M, or X) is based on the peak flux in  soft X-rays as measured by GOES satellites. The energy released by solar flares can vary over several magnitudes, roughly between 10$^{27}$--10$^{32}$ ergs \cite{Crosby1993}. }
However, in the case of stellar flares, there is no such elaborate classification. The distinction between flares and superflares is not made on a physical basis, but it is merely a matter of terminology. In general, flares with radiative energy of $\sim$10$^{33}$\,erg, or more are labeled with the prefix "super" (see, e.g., \citet{Shibata2013}), so in our paper, we follow this custom in nomenclature as well. In this sense, the most intense solar flare of $\sim$X45 class recorded to date, associated with the Carrington event, with its estimated energy of $\sim$5$\times$10$^{32}$\,erg \citep{Cliver2013JSWSC...3A..31C} was still {below the "super" category. }



In the most common solar flare model (e.g. \citet{1976SoPh...50...85K}), the magnetic energy release is induced by the reconnection of the magnetic field lines. There are a few known possible mechanisms capable of inducing reconnection. One of these possible mechanisms is flux emergence \cite{1977SoPh...53..255H}, where the interaction between separate magnetic structures can trigger magnetic reconnection. \citet{1995JGR...100.3355F} has proven this hypothesis by observing filaments releasing energy and mass (and hence, resulting in flares accompanied by coronal mass ejections) after the destabilization of their magnetic structure by another rising magnetic flux bundle. Another possible mechanism is the "kink instability" proposed by \citet{1991SoPh..136..133T}. In this case, the emerging flux bundle is already twisted and kinked, and further twisting induces reconnection and energy release. 
Solar flares are known to be accompanied by several physical processes in the corona. 
\citet{1976SoPh...50...85K} proposed that rising loop prominence systems quite frequently seen after solar flares and other coronal transients are the results of magnetic field reconnection as well. With the magnetic field lines opening up, the solar wind carries mass to the prominence system, where it is captured by the closing lines during reconnection. Modern textbooks (e.g. \citet{2010hssr.book..159F} or \citet{2014masu.book.....P}) on the topic refines the picture of the "standard solar flare model" further with several previously observed processes. Reconnection accelerates particles along the magnetic field lines, heating the chromosphere at the footpoints of the magnetic loop and evaporating plasma into the corona. The process can be observed in a wide range of the electromagnetic spectrum from radio to ultraviolet radiation to soft and hard X-rays, resulting in observables, e.g. fast evaporation upflows, hard X-ray source regions or X-ray loops caused by the bremsstrahlung of fast electrons captured by the magnetic field. \citet{2020ApJ...896...97R} provided a model that connected multi-dimensional MHD description of the plasma with an analytic fast electron treatment, describing these phenomena.

For now, it appears that flares observed in a wide range of late-type stars are manifestations of the same mechanism, that is, stellar flares are believed to originate from magnetic reconnection which, first of all, presumes an underlying magnetic dynamo. In this regard, therefore, we expect that the more magnetic energy a star can accumulate, the more energetic flares will be produced. This scaling idea came up several times, not so long ago, e.g. in the paper by \citet{Balona2015MNRAS.447.2714B}, who studied \emph{Kepler} flare stars of different luminosity classes and found that higher luminosity class stars generally have higher energy flares. 
\vidak{This correlation with radius -- possibly connected with 
a correlation with the typical magnetic field associated with the flares -- was explained by the typical flare loop length-scales increasing with the stellar radius. }
This is further supported by the finding of \citet{He2018ApJS..236....7H} who concluded that both magnetic feature activity and flare activity are influenced by the same source of magnetic energy (i.e. the magnetic dynamo), similar to the solar case \citep[see][]{Hathaway2015LRSP...12....4H}. In accordance with all these, a common dynamo scaling in all late-type stars was proposed recently by \citet{Lehtinen2020NatAs...4..658L}.
Moreover, this scaling idea can also be reconciled with the so-called "avalanche models" of solar flares, which regard flares as avalanches of many small reconnection events \citep[for a detailed review see][]{Charbonneau2001SoPh..203..321C}. This statistical framework could consistently be extended to provide a common base for solar flares to stellar flares to superflares over many orders of magnitude in energy 
\citep[cf. the conclusion in][]{Kovari2020A&A...641A..83K}.

\begin{figure}
    \centering
    \includegraphics[width=0.5\textwidth]{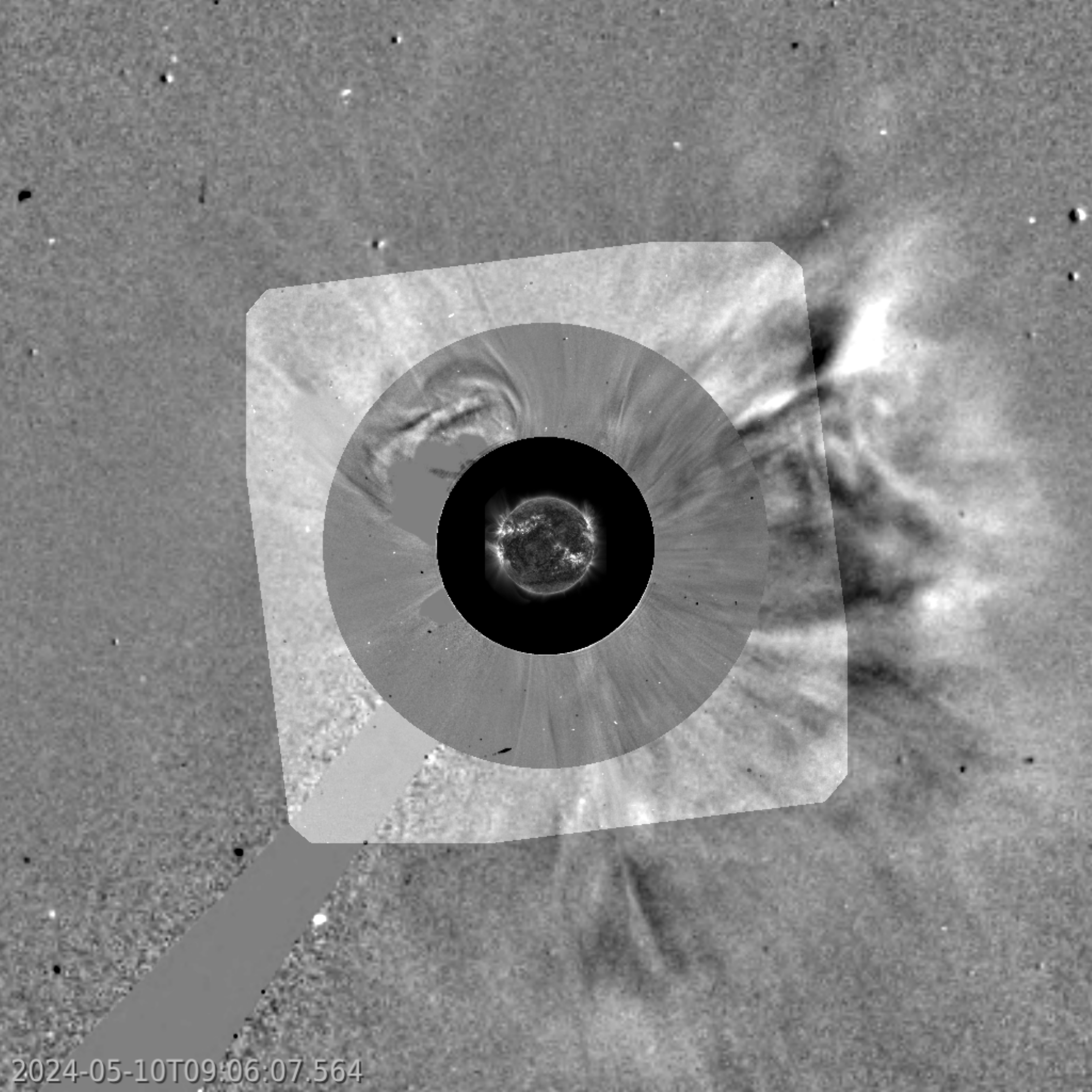}
    \caption{Coronal mass ejection on the Sun on 2024 May 10 as detected by SOHO/LASCO C2 and C3 instruments. The plot shows a running difference between consecutive frames. In the middle, SDO/AIA 171 observation of the Sun is shown for reference. }
    \label{fig:lasco-cme}
\end{figure}
On the Sun, an energetic activity phenomenon is often closely related to flares, namely coronal mass ejections (CMEs): these are magnetized plasma clouds expelled into the heliosphere. In many cases, an erupting filament/prominence represents the early evolutionary stage of a CME. Filaments and prominences are the same phenomena seen either against the solar disk as dark elongated features (filaments) or seen on the solar limb as a bright arching structure (prominences). Filaments/prominences are magnetic loops that are filled with cool chromospheric plasma located at a mean height of $\sim$25~Mm \citep{Wang2010}. The principle picture of a filament/prominence eruption is closely related to the standard flare model of a two-ribbon flare \cite{Martens1989} -- here, the magnetic field lines forming a loop are anchored to the solar photosphere through their footpoints. As the photosphere is not a static medium, its plasma motion can lead to footpoint motions bringing the magnetic field lines of the loop close together and thereby possibly forming a reconnection region yielding the conditions for an erupting flare event. If the magnetic loop above the reconnection region confines solar plasma, this can be ejected from the Sun if the above-lying magnetic field \vidak{opens up, e.g., by a reconnection of the magnetic field lines}. 
Such a filament/prominence eruption propagates further through the corona, drags coronal material with it, and forms a CME  with its typical three-part structure: the core, the cavity, and the leading edge. In images of the Large Angle Spectroscopic Coronagraphs (LASCO) this structure can be seen very well {(see Fig.~\ref{fig:lasco-cme})}.  

CMEs are, as they share a similar generation mechanism, closely connected to flares. For the energetic so-called X-class flares (GOES classification scheme) the association rate of flares and CMEs on the Sun reaches unity \citep{Yashiro2009}, or in other words, every X-class flare has an accompanying CME. However, recent observations revealed a large active region that produced several X-class flares without CMEs \citep[e.g. ][]{Thalmann2015}. This led to the hypothesis that the flare--CME association rate is both a function of flare magnitude and the magnetic flux of the active region \citep{Li2020}. The stronger the magnetic field the stronger the flare needs to be to erupt. This aspect of the flare--CME association rate considers magnetic confinement, which is believed to be one of the reasons \vidak{(besides observational reasons; that is, the lack of spatial resolution, the sensitivity of the instrument, etc.)} why on stars so few signatures of CMEs or erupting filaments/prominences have been found so far. 

CMEs have several other accompanying phenomena which are important tools to be used to \vidak{search for CMEs on stars}\petra{, such as radio bursts and coronal dimmings.}
Type II radio bursts (slowly drifting herring-bone-like structures as seen in dynamic spectra) are the signature of a shock wave driven by CMEs and produced by electrons that are accelerated at the shock front. 
\vidak{Type IV bursts are characterized by an outward-moving continuum source that is often preceded by a Type II burst in association with a coronal mass ejection (CME). 
Type IV bursts have been classified into stationary and moving categories with the moving component attributed to energetic electrons trapped in the CME, emitting plasma emission, gyrosynchrotron or synchrotron, or sometimes electron cyclotron maser emission \petra{(see the review of \citet{Carley2020})}.} CMEs are also frequently accompanied by coronal dimmings \citep{Dissauer2019}. Coronal dimmings are evacuated regions in the corona, obvious in extreme ultraviolet (EUV) and soft X-ray images of the Sun (e.g. Yohkoh or SDO), seen during/after CMEs. 

On stars also other methods have been established to identify CMEs. For instance, continuous X-ray absorptions in flares were interpreted as stellar CME plasma temporarily obscuring the flaring region \citep{Moschou2019}. Another method that does not rely on the solar--stellar analogy and which is a more direct measure of plasma being ejected from a star is the Doppler-shifted emission/absorption in spectral lines caused by moving plasma \citep[e.g.][]{Houdebine1990, Leitzinger2011a, Vida2016, Argiroffi2019, Namekata2021}. All of these methods have been applied to stellar observations and have their advantages and drawbacks. In section~\ref{sect:CMEs} we discuss in more detail observational and modeling aspects of stellar CMEs \citep[for further reviews on stellar CMEs see also][]{Moschou2019, Leitzinger2022c, Osten2023, Tian2023}.

\begin{figure}
    \centering
    \includegraphics[width=0.9\textwidth]{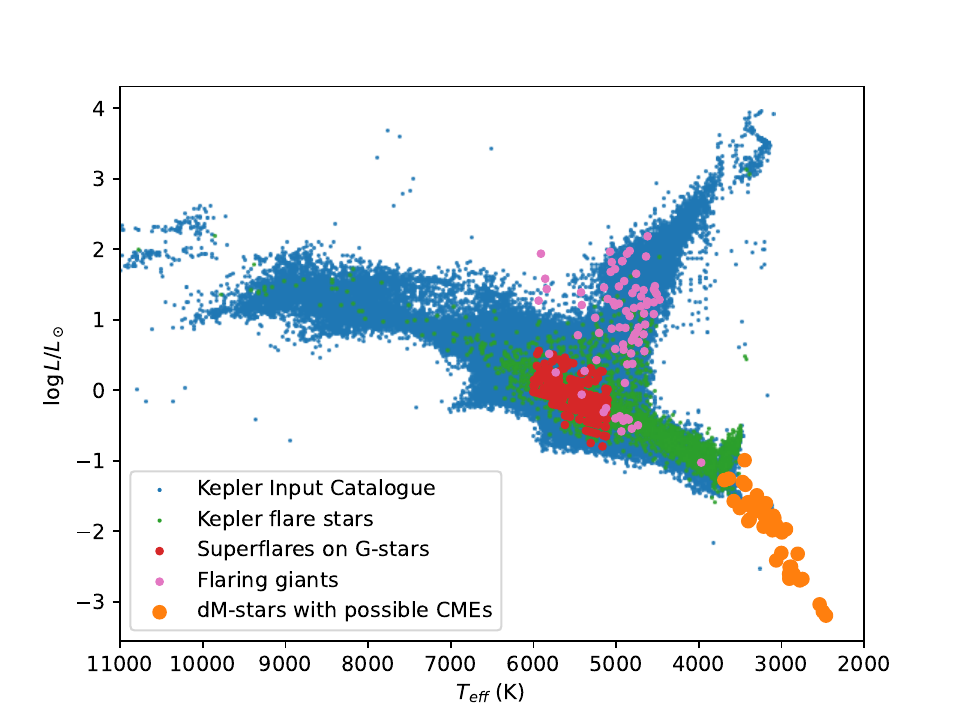}
    \caption{Flaring stars across the HRD. Blue points show the \emph{Kepler} input catalogue (KIC) \cite{KIC2011AJ....142..112B}, 
    green points denote the \emph{Kepler} flare catalogue of \citet{Yang2019ApJS..241...29Y},
    red points show  G-type stars showing superflares \cite{Shibayama2013ApJS..209....5S}, and
    pink points mark flaring giant stars \cite{Olah2022A&A...668A.101O}.
    Orange points show stars with \petra{possible detections of flare-related} coronal mass ejections \cite{Vida2019A&A...623A..49V} (note, that in the latter case, physical parameters are not from the \emph{Kepler} catalogue, as that cuts off at 3100K. 
    }
    \label{fig:hrd-flare}
\end{figure}

Our Fig.\,\ref{fig:hrd-flare} shows the distribution of different types of stars showing flares/CMEs on the Hertzsprung--Russell diagram (HRD). This shows that the flaring stars are mainly of G--K--M spectral type. These stars have a deep convective zone in which the dynamo effect can generate a strong magnetic field on both the main sequence (MS) and in evolved stars \citep[cf.][]{Lehtinen2020NatAs...4..658L}. At the same time, moving towards the earlier spectral types, i.e., F and A on the main sequence, the convective zone gradually thins, so we would expect fewer flares due to the weakening magnetic field, if at all. That is why the finding in \citet{Balona2012MNRAS.423.3420B}, according to which large energetic flares were also found on A-type stars in the \emph{Kepler} field, seemed very surprising. However, a closer look at these supposedly flaring A-stars revealed that there are possible alternative explanations for $\approx$60\% of these targets (e.g., unresolved cool companions), while the remaining $\approx$40\% of the observed targets in question are not really convincing to support the hypothesis of flaring A-type stars \citep{Pedersen2017MNRAS.466.3060P}.




\section{Flare detection in space-borne photometry}

With the long, uninterrupted photometric monitoring of a sizable portion of the sky provided by the \emph{Kepler} and \emph{TESS} satellites, the study of stellar flares has entered the Big Data era. With the vast amount of light curves available, the identification of stellar flares is no longer a trivial task.
There are several approaches to detecting flares in light curves besides the obvious, manual method. These methods are often based on smoothing the light curve and detecting the outliers  \cite{Davenport2016ApJ...829...23D, Stelzer2016MNRAS.463.1844S}, or other outlier detection methods \cite{Vida2018A&A...616A.163V.FLATWRM}, but recently there are also flare-detection methods available that are based on different neural network architectures
\cite{Feinstein2020AJ....160..219F, Vida2021A&A...652A.107V.FLATWRM2}.
The main challenge of flare detection is the trade-off between high completeness and low false alarm rate, as a large variety of astrophysical and instrumental phenomena can mimic the appearance of flares, while real flares have amplitudes and durations ranging orders of magnitudes.

The typical time scale of flare events is a few minutes to a few hours, but in some cases, even day-long events have been reported \cite{Kuerster1996A&A...311..211K, Vida2009A&A...504.1021V}. The light curve shape of a typical, single peaked flare is well described by a short rise, and a more gradual, exponential decay phase \citep{Davenport2014ApJ...797..122D, 2022AJ....164...17M}. However, many flares show more complex light curves, especially when observed with higher cadence \citep{2022ApJ...926..204H}, including multiple peaks, bumps, or quasi-periodic oscillations.




\section{Superflares on solar-like stars}
\label{sect:solar-like}
Large flares on the Sun and solar-like stars are particularly interesting to researchers due to their potential impact on both astrophysics and space weather.
We know that solar flares follow a power-law distribution \citep[see][and references therein]{Cliver2022LRSP...19....2C}. The largest recent flare was the Carrington event in 1859 with an estimated energy in the order of $\approx$10$^{32}$\,erg, that was orders of magnitudes smaller than the largest events observed on other stars, mainly rapidly rotating M dwarfs -- but is it possible that there were larger events on the Sun in the past?
There are documented events from East Asia in 1770 describing low-latitude aurorae --
manifestations of magnetic storms \citep{Hayakawa2017ApJ...850L..31H}.
One of the most interesting such reports is a description of two sunspots seen in 1128 from Worcester, England, which sighting coincided with the appearance of the aurora 5 days later in Korea \citep{Willis2001AnGeo..19..289W}.
Longer-term increases in solar activity can be traced using the variation of traces of
$^{14}$C in tree rings  \citep[e.g.,][]{Miyake2012Natur.486..240M,Miyake2013JGRA..118.7483M, Usoskin2013LRSP...10....1U} 
or 
$^{10}$Be in ice cores \citep{Mekhaldi2015NatCo...6.8611M} that show 
increased concentration in the years 774/5 and 993/4.
In these cases, however, the exact origin of the high-energy particles can not be narrowed down to the Sun, but if solar flares cause these, their estimated energy was at least five times stronger than any instrumentally recorded solar event.

Given the limited amount of precise historical solar observations, we need another approach to find out if superflares can occur on the Sun -- a good proxy could be observing a large number of solar-like stars, e.g., using the light curves of the \emph{Kepler} space observatory. The \emph{Kepler} light curves provide a database of approximately four-year-long light curves of $\approx$80\,000 solar-like stars, including more than 2000 superflares on more than 250 G-type dwarfs; cf. Fig.\,\ref{fig:hrd-flare} \citep[and see also][]{Maehara2012Natur.485..478M, Shibayama2013ApJS..209....5S, Okamoto2021ApJ...906...72O}. Their energy distribution proved to be similar to solar flares
following a $dN/dE \propto E^{-\alpha}$ distribution with $\alpha\approx2$. The physical reason behind the higher superflare rates is not clear yet, but their occurrence rate was higher on faster-rotating stars (i.e., on younger stars). At the same time, the proposed connection of flaring and hot Jupiters was not confirmed. The studies \vidak{of the \emph{Kepler} light curves of solar-like stars} concluded that the Sun could have a superflare with an energy of $\approx 7\times10^{33}$\,erg ($\sim$X700 class) once every 3000 years, and $\approx 1\times10^{34}$\,erg ($\sim$X1000 class) once every 6000 years \cite{Okamoto2021ApJ...906...72O}. \vidak{\citet{Karoff2016} reached a similar conclusion based on LAMOST spectroscopic data.}

\begin{figure}
    \centering
    \includegraphics[width=0.9\textwidth]{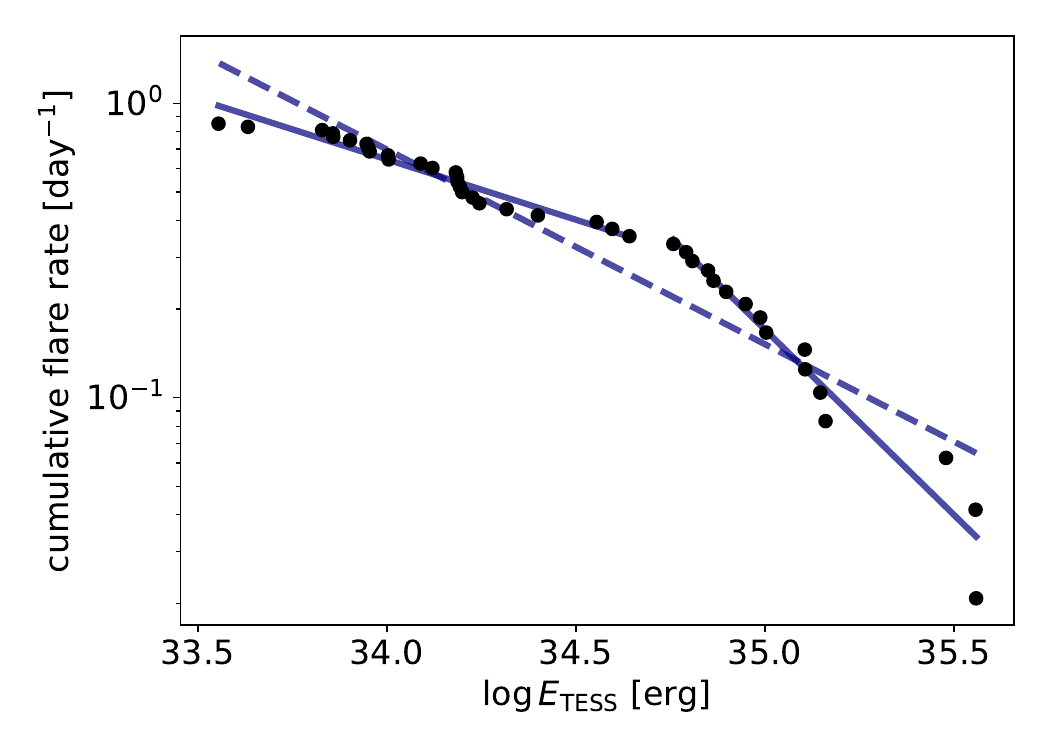}
    \caption{
    Flare frequency distribution for EI\,Eri following a broken power law shape \citep{Kriskovics2023A&A...674A.143K}. The dashed line denotes a fit for all of the points with a power law index of $1.66\pm0.04$, while the solid lines show a two-component fit with indices of $1.41\pm 0.02$ and $2.26\pm0.08.$
    }
    \label{fig:eieri-ffd}
\end{figure}

\section{Flares and superflares of cool dwarf stars}
\label{sect:late-type}

Flares on low-mass stars have been extensively studied, beginning with early the works of \citet{Gershberg1972Ap&SS..19...75G} and \citet{Lacy1976ApJS...30...85L} which presented the energy distribution of flares resulting from the extended ground-based photometric monitoring of several targets, showing that flares occur more frequently but with lower energy on mid-M dwarves (M3--M5) compared to earlier-type (M0--M2) stars. \citet{Shakhovskaia1989SoPh..121..375S}, however, argued that the flare energy distribution depends more on the age of the star than its mass.

From a volume-limited (20 pc) survey of M--L dwarfs \citet{Schmidt2007AJ....133.2258S}  suggested that 78\% and 23\% of M and L dwarfs show chromospheric activity, respectively, with the activity level peaking around M7 and declining through mid-L objects. The \vidak{flare duty cycle} (i.e., the time ratio the star spends with flaring) was estimated to be around 5\% for late-M dwarfs and 2\% for L dwarfs.

The advent of space photometry opened up a new horizon to study flares with the huge databases of the \emph{Kepler} and \emph{TESS} observatories: one of the first such studies was done by \citet{Hawley2014ApJ...797..121H} who selected the most active and brightest late-type single M dwarfs in the \emph{Kepler} field and concluded that a power law fits well the energy distribution of flares with energies $ E > 10^{31} $erg, 
but with a steeper power-law slope compared to those found from ground-based data, suggesting that the energy distribution of flares may approach $\alpha = 2$ ($\alpha$ describing the slope), implicating coronal heating. Furthermore, they found that the flare amplitude, duration, and energy are all strongly correlated: high-energy flares typically have both long duration and high amplitude, and complex flares (having more than one peak) have longer duration and higher energy at the same amplitude. In a follow-up study, \citet{Davenport2014ApJ...797..122D} studied the most active star from the \emph{Kepler} sample, GJ 1243, and generated an analytical flare template consisting of two exponential cooling phases that are present during the white-light flare decay. The template describes the most simple flare shapes and has been widely used since then by the scientific community. 
\vidak{Recently, using the light curves of the \emph{Kepler} and \emph{TESS} satellites  combined with the LAMOST low- and medium-resolution spectral survey
\citet{Zhang2020MNRAS.495.1252Z, Zhang2023ApJS..264...17Z}
confirmed the increase of the fraction of flaring stars with spectral types from F to M type. They also found a positive correlation between the flare energy and chromospheric activity
indicators.}

For most stars, the cumulative distribution of flare energies -- the flare frequency distribution (FFD) -- is a simple power law. Due to observational constraints, this power law is limited to a given energy range. In the low-energy part of the FFD, flares become increasingly harder to detect in the presence of photometric noise, making the observed histogram incomplete. The level of incompleteness can be estimated for a given dataset using injection-recovery tests \citep[see, e.g.,][]{Gunther2020AJ....159...60G,Kovari2020A&A...641A..83K,Gao2022AJ....164..213G}. On the other hand, flares with higher energies become increasingly rare, making the high-energy end of the FFD noisy.
In some cases, the FFD shows a broken power law behavior (see Fig. \ref{fig:eieri-ffd}). According to \citet{2018ApJ...854...14M}, this occurs at a critical flare energy $E_{f_c}$, where the size of the flaring loop exceeds the local scale height. However, $E_{f_c}$ depends on local densities and local field strengths, thus it varies from star to star. This also suggests that the breakpoint for a given star is not necessarily in the observed range of flare energies, so we do not see this behavior in the flare energy distribution for all stars.
\vidak{Observational constraints, such as the fact that lower-energy flares are increasingly difficult to detect, inherently delimit the observable energy range, and $E_f$ may fall outside this range. FFDs can also show an asymptotic behavior towards the higher energy regime. This could either mean that there is a maximal energy released by a flare on a particular star, or this can be an observing bias due to the rarity of these highly energetic flares. Nevertheless, this may also affect the shape and bimodality of these FFDs and limits the observed energy range -- currently it is not clear if this break in the FFD is merely an observational effect or has a physical background. }

A study of 2-min-cadence \emph{TESS} light curves of almost 25,000 stars identified 
 more than 1200 flaring stars with half of them being M-dwarfs with events having a bolometric energy range of $\approx10^{31}-10^{37}$erg  \cite{Gunther2020AJ....159...60G}. Fast-rotating M-dwarfs were most likely to show flares, but the flare amplitude was found to be independent from the rotation period of the stars. From a larger sample of 330,000 stars and 39 \emph{TESS} sectors a similar conclusion was drawn: flare energies ranged between $10^{31}-10^{36}$ erg, and 7.7\% of the total sample showed flaring activity. 
\vidak{Flares detected on cool stars were found to have, on average, higher amplitudes due to the greater contrast between the flares and the star's surfaces \cite{Pietras2022ApJ...935..143P}. }
 
  The flaring ratio was more than 50\% among the M-dwarfs, confirming previous findings of e.g. \citet{Gunther2020AJ....159...60G}, although it is worth mentioning that the sample of stars having a 2-min cadence light curve is not random and introduces a sampling bias, given that only those targets will have short-cadence light curves that were included in an observing proposal for some reason.  
\vidak{
Similarly, \citet{YangZ2023} find that the proportion of flaring stars increases from M0 to M5, and decreases from M5 to M7.
}

Even though their quiescent luminosity is approximately two orders of magnitudes lower than the luminosity of solar-like stars, late-type stars are known to show superflares as well. Some of the prominent examples were found on the dM3- type AD Leo ($10^{34}$ erg in $U$ and $B$ bands 
\cite{Hawley1991ApJ...378..725H, Dal2012NewA...17..399D}),
the dM4-type 2MASS J00453912+4140395 ($10^{35}$erg in $B$ and $V$ bands) \cite{Kovari2007AN....328..904K}), 
the exoplanet-hosting F-type TOI 837 ($10^{34}$ erg in \emph{TESS} passband)\cite{Savanov2022AstBu..77..431S} 
or 
the dM0+dM5 binary V405 And ($10^{35}$ erg in $B, V, R, I$ bands).

An analysis of 402 stars \citep{Roettenbacher2018ApJ...868....3R} with spectral types ranging from late-F to mid-M concluded that the strongest flares do not appear to be correlated to the largest starspot group present, but are also not uniformly distributed in phase with respect to the starspot group. The weaker flares, however, do show an increased occurrence close to the starspot groups.
{A similar result was found in the case of Kepler-411, where the timing of flares and superflares were correlated with spot locations derived from planetary transit mapping \cite{Araujo2021ApJ...922L..23A}.}
One explanation for this phenomenon could be that on stars with stronger magnetic fields, reconnections between different large active nests are more frequent -- unlike on our Sun, where flares are detected mainly in bipolar regions\petra{, with stronger flares occurring preferably in regions with higher complexity \citep{Lin2023}}.
Alternatively, it is possible that the strongest flares can be observed not only when the flaring regions are located on the hemisphere facing us, but also over the stellar limb, or the largest flares could be associated with polar spots that are visible regardless of the rotational phase.
A study of periodic flaring in \emph{TESS} data found only significant periodicity in only a few percent of the studied sample and these periods seemed to be connected to the rotation period only in some cases \cite{Howard2021ApJ...920...42H}. 
However, flaring could be still connected to spots and still not show periodicity if the flaring is preferentially polar, as in the case of CM\,Dra
-- this could have positive implications for the habitability of planets orbiting M dwarfs \cite{Martin2024MNRAS.528..963M}.

While the flaring rate on our Sun is known to be correlated with the activity cycles, there is only incidental evidence for that in the case of stars, mainly due to the lack of continuous observational data -- ground-based observations are typically too sparse, and space-borne observations lack the time base needed: even the \emph{Kepler} observatory with its $\approx$4 year-long datasets is shorter than most stellar activity cycles. Nonetheless, there are some indications that there could be a similar connection on other stars: on EV\,Lac \citet{Mavridis1986A&A...154..171M} found anticorrelation between the quiescent luminosity level of the star and flaring activity during 1972--1981, and a similar indication was found later from fast-photometric observations during 1996--1999 \citep{Alekseev2000ARep...44..689A, Alekseev2005Ap.....48...20A}. A cyclic variation in flaring activity was also observed in two Pleiades members, but in this case, this was not connected to other detection of activity cycles \citep{Akopian2010Ap.....53..544A}. The method was also proposed as a possibility for finding activity cycles in \emph{Kepler}/\emph{TESS} data, however in the case of GJ1243 the constant spot- and flare activity revealed no sign of activity cycles over the studied ten years of data \citep{Davenport2020AJ....160...36D}.

\emph{Kepler} data of G-, K-, and M-type stars showed that superflares (with $E>10^{34}$ erg) are more common in cooler stars
\vidak{--  as they are more active, they have a higher number of high-energy flares in a given time compared to hotter stars, even though the latter group typically shows stronger flares --}
, and fast-rotating stars have both higher spottedness and higher (super)flare rates 
\citep{Candelaresi2014ApJ...792...67C}.
However,  while fast rotation is associated with higher activity -- the flaring activity drastically increases for stars with $P_\mathrm{rot} < 10$ days --, the highest flare rates are not found among the fastest rotators, and the largest flares were not found on the most flaring M-type stars \citep{Raetz2020A&A...637A..22R}.
The superflare frequency ($E \geq 5 \times 10^{34}$ erg in this latter study) for the fast-rotating M stars was shown to be twice higher than for solar-like stars in the same period range. The slope of the flare energy distribution was consistent with solar-like stars ($\alpha=1.84$).

Ultracool dwarfs are the lowest mass objects on the main sequence, with spectral types M7 or later. Some ultracool dwarfs are known to show (super)flares with energies up to $10^{33}$ erg, with an energy distribution similar to solar-like stars, i.e. with a power-law slope of $\alpha = 1.8$ 
\citep{Gizis2017ApJ...838...22G, Gizis2017ApJ...845...33G}. Different authors have reached the same conclusion, that although ultracool dwarfs -- favored targets for habitability studies -- flare less frequently than earlier M dwarfs, the slopes of their FFDs are similar \citep{2022MNRAS.513.2615M, 2024MNRAS.527.8290P}. Among ultracool dwarfs, one notable object is TRAPPIST-1, an M8 star that hosts seven rocky exoplanets. Its flaring behaviour was studied with \emph{Kepler K2} \cite{Vida2017ApJ...841..124V, 2018ApJ...858...55P}, the ground-based Evryscope array
\cite{2020ApJ...900...27G}, and even with the James Webb Space Telescope \cite{2023ApJ...959...64H}. Using \emph{TESS} light curves of stars similar to TRAPPIST-1 in color and absolute magnitude, the flaring activity of TRAPPIST-1 appears to be typical for its spectral type \cite{2021A&A...650A.138S}.

\section{Flares on giant stars}
\label{sect:giants_flares}

\begin{figure}
    \centering
    \includegraphics[width=0.6\textwidth]{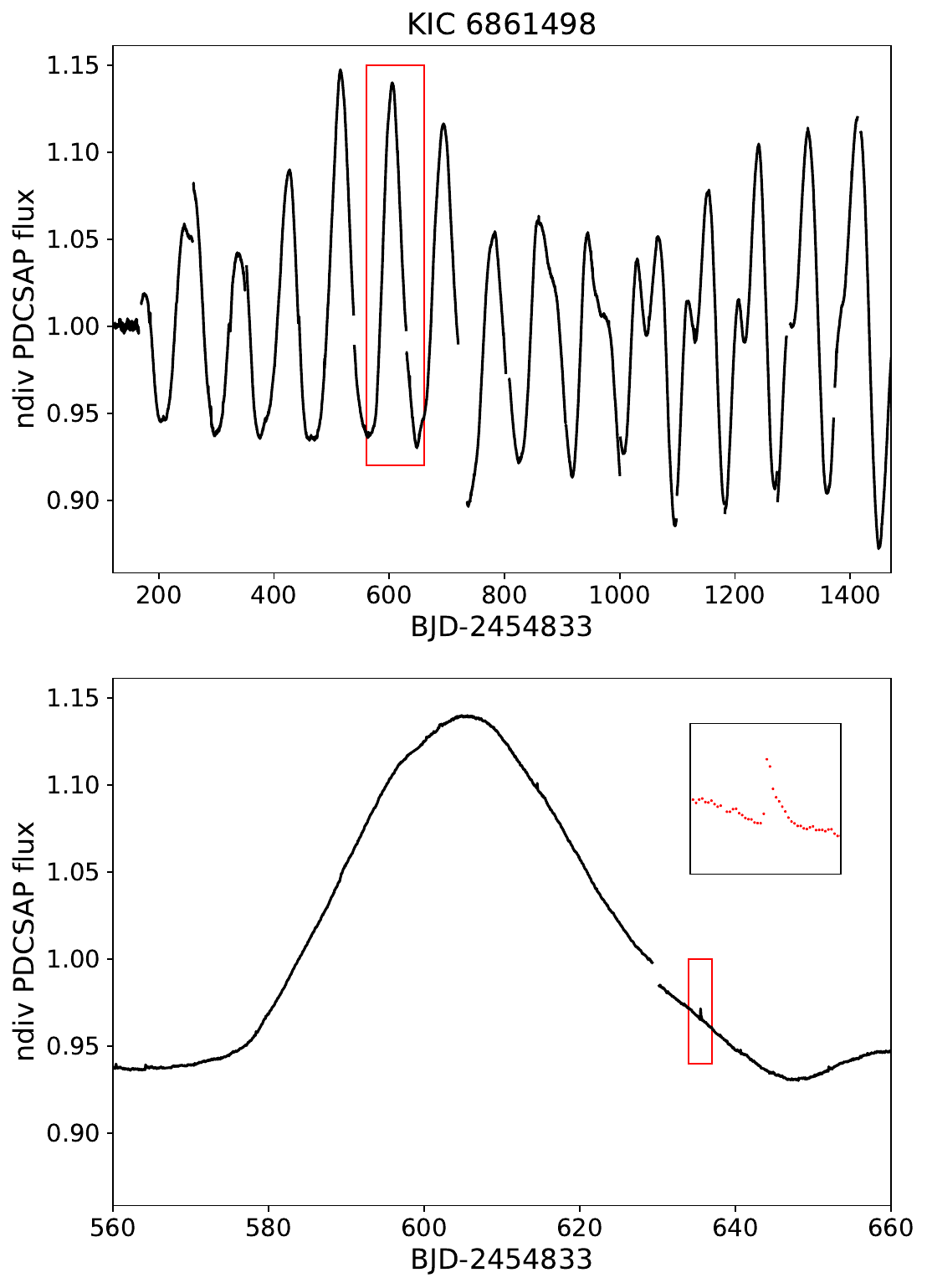}
    \caption{Example of \petra{a flare observed by \emph{Kepler}} on the giant star KIC\,6861498, which, at first glance, is not evident on the fairly high amplitude light curve. The part framed in red in the upper panel is shown enlarged in the lower panel, on which the flare has been zoomed in even further to make it visible.
    }
    \label{fig:KIC6861498_flare}
\end{figure}

The flaring activity of giant stars is known from the beginnings of systematically studying stellar activity, but their observed (few) flares were mostly detected in X-rays. The reason for this originates partly from the fact that, due to their fast evolution, there are much less active giant stars compared to their main sequence counterparts. Observationally, in the visible spectral regions, flares are not easy to detect above the higher background brightness of the giant stars; see Fig.\,\ref{fig:KIC6861498_flare} for an example. With the exception of \vidak{the Earth's polar regions} \citep[see][]{2008A&A...490..287S}  continuous monitoring of stars is impossible from the ground, and the limited signal-to-noise ratio puts another limit to the detections. However, space photometry overtook these two factors with high precision continuous monitoring of stars from weeks (\emph{TESS}) to years (\emph{Kepler}). So, the systematic search for flares included giant stars as well in some attempts. But another feature, oscillation, is widespread among red giants, and their amplitudes are in the range of the smaller amplitude flares. Therefore, less attentive examination of the search result could give rise to many false positive flares on giants and, consequently, higher flare incidence than the reality. This question is thoroughly investigated by \citet{Olah2021A&A...647A..62O}. 

In the \emph{Kepler} field the incidence of flaring stars is about 0.3\% of the giant population \citep[see][]{Olah2021A&A...647A..62O}, this means altogether 61 giant stars (with log~$g\le 3.5$). The number of oscillating red giants, on the other hand, is about ten times higher. Oscillations and flaring may appear together, and eight such cases are listed in \citet{Olah2021A&A...647A..62O}. We note that the magnetic activity may suppress the oscillations on the giant stars in close binaries as is discussed by \citet{2014ApJ...785....5G}, see also Section~\ref{sect:binaries}.

\begin{figure}
    \centering
    \includegraphics[width=0.5\textwidth]{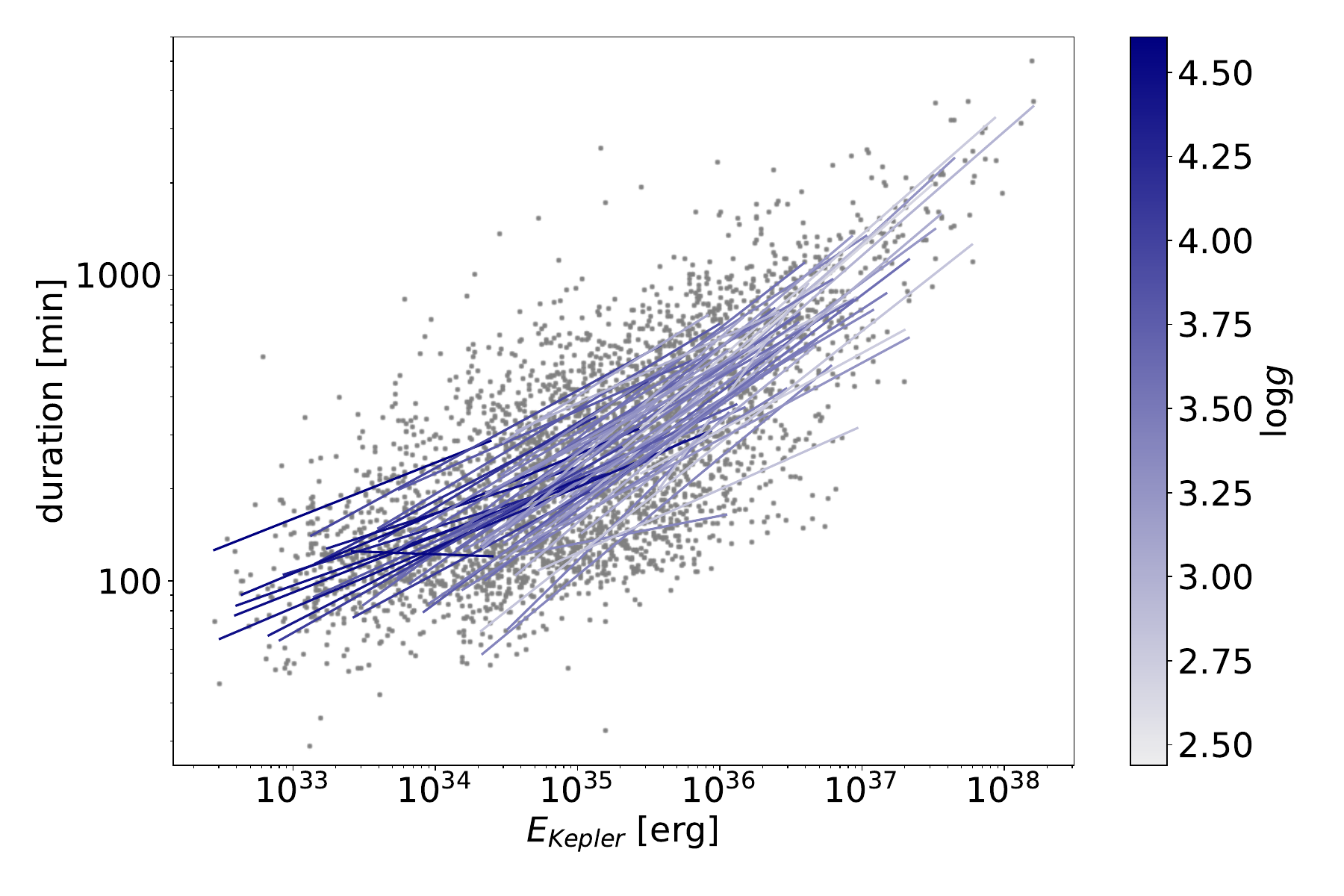}\includegraphics[width=0.5\textwidth]{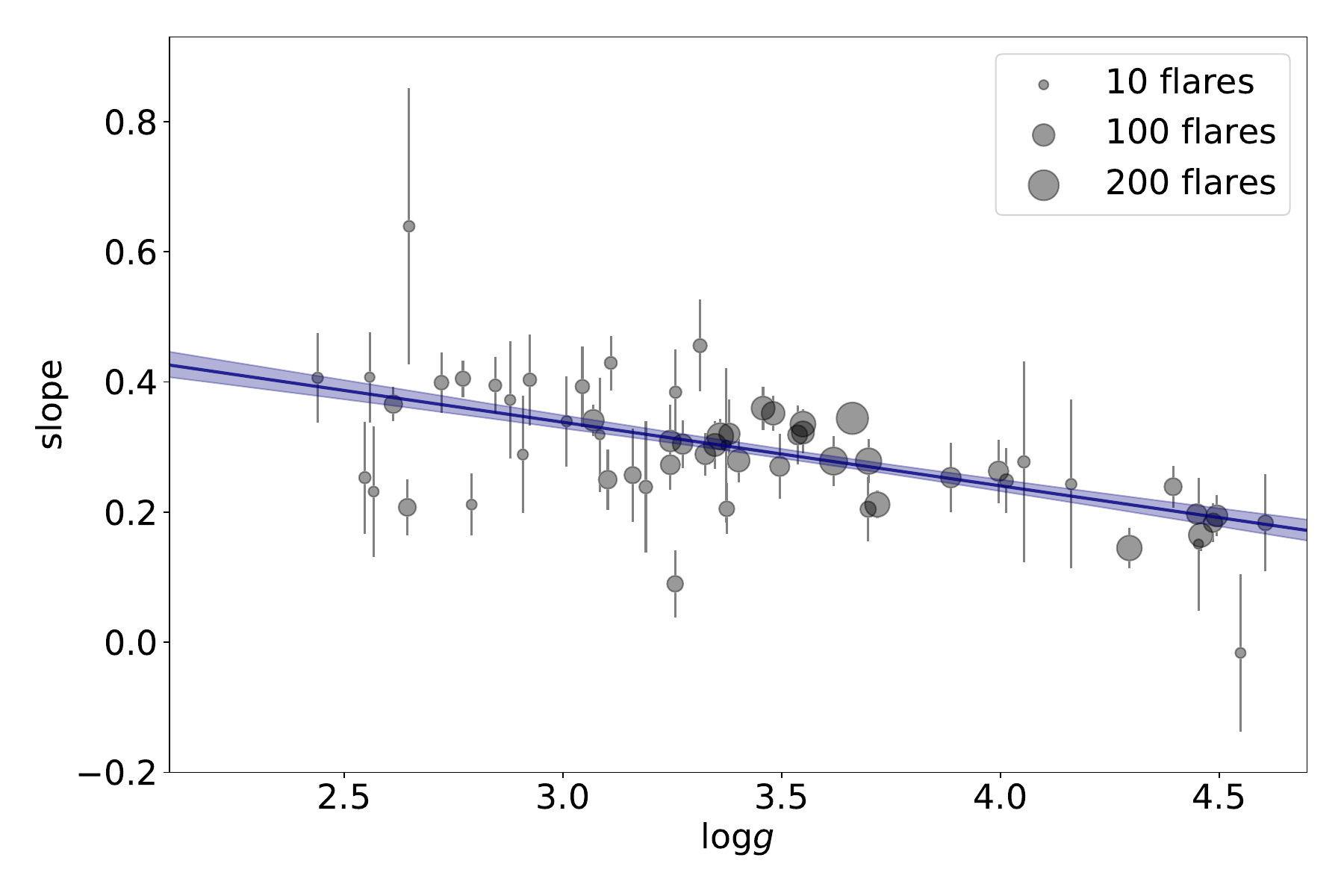}
    \caption{Flare duration vs. flare energy for a sample of stars with different surface gravity values in the \emph{Kepler} field \citep{Olah2022A&A...668A.101O}. Left: flare duration values for flares in the \emph{Kepler} 30-min sample as a function of flare energy on a log-log scale with linear fits to the values of individual stars. Right: slopes of the fitted lines from the left panel as a function of $\log g$. Error bars show the uncertainties of the slopes, and the point sizes represent the number of flares observed on each star. The blue line shows a linear fit to the slopes.
    }
    \label{fig:flare_energy_duration}
\end{figure}

The shapes of the flares on giant stars are similar to the ones observed on the dwarfs, while their energies cover the range of about a few times $10^{32}-10^{38}$ ergs in the \emph{Kepler} bandpass, are shown by \citet[][see their Figs.\,5--6]{Olah2022A&A...668A.101O}, that is, they release more energy that the most energetic solar flares. As it is also shown in the cited paper, the relation between the energy and duration of the flares on giants seems to be not linear as is generally thought in the case of main sequence stars \cite{2015EP&S...67...59M}. Instead, a dependence on surface gravity is suggested for the scaling law of the energy--duration relation of \citet{Olah2022A&A...668A.101O}  
(Fig.\,\ref{fig:flare_energy_duration}).

\section{Flares in binary systems}
\label{sect:binaries}

In the case of close binaries (e.g., BY\,Dra or RS\,CVn type systems), the role of binarity in the stellar activity, and thus in the formation of flares, cannot be ignored in any way.
Tidal forces in close binaries are known to affect magnetic activity in several ways. The first and most important effect is the maintenance of rapid rotation for a longer time \vidak{than for} single stars during stellar evolution, which is known to be one of the key elements of magnetic activity at higher levels. The synchronicity in close binaries counteracts magnetic braking, which otherwise effectively slows down the rotation of single stars. On the other hand, the rotation of the star also slows down during stellar evolution (i.e., when turning off the main sequence to become a red giant), but synchronicity also works against this. The "long-period" group of RS\,CVn binaries typically consists of systems with such evolved but still rapidly rotating (sub)giant components.

\begin{figure}
    \centering
    \includegraphics[width=1.0\textwidth]{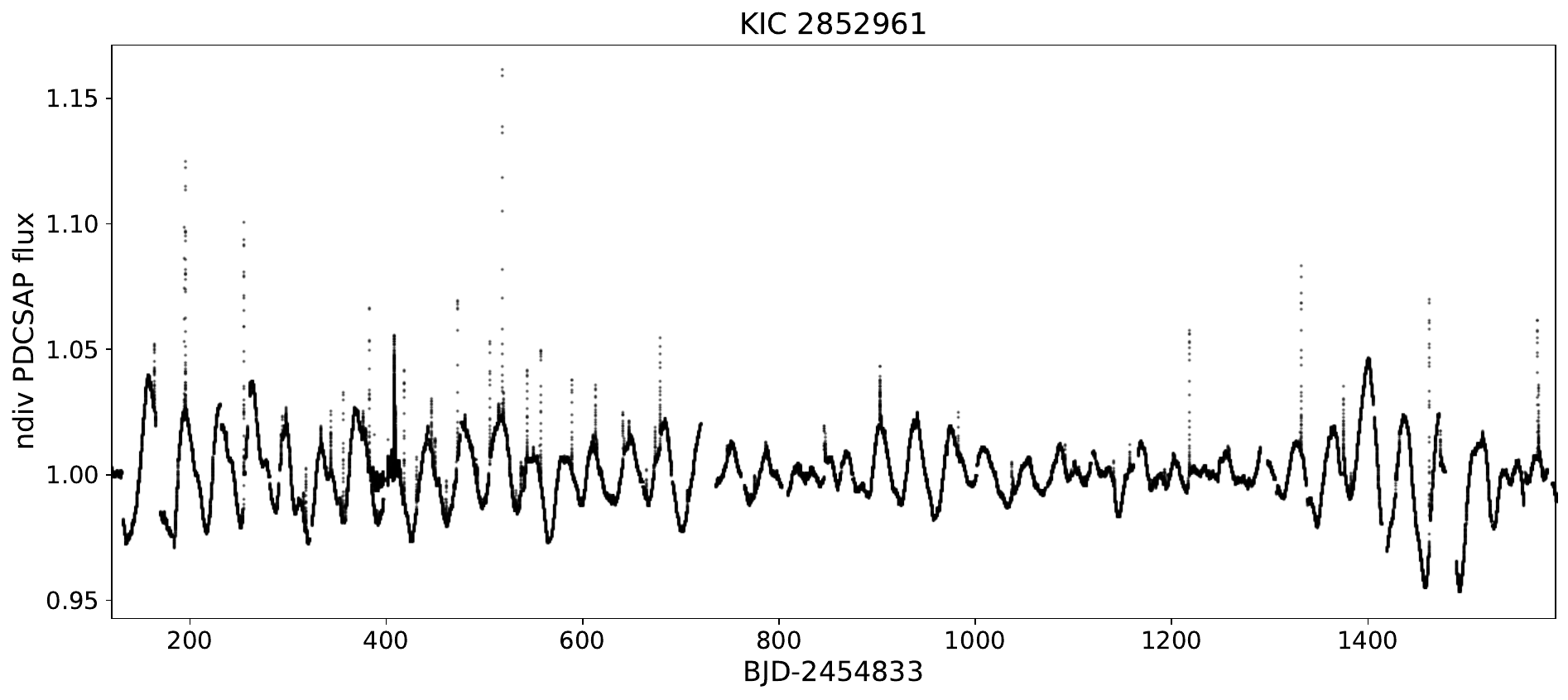}
    \caption{Fifty-nine confirmed flare events \vidak{-- sharp increases of intensity on the timescale of a few minutes--hours --} on the $\sim$K0 giant primary of the SB1 binary KIC\,2852961, based on the long-cadence \emph{Kepler} data \citep{Kovari2020A&A...641A..83K}. Note the changing amplitude of the rotational light curve due to the constantly evolving spotted surface.
    }
    \label{fig:kic2852961_lc}
\end{figure}

The difficulties of observing flares on active giants have already been mentioned in Sect.\,\ref{sect:giants_flares}. A significant advance was the rise of space photometry, which was able to prove that flares are also common on red giant stars; see Sect.\,\ref{sect:giants_flares} in this paper. In a recent study, \citet{Olah2021A&A...647A..62O} found that the majority of flaring red giants in the \emph{Kepler} database likely belong to close binary systems. Hence, the presence of flares on a fast rotating ($P_{\rm rot}$<50\,d) red giant star may be a further indication that the star is a member of a close binary system, as suggested by \citet{Gehan2024arXiv240113549G}. In the following, through some examples, we focus on those cases that can be specifically associated with the close binary nature.

\begin{figure}[t]
    \centering
    \includegraphics[width=0.7\textwidth]{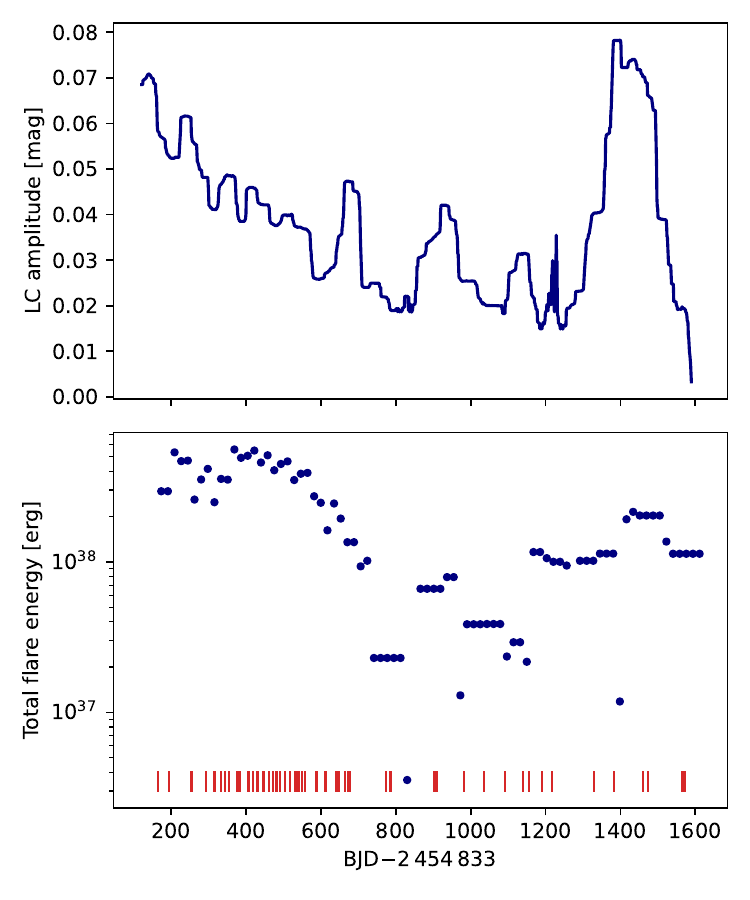}
    \caption{Correlation between the light curve amplitude variation of KIC\,2852961 (shown in Fig.\,\ref{fig:kic2852961_lc}) and the total energy released by flares according to \citet{Kovari2020A&A...641A..83K}. The top panel shows the moving average of the amplitude within 3$P_{\rm rot}$ time interval, while the bottom panel indicates the total flare energy release within the same interval. Tick marks at the bottom are the peak times of the flares.
    }
    \label{fig:kic2852961_ampen}
\end{figure}

Based on the paradigm of the dynamo mechanism, we expect the increased activity on such stars to result in a stronger magnetic field, that is, more spots and therefore, a higher rate of flare activity as well. The connection between spot activity and the frequency (or total energy) of flaring is well exemplified by the finding of \citet{Kovari2020A&A...641A..83K}, according to which there is a clear correlation between the amplitude of the rotational brightness variability and the overall energy release by flares; see Fig.\,\ref{fig:kic2852961_ampen}. According to this result, the rotation amplitude of KIC\,2852961 as an indicator of the level of magnetic activity is changing together with the overall magnetic energy released by flares within the same period. This result tells us that larger active regions on the stellar surface produce more and/or higher energy flares, while when there are fewer and/or smaller active regions, the flare activity decreases as well.

One of the most remarkable manifestations of the effect of binarity on activity is the appearance of active longitudes (i.e., spot concentrations) on one or both components of the system at certain phases bound to the orbit \citep[e.g.,][]{Olah2006Ap&SS.304..145O}. Active longitudes require breaking the axial symmetry. This can be explained by the fact that non-axisymmetric dynamo modes are expected to develop due to tidal effects in the binary system \citep{Moss1997A&A...321..151M}. Another condition for the excitation of stable, non-axisymmetric fields is that the differential rotation is not too strong \citep{Moss1995A&A...294..155M}. We note that in the case of close binaries, this condition is usually given because in such systems the differential rotation is confined anyway due to tidal forces \citep{Kovari2017AN....338..903K}.

\begin{figure}[t]
    \centering
    \includegraphics[width=0.6\textwidth]{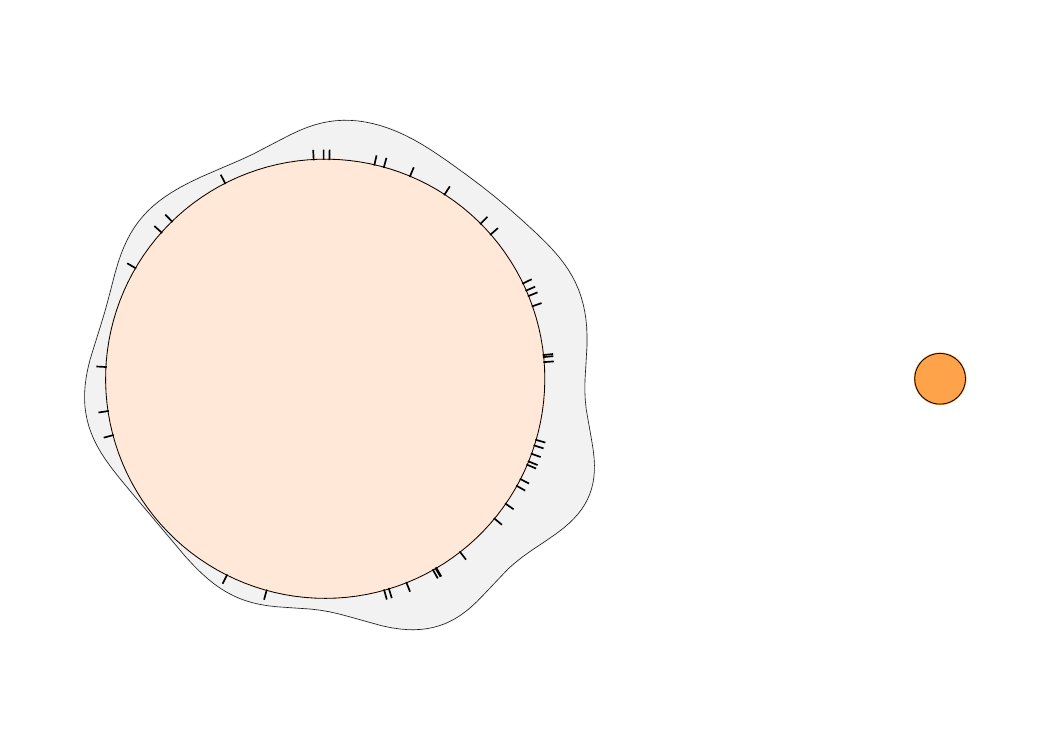}
    \vspace{-0.6cm}
    \caption{Distribution of flares detected along the orbital phase in the single-lined spectroscopic binary EI\,Eri from \citet{Kriskovics2023A&A...674A.143K}. The active G5 subgiant component is shown on the left, and the unseen companion on the right. The tick marks around the G5 star show the phase values of the individual flare detections. \kovari{The polar distribution function of the detections, plotted as a solid line above the surface indicates that the spatial distribution of the flares is not random.}
    }
    \label{fig:eieri_flares}
\end{figure}

Although it is essentially impossible to \vidak{infer the stellar surface position of the flares} based on photometric time series, we can make estimates of their longitudes.
First of all, we can assume that the source of a given flare is an active (spotted) region currently present on the visible hemisphere of the star \citep[see, e.g.][]{Pi2019ApJ...877...75P}. And secondly, we can also assume that the rotation phase of a given flare event corresponds statistically to the longitudinal coordinate of the source active region. With the appearance of active longitudes and according to the correlation between active regions and flares, it follows that in such cases the distribution of flares along the rotational phase is not uniform. We present such an example in Fig.\,\ref{fig:eieri_flares}, which indeed shows a non-uniform phase distribution of the flares of the active close binary EI\,Eri, a synchronized system of an active G5 subgiant and a faint M dwarf, based on \emph{TESS} observations \citep{Kriskovics2023A&A...674A.143K}. In this case, it seems that the position(s) of the active flaring region(s) are related in some way to the orbit, meaning that the close companion influences the magnetic-flux emergence, ultimately the occurrence of flares on the G5IV star \citep{Holzwarth2003A&A...405..291H,Holzwarth2003A&A...405..303H}.

\begin{figure}[t]
    \centering
    \includegraphics[width=0.7\textwidth]{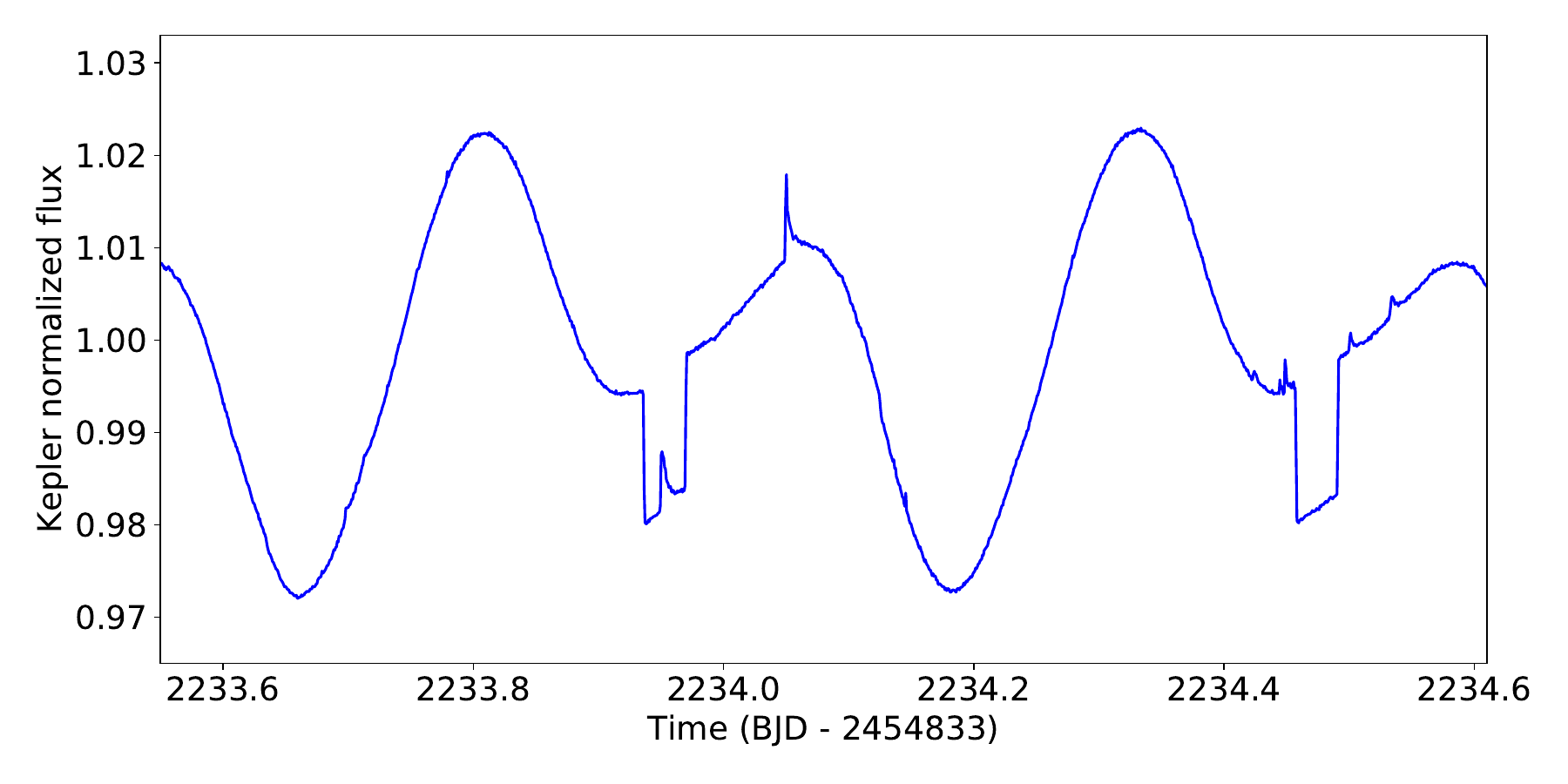}\\
    \includegraphics[width=0.7\textwidth]{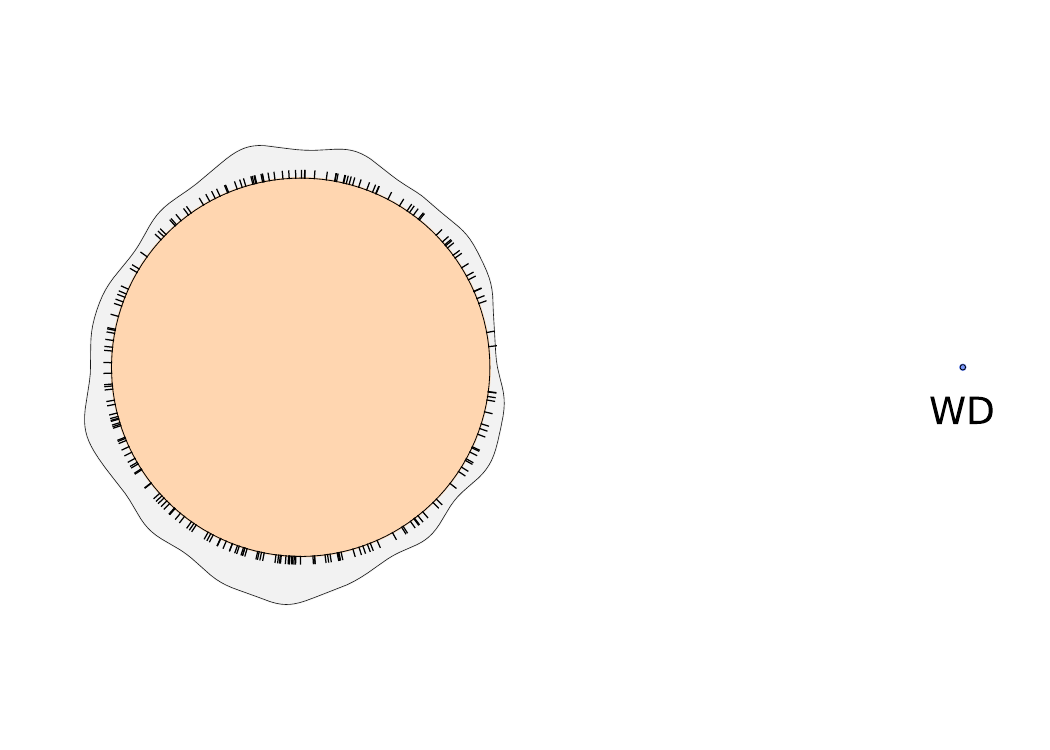}
    \caption{Flare detections in the red dwarf+white dwarf eclipsing binary system V471\,Tau from the \emph{Kepler} K2 mission data \citep{Kovari2021A&A...650A.158K}. In the top panel, as an example, we show a segment of the \emph{Kepler} K2 light curve, on which a flare occurs just when the white dwarf component is obscured. In the lower panel the scaled diagram of the system is shown with the phases of the 198 flare observations marked around the red dwarf component. Although the \kovari{polar distribution function above the surface} \vidak{is not definitive}, it rather suggests that the occurrence of flares is random.
    }
    \label{fig:v471tau_flares}
\end{figure}

At this moment it is not yet clear to what extent the overall flare occurrence depends on the tidal force of the companion, and to what extent it is influenced by other circumstances such as orbital eccentricity, (a)synchronism, the strength of the differential rotation, mutual activity or, for instance, interbinary magnetic coupling, etc. The case of V471\,Tau is an example of how complex the problem could be. The close binary system of 0.52\,d orbital period is formed by an active K5 dwarf and a white dwarf. The synchronously rotating K5V star has a permanent active region that has been observed for many years; a dominant cool spot that is mostly present at longitudes facing the white dwarf \citep[see][and their references]{Kovari2021A&A...650A.158K}. Despite having this permanent active longitude, the phase distribution of the flare observations shown in Fig.\,\ref{fig:v471tau_flares} does not seem to show any remarkable asymmetry. Moreover, even based on the latest \emph{TESS} data, it does not seem as if there is any kind of phase dependence. However, a possible explanation for this could be, on the one hand, a purely geometric effect, according to which the permanent active region also extends to the pole, so it is mostly visible continuously. Another possibility is that the distribution of spots (activity nests, i.e. sources of flares) along the rotation is roughly uniform, but this is in contrast to the observational fact of the permanent active region facing the white dwarf. Nevertheless, there are other observations that show phase independence of flare occurrence \citep[e.g.][]{Smelcer2017MNRAS.466.2542S}. In order to learn more thoroughly to what extent the phase distribution of flares is influenced by a nearby companion star, it is obviously necessary to examine additional systems.
\vidak{We note, that planetary systems may also influence the activity of stars as is suggested in the case of the Sun, where activity cycles were suggested to be influenced by planetary motions (see, e.g., the works of \citet{TanCheng2013, Stefani2024}).}

A special group of active binaries are those close binaries whose both components are magnetically active, so that their magnetospheres can interact directly with each other. In such cases, the rotation and orbital motion of the stars can lead to a very slow overall winding of the coupled magnetic fields, allowing a gradual accumulation of magnetic energy that can eventually be released as an "interbinary flare" through instability \citep[][]{Cherkis2021ApJ...923...13C}. In the case of the \petra{RS\,CVn-type system} UX\,Ari (G5V+K0IV), something similar presumably happened when the corotating, giant magnetic loops present on both stars interacted and flux tubes temporarily connected the two stars \citep{Simon1980ApJ...239..911S}. Furthermore, such connecting flux loops can pave the way for mass transfer between the two stellar components \citep[see, e.g.][]{Ferreire2005A&A...433.1055F}, which can even be observed in the form of orbital period changes \citep[cf.,][]{Decampli1979ApJ...230..815D}. Traces of this process have been observed in many cases, see the paper by \citet{Hall1980AcA....30..387H}, which lists several such binaries.

An interesting scenario is represented by the case of DQ\,Tau, a double-lined spectroscopic binary \petra{consisting} of two $\sim$0.65$M_{\odot}$ mass pre-main sequence stars on \petra{an} eccentric orbit \citep{Mathieu1997AJ....113.1841M}. The stars are therefore not constantly close enough to each other, however, at periastron encounters the magnetospheres of the components collide and trigger each other, resulting in magnetic reconnection and so recurring flares in these orbital phases \citep{Salter2010A&A...521A..32S}.

Close but detached binaries in which the coronae of the components permanently overlap each other deserve special attention. The presence of such interbinary coupling in the form of large coronal loops, comparable in size to the binary separation, is supported by the enhanced X-ray emission of a significant proportion of RS\,CVn type binaries \citep[e.g.,][]{Walter1980ApJ...236..212W,Pres1995MNRAS.275...43P,Siarkowski1996ApJ...473..470S}. 
It is still not completely clear what the role of these interbinary magnetic fields is in the flare activity. However, it is very likely that these fields are not only responsible for the total quiescent X-ray luminosity, but may also be the source of energetic bursts. The coronal connection can indeed build complex magnetic topologies, thus enabling the occurrence of X-ray flares, as suggested in their recent study by \citet{Singh2022ApJ...934...20S}.

Finally, special mention should be made of the cases where the closely interacting object is a substellar object, e.g., a hot Jupiter. In such systems, similarly to star--star systems, the planetary companion also acts on the star, although to a lesser magnitude. Therefore, planet-triggered flares are expected to correlate with the planet's orbital period. Such a possible clustering of stellar flares along the orbital phases of the planets was investigated by \citet{Ilin2024MNRAS.527.3395I} based on \emph{Kepler} and \emph{TESS} time series. In this tentative study, both tidal and magnetic interactions were considered, and magnetic interaction was found to be more dominant in younger systems, while tidal interaction was more dominant in older ones. Among the 25 systems examined in detail (including TRAPPIST-1, AU\,Mic, Proxima Cen etc.), a young, hot Jupiter host system, HIP\,67522, showed the most prominent clustering of flares along the orbital phase, consistent with remarkable magnetic star--planet interaction, with tidal interaction being less dominant. In accordance with the case of HIP\,67522, the magnetic interaction is expected to be the most significant, when the planet is deeply embedded in the sub-Alfv\'enic zone of the host-star, i.e. within the Alfv\'en radius, at which the stellar wind velocity exceeds the Alfv\'en velocity of the magnetized plasma.
Within this region, a close-in planet can have a significant effect on the structure of the stellar corona, and the resulting magnetic interaction can even generate flare events \citep[e.g.,][]{Shkolnik2008ApJ...676..628S,Lanza2009A&A...505..339L,Cohen2011ApJ...733...67C}.

\section{Stellar CMEs}
\label{sect:CMEs}
As mentioned in the introduction, there are several methods how to detect stellar CMEs. In principle, those can be distinguished into methods that are based on the solar--stellar analogy, such as radio bursts and coronal dimmings, into direct methods, such as the Doppler shifted emission/absorption, and other indirect methods such as continuous X-ray absorption.

The Doppler-shifted emission/absorption is a direct signature of plasma moving away from a star. This Doppler-shifted emission/absorption in optical spectra is a signature of an erupting prominence/filament on the Sun. Already \citet{Den1993} and \citet{Ding2003} have shown that erupting filaments can be recognized in H$\alpha$ spectra as absorption features. A spatially integrated  H$\alpha$ spectrum of an erupting filament is shown in \citet{Ichimoto2017}. Only recently, the Sun-as-a-star signature of erupting filaments has been investigated further using data from SMART/SDDI \citep{Namekata2022b, Otsu2022, Otsu2024} and \petra{Mees} MCCD \citep{Leitzinger2021}, showing spatially integrated H$\alpha$ signatures of selected erupting filaments and prominences, surges, and flare-related plasma motions. 

\begin{figure}[t]
    \centering
    \includegraphics[width=0.53\textwidth]{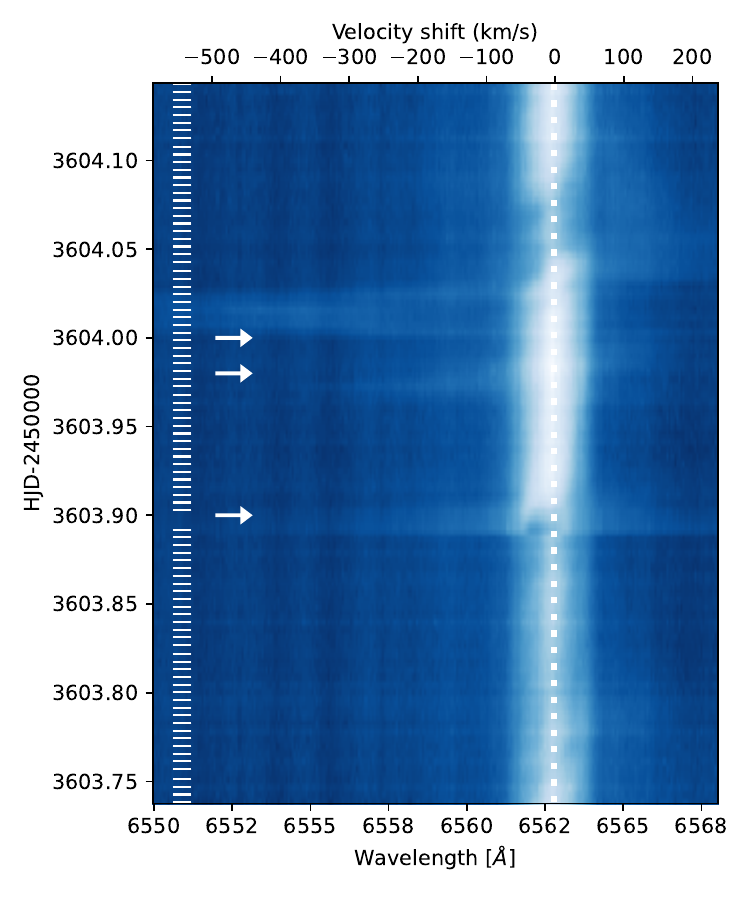}
    \includegraphics[width=0.40\textwidth]{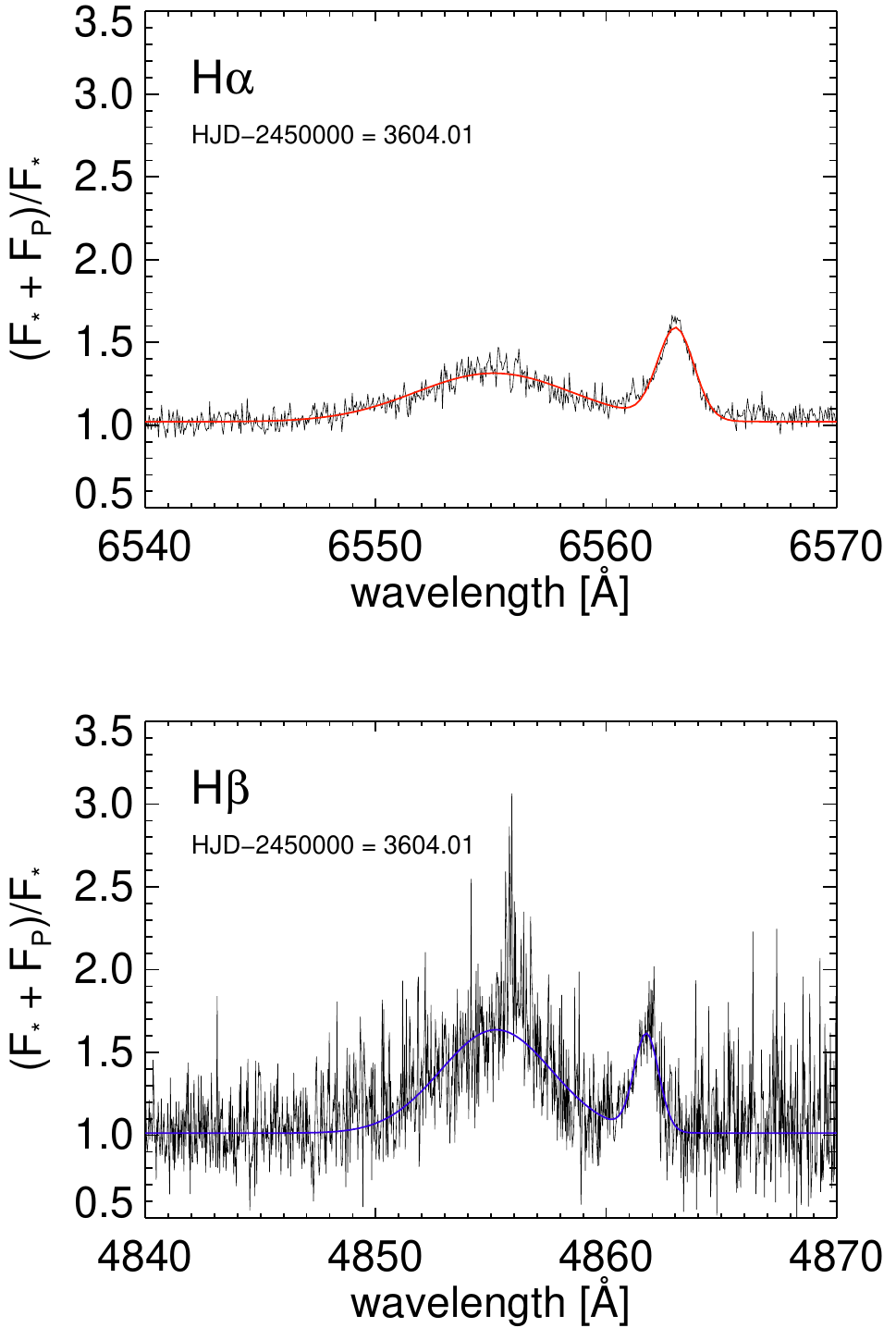}
    \caption{Left panel: Complex coronal mass ejection event on V374 Peg -- the event can be seen as a distortion of the blue wing in the H$\alpha$ line \cite{Vida2016}. Marks on the left show the times of the spectra, and arrows note the times of the outbreaks. Right panel: Pre-event normalized spectra from HJD-2450000=3604.01 in H$\alpha$ and H$\beta$ \citep[see][]{Leitzinger2022} corresponding to the fastest event detected on V374~Peg shown in the left panel \vidak{around HJD 2453604.00, marked with the top arrow}.}
    \label{fig:v374peg_cme}
\end{figure}

On stars, a Doppler-shifted emission signature has been interpreted for the first time as a mass ejection in \citet{Houdebine1990} \petra{when a sequence of broad extra emissions occurring on the blue side of the H$\gamma$ line at the onset of a strong flare} was found on AD~Leo. This is still the fastest event ever detected using the method of Doppler-shifted emission/absorption, with a projected bulk velocity of $\sim$3000~km~s$^{-1}$ \cite{Leitzinger2022c}. The related mass was estimated to be in the order of $\sim 10^{18}$~g. Another event was then presented by \citet{Guenther1997} on an M-type weak-line T-Tauri star, showing a projected bulk velocity of $\sim$600~km~s$^{-1}$ and a prominence mass in the order of $10^{18}$--$10^{19}$~g. Both the \citet{Houdebine1990} and \citet{Guenther1997} events were certainly events which left the star, as both were above the stars' escape velocities. A 250~km~s$^{-1}$ emission event was detected on the dM star AT Mic \petra{\citep{Gunn1994}} but interpreted by the authors as an evaporation event, not as a prominence eruption. A fast ($\sim$400~km~s$^{-1}$) and more complex event was presented in \citet{Vida2016} on V374~Peg, an active dMe star (Fig. \ref{fig:v374peg_cme}), with a mass in the order of 10$^{16}$~g. Up to then, only emission features on predominantly dM stars have been presented in the literature. \martin{\citet{Senavci2018} present activity investigations of the RS\,CVn eclipsing binary SV\,Cam (F9V+K4V) and find excess absorption near the secondary minimum which the authors interpret as cool plasma (filaments/prominences) obscuring the primary component.} Then \citet{Namekata2021} present the first detection of an absorption signature on a solar analog occurring in the blue wing of H$\alpha$ on the solar analog EK~Dra, a nearby star that resembles the young Sun at an age of $\sim$100~Myr. The presented Doppler-shifted absorption shows a maximum bulk velocity of $\sim$510~km~s$^{-1}$ and a mass of $\sim$10$^{18}$~g. The event also involves back-falling material which is observed also on the Sun for many erupting filaments. Complementary, the final evolution of this event is presented in \citet{Leitzinger2024}, revealing that the back-falling filament reaches zero velocity until it dissolves. \petra{Later} also erupting prominences have been reported on EK~Dra \citep{Namekata2024}, where the maximal projected bulk velocity is close to the escape velocity of the star. There are coordinated X-ray observations of \petra{one of these optical events}, revealing a possible post-flare dimming.

\begin{figure}
    \centering
    \includegraphics[width=\textwidth]{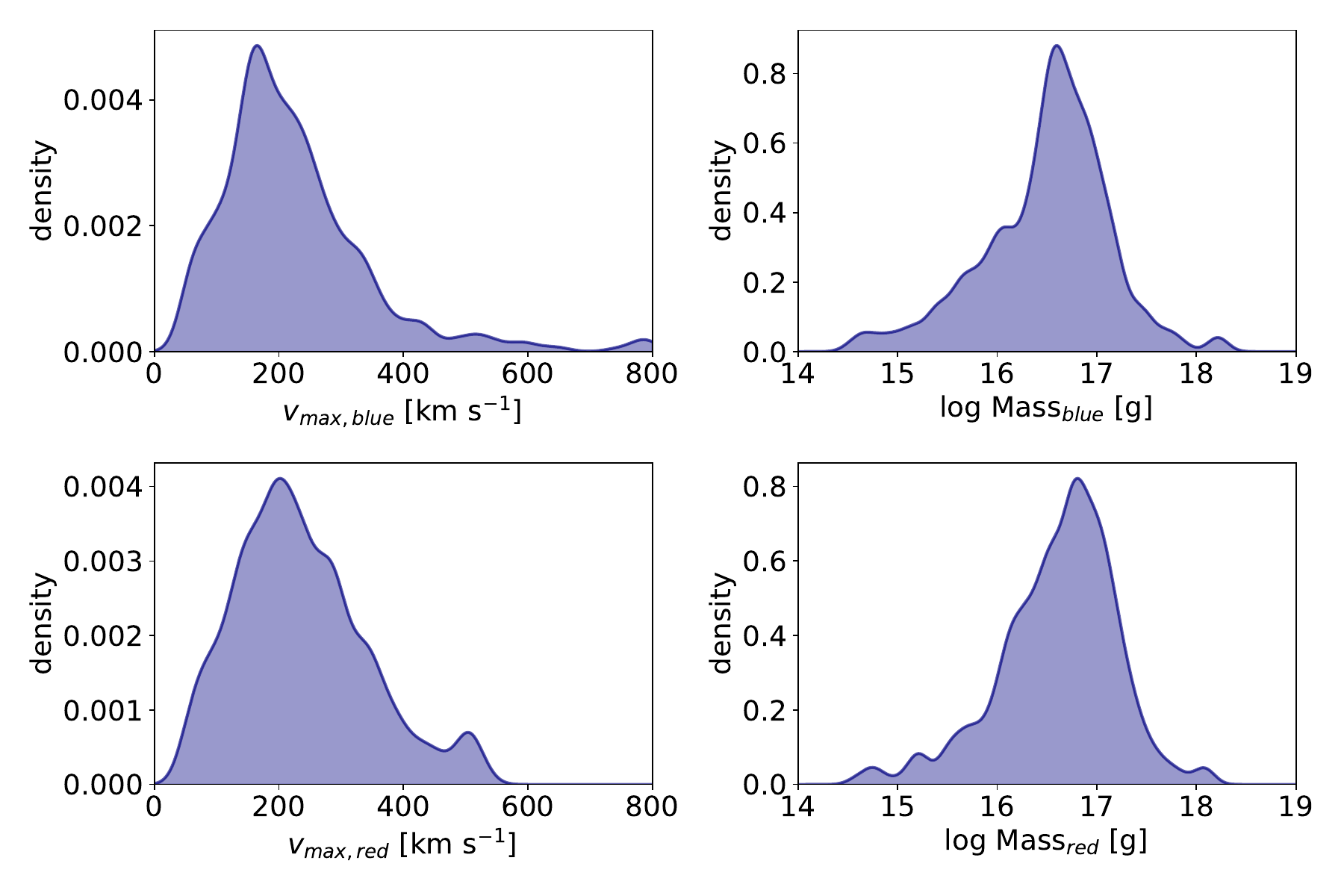}
    \caption{
    \seli{Parameters from the \petra{blue and red enhancement of the H$\alpha$ line asymmetries possibly related to} stellar coronal mass ejections from \cite{Vida2019A&A...623A..49V}. Each panel shows a Gaussian kernel density estimate, with 20 km\,s$^{-1}$ bandwidth for the velocity shifts, and 0.1 for $\log$\,Mass.
    \vidak{The blue and red enhancements are probably caused by the Doppler-shifted emission of the ejected/falling back material, respectively.}
}}
    \label{fig:cme_parameter_dist}
\end{figure}

There are many more investigations that report on rather slow blue- or red-wing asymmetries found in optical spectroscopic observations of late-type stars which can be only suspected to be erupting events. Such events have been detected on dM stars \citep{FuhrmeisterSchmitt2004, Honda2018, Muheki2020a, Muheki2020b, Maehara2021, Johnson2021, Wang2021, Wang2022}, dK stars \citep{LopezSantiago2003}, and weak-line T-Tauri stars \citep{Hill2017}. 
Up to now, we have mentioned studies that used a single-star observational approach. To increase efficiency, observing several stars simultaneously is an alternative approach. Such multi-object observations have been performed by \petra{\citet{Guenther1997, Leitzinger2014, Korhonen2017} and \citet{Vida2024b}}. Furthermore, also dedicated searches in data archives have been performed. The Polarbase archives have been searched for optical signatures on dM \citep{Vida2019} and solar-like stars \citep{Leitzinger2020}. 
\vidak{These revealed no events on solar-like stars, but detected more than 400 line asymmetries on 25 dM stars.
In most cases the detected velocities did not reach surface escape velocity ($\approx$580\,km\,s$^{-1}$): the typical observed maximum velocities are on the order of 100--300\,km\,s$^{-1}$, while the typical masses of the ejecta were on the order of $10^{15}-10^{18}$\,g (see Fig.\,\ref{fig:cme_parameter_dist}). The line asymmetries were found to be more frequent on cooler stars with stronger chromospheric activity.
}
\citet{Fuhrmeister2018} used the archive of ``Calar Alto high-Resolution search for M dwarfs with Exoearths with Near-infrared and optical Echelle Spectrographs'' (CARMENES), \citet{Koller2021} used the Sloan Digital Sky Survey (SDSS) Data Release 14, and \citet{Lu2022} used the LAMOST Medium-Resolution Spectroscopic Survey (LAMOST-MRS) to search for signatures of erupting filaments/prominences in optical data. The all-sky data archives of SDSS and LAMOST revealed a handful of events each, whereas the Polarbase and CARMENES archives showed more events. 

The method of Doppler-shifted absorption/emission has also be used at other wavelengths than the optical. \citet{Leitzinger2011a} searched the archive of the Far Ultra-violet Spectroscopic Explorer (FUSE) for Doppler-shifted absorptions/emissions. One slow event (projected velocity of 84~km~s$^{-1}$) on AD~Leo was detected in the OVI doublet at 1032~\AA{} in one spectrum after a flare. \citet{Argiroffi2019} identified in Chandra data a similar slow event in the OVIII line at X-ray wavelengths on the magnetically active giant star HR 9024 (G1 III) also after a flare, which the authors interpreted as a possible CME. They also identified up and down flows in flare loops with projected velocities in the range of 100--400~km~s$^{-1}$. The flare event presented by \citet{Argiroffi2019} shows an exceptional duration of $\sim$1~day. Also using Chandra data, \citet{Chen2022} investigated the young and active dMe star EV~Lac. In several flares, the authors also measured up and down flows of $<$130~km~s$^{-1}$. In one flare a decreasing plasma density was found which the authors interpreted as a possible filament eruption.

There have been also investigations of the Sun-as-a-star signature of erupting filaments in the EUV. \citet{Xu2022} report on spectral line asymmetries seen in OIII, OV, and OVI during a solar eruption using SDO/EVE observations, i.e. EUV Sun-as-a-star observations. Using imaging information the authors could relate the asymmetries with a CME on the Sun. \citet{Lu2023} report on eight CME and filament eruption events where "blue-wing" asymmetries {(i.e., asymmetries towards the shorter wavelengths)} were found in the OIII spectral line. Also \citet{Otsu2024} report on similar events observed in addition in H$\alpha$ and present "blue-wing" asymmetries in the OV and OVI spectral lines related to a filament eruption. This demonstrates the potential for stellar CME detections with EUV spectroscopy. 

Based on the solar--stellar analogy, there are two types of radio bursts that are routinely used to search for CMEs on stars. These are the radio type II and moving type IV bursts. To identify those, radio multi-channel receiver observations are needed, as only in dynamic spectra the frequency-dependent structure can be seen. But already in the 1990s single channel receivers \petra{observing} at very few frequencies were used to identify stellar activity from the prominent \petra{active star EV~Lac} using the Ukrainian T-shaped radio array second modifications (UTR-2) \petra{\citep[e.g.][]{AbdulAziz1995, Abranin1998}}. Some years later\petra{, several studies} \citep{Leitzinger2009, Boiko2012, Konovalenko2012} repeated those observations but this time with multi-channel receivers, resulting in a number of bursts, the majority being similar to solar type III bursts and some having a high probability to be of stellar origin. Again a few years later the search for type II bursts continued, but either no bursts were detected \citep{Crosley2016} or the deduced parameters did not match expected parameters from coronal models which led the authors to conclude that CMEs on other stars than the Sun are rare \citep{Crosley2018a, Crosley2018b}. Following that, \citet{Villadsen2019} also searched for type II radio bursts in active stars, and although they detected several bursts, they state that none of them are type II-like. \citet{Mullan2019} hypothesize that on dM stars, due to the powerful magnetic fields, CMEs must have unrealistic high velocities to produce type II emission. One year later \citet{Zic2020} presented a radio burst on Proxima Centauri which the authors assign to be a stellar analog of the solar radio type IV burst because of its polarization and temporal structure. Already in the eighties of the last century \citet{Kahler1982} presented radio light curves suspecting a type-IV-like burst associated with a flare. Only recently \citet{Bloot2024} presented a 250-hour-long investigation of the planet-hosting young and active star AU~Mic using the Australia Telescope Compact Array in a frequency band between 1.1 and 3.1~GHz. These authors found a variety of bursts on AU~Mic and present a unique classification, but confirm no analogs of solar type-II or type-IV occurrences. \petra{Even} more recently, \citet{Mohan2024} observed AD~Leo for only 8~hours with the uGMRT and found a type III burst and in the post-flare phase a highly polarized type-IV burst which the authors suspect to be potentially correlated to a CME.

\begin{figure}[t]
    \centering
    \includegraphics[width=0.7\textwidth]{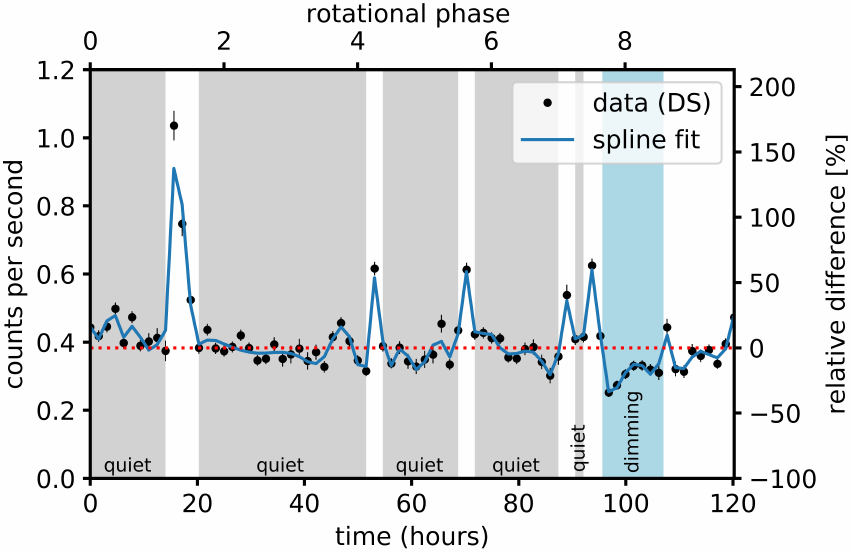}\\
    \caption{EUVE DS light curve of the active K-dwarf AB~Dor showing several flares and one dimming occurring at the end of the light curve after the last larger flare \petra{(adapted from \citet{Veronig2021})}. Grey-shaded areas denote data points from which the quiescent level of AB~Dor was determined.}
    \label{fig:abdor_euve}
\end{figure}

Coronal dimmings -- evacuated \petra{coronal} regions after large flares -- are a rather recently developed and applied methodology as CME tracers. \citet{Veronig2021} established the Sun-as-a-star signature of solar EUV dimmings using Extreme ultraviolet Variability Experiment (EVE) data using a sample of $\sim$60 solar events. Based on \petra{the resulting close relation between dimmings and CMEs,} these authors used stellar X-ray (\petra{X-ray Multi-mirror Mission/XMM-Newton} and Chandra) and EUV (Extreme UltraViolet Explorer/EUVE) databases to search for flares that possibly show coronal dimmings. This search yielded 21 events that showed a \petra{significant} flux depression after the flare (see Fig.~\ref{fig:abdor_euve}). \citet{Loyd2022} performed a coronal dimming search on the dK2 star $\epsilon$~Eri in three FUV flares in Hubble Space Telescope (HST) data, but found no convincing dimming event and therefore derived upper limits of masses of possibly associated CMEs \petra{of a few $10^{15}$~g for 1~MK plasma}.

The method of continuous X-ray absorptions dates back to an interpretation of increased hydrogen density during an X-ray flare having occurred on Proxima Centauri \citep{Haisch1983}. These authors interpreted the increase in hydrogen density as the passage of an erupting filament obscuring the flaring region. Since then there have been several detections of this phenomenon not only on main-sequence stars \citep[Proxima Cen, V773 Tau][]{Haisch1983, Tsuboi1998, Briggs2003} but also on binary stars \citep[e.g. Algol][]{Ottmann1996, Favata1999} and RS CVn systems \citep[e.g. UX~Ari][]{Franciosini2001, Pandey2012}. The event on Algol was reinvestigated by \citet{Moschou2017} who determined the parameters of a possible related CME by applying a geometrical model. These authors derive a CME mass in the range of 10$^{21}$--10$^{22}$~g. 

\citet{Bond2001} investigated HST spectra of the pre-cataclysmic binary V471~Tau, which consists of a hot white dwarf and a cool red K-dwarf. The authors obtained Ly$\alpha$ time series and detected the sudden appearance of the SiIII line in absorption for several consecutive spectra. This appearance was interpreted by \vidak{the authors as CMEs from the cool K-dwarf} passing through the light of the white dwarf. According to their detections, these authors derive a CME rate of 100--500 day$^{-1}$ from the K-dwarf. \citet{Kovari2021A&A...650A.158K} investigated the same system using data from CFHT/ESPaDOnS and created Doppler images. The Doppler images have shown a dominant spot region at higher latitudes. These authors used around 500 spectra for their analysis summing up to one and a half days of total on-source time. As the data set used in \citet{Kovari2021A&A...650A.158K} includes the Balmer lines, one should have seen signatures of erupting filaments/prominences using the method of Doppler-shifted emission/absorption, but none were found.

Another approach to estimating CME frequencies on other stars is to use their observed flare frequencies and combine them with solar flare--CME relationships. The only missing link to CMEs, the CME--flare association rate, is usually a critical point in such investigations as there is no determined stellar flare--CME association rate from any star known up to now. \citet{Aarnio2012} used their own relation of X-ray flare energy and CME mass, derived for solar events, which they related with an X-ray flare energy distributions from stars in the Orion Nebula Cluster (ONC), to estimate mass-loss rates of those stars. For their approach, they assumed a CME--flare association rate of 100\%. As stars selected in the ONC are very young, as it is a star-forming region, the derived mass loss rates were one to five orders of magnitudes higher than for the present-day Sun. One year later, \citet{Drake2013} used a similar approach. They utilized the flare energy--CME mass relation and flare--CME association rate from \citet{Yashiro2009} and combined them with theoretical flare energy distributions that were scaled with the stars' X-ray luminosities. Based on this approach these authors found that the CME mass-loss rate scales with increasing \petra{X-ray luminosity}, where for the most active stars the mass-loss rates are up to four orders of magnitude higher than the solar mass-loss rate. These mass-loss rates are comparable with the ones from \citet{Aarnio2012}, however, the related kinetic energy losses are too high and would require unrealistically high \petra{fractions of the stellar energy budget}. Therefore \citet{Drake2013} concluded that solar flare--CME relations can not be extrapolated easily to the the most active stellar cases. \citet{Odert2017} refined this method by using a power law derived from stellar EUV observations to estimate the stellar FFDs depending on stellar X-ray luminosity. The determined CME mass-loss rates are one to two orders of magnitude lower than the maximum values derived in \citet{Drake2013}. These authors also cross-checked their CME mass-loss rates with the available total mass-loss \petra{rate measurements} presented in \citet{Wood2005} \citep[updated in][]{Wood2021}. The total stellar mass-loss rates also include the mass-loss by CMEs, and therefore the modeled CME mass-loss rates from \citet{Odert2017} should be below the observed total mass-loss rates \citep{Wood2005, Wood2021}. It showed that up to an FFD power law index of $\alpha=2.2$, the CME-related mass loss rates in \citet{Odert2017} lie below the total observed mass-loss rates derived in \citet{Wood2005, Wood2021} and are therefore in agreement. Another approach in this category was presented by \citet{Osten2015}. Here, energy equipartition between bolometric flare radiation and kinetic CME energy is \petra{assumed} instead of using solar flare energy--CME mass distributions. The application of this method to observed stellar flare statistics revealed comparable CME mass-loss rates as presented in studies before. \citet{Cranmer2017} presented a different approach to assess stellar CMEs, namely they used a relation between surface-averaged magnetic flux with the mean kinetic energy flux of CMEs on the Sun. Their conclusion is that on young solar-mass stars, the mass loss is likely CME-dominated.

\begin{figure}[t]
    \centering
    \includegraphics[width=\textwidth]{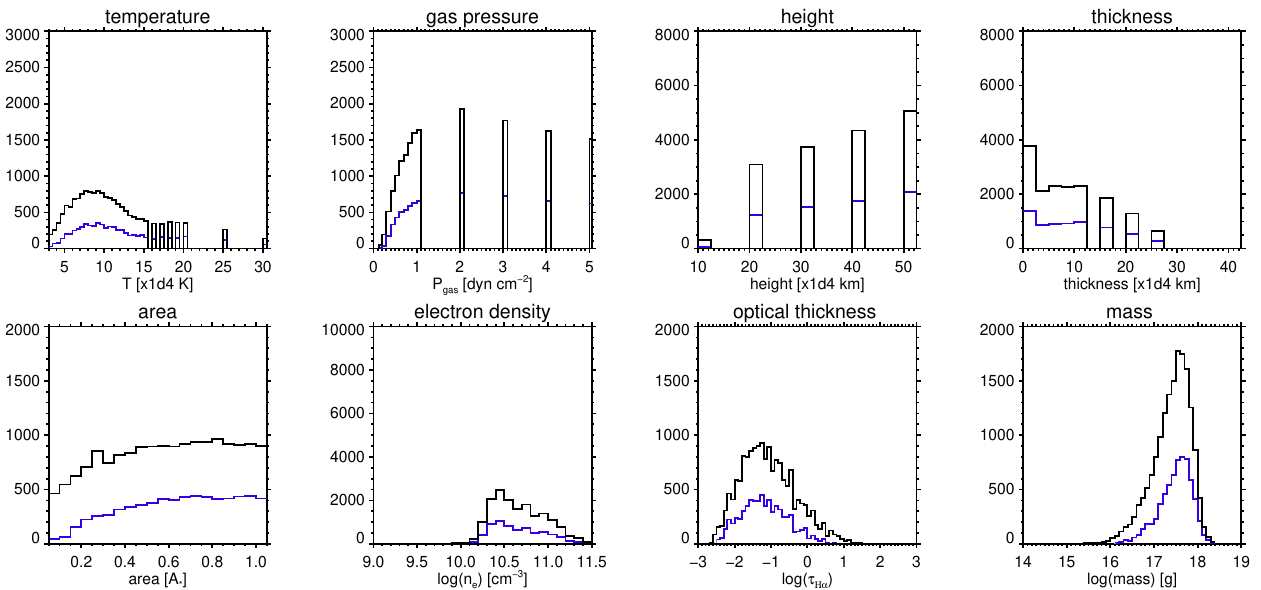}\\
    \caption{Shown are parameter distributions, both physical and geometrical, obtained from NLTE modeling \citep{Leitzinger2022} of an erupting filament scenario, applied to the event detected in \citet{Vida2016}. These distributions were derived for the event shown in the right panels of Fig.\ref{fig:v374peg_cme}. Black solid lines refer to 2-$\sigma$ results whereas blue solid lines refer to 1~$\sigma$ results.}
    \label{fig:cloudmodel}
\end{figure}

In recent years also modeling of stellar CMEs and their related signatures has been established, yielding highly interesting new aspects. Signatures of erupting prominences/fila\-ments have been modeled by \citet{Odert2020}, who applied a simplified approach based on radiative transfer equations aimed at predicting signatures in Balmer lines. These authors have shown that the later the spectral type of the star the less signal-to-noise is required to detect the signature of an erupting prominence/filament in Balmer lines due to the contrast effect in terms of the continuum level of the star. \citet{Leitzinger2022} applied a 1D NLTE model originally developed for the Sun \citep{Heinzel1995, Heinzel1999} to a prominent case in the literature, namely the complex prominence/filament eruption on the active dMe star V374~Peg \citep[presented in][]{Vida2016}. Solutions were found for both filament (see Fig.\ref{fig:cloudmodel}) and prominence geometry, indicating that on dMe stars an emission signature can also be seen in filament geometry. This can not be explained with scattering only, as the emissivity of the prominence/filament needs to be taken into account in the formalism. \citet{Ikuta2024} applied a 1D hydrodynamical simulation of the plasma flow to the possible filament eruption seen in H$\alpha$ from \citet{Namekata2022}. The authors find that the variations (absorption) seen in the stellar H$\alpha$ profile can be explained in terms of a failed filament eruption.

Potential diagnostics of UV signatures for stellar CME detection have been investigated by \citet{Wilson2022}. From solar UVCS observations, they find that CIV1550\AA{}, OVI1032\AA{}, and CIII977\AA{} are predicted to be the most favorable spectral lines to be searched for massive stellar CMEs. An analytical CME model is used to investigate the detectability of CMEs on solar-type stars in EUV spectra in the study of \citet{Yang2024}. The authors show that it is feasible to detect stellar CME signatures in EUV spectra of moderate spectral resolution and S/N. \citet{Cully1994} model the EUV light curve of a long-duration flare on AU~Mic by applying the scenario of an expanding CME with a mass of 10$^{20}$~g. Contrary to this explanation of the EUV light curve, \citet{Katsova1999} presented another explanation involving post-eruptive energy release without any CME.

Type II radio bursts have been searched for a long time \petra{to identify possible stellar CMEs}. \citet{Mullan2019} suggested that on dMe stars very high CME velocities are needed to produce type II bursts due to their strong magnetic fields. \citet{Alvarado2020} used numerical modeling to show that the emitting regions for type II bursts are more distant from the star, making the signature appear at different frequencies than on the Sun which may lie below the ionospheric cut-off, and are therefore not accessible with ground-based radio observatories. \citet{OFionnagain2022} also focused on type II bursts and predicted their occurrence for $\epsilon$~Eri, as \petra{this star} has no such strong magnetic field as other younger and more active stars, and the CMEs may not be confined. These authors also stress that the location of the radio burst is relevant in terms of \petra{observed} intensity and duration.

Also the signature of coronal dimmings, as another signature of CMEs, was subject to theoretical investigations. \citet{Jin2020} studied a flux rope ejection in a magnetic field and found that for more active stars than the Sun, coronal dimmings may appear at higher temperatures.

\citet{Drake2013} suggested magnetic confinement as a possible reason for the so far sparse detection of CMEs on other stars. This confinement scenario was investigated by \citet{Alvarado2018} by placing a flux rope in a coronal magnetic field. For a 75~G magnetic field, it was shown that solar-like CMEs may not erupt and remain confined. Only more energetic events may erupt. Also, the application of such simulations to the M-dwarf regime \citep[Proxima Cen, ][]{Alvarado2019} reveals similar results. Associated coronal dimming signatures were also determined even when the events were confined. Both, the global overlying and also the local magnetic field are relevant for magnetic confinement. The local magnetic field was taken into account by \citet{Sun2022} who investigated the case of torus instability above star spots as a driver for CMEs. An MHD model was used to model a superflare on the solar analog \petra{$\kappa^1$~Cet \citep{Lynch2019}}. The results have shown that the global-scale shear concentrated near the radial-field polarity inversion line may yield the energy to initiate an eruptive superflare similar to the Carrington event which happened on the Sun in the 19th century.

Stellar CME propagation has also been investigated. \citet{Kay2016} investigated CME deflection by the magnetic field leading to changing CME trajectories on the young and active dMe star V374~Peg, using an adapted solar CME deflection model. As on dMe stars the magnetic field is much stronger than on the present-day Sun, these authors found also much stronger deflections towards the astrospheric current sheet. Accordingly, the authors suggested that CME impact on planets is maximal if the planetary orbit is not inclined relative to the astrospheric current sheet. Applying the model to the solar analogue $\kappa$~Cet \citep{Kay2019} revealed that CMEs on that star are deflected towards the astrospheric current sheet which may result in high impact rates on orbiting planets around $\kappa$~Cet. The initial parameters of stellar CMEs and their influence on the CME trajectories have been investigated by \citet{Menezes2023} for Kepler-63 and Kepler-411, two young ($\sim$200~Myr) G-type stars. The results indicate that deflections of CMEs decrease with their rotational velocity and increase with ejection latitude, and also that stronger magnetic fields cause greater deflections.

\section{\kovari{Final thoughts and outlook}}
More than half a century has passed since the XIII IAU General Assembly organized the Working Group on Flare Stars to 
coordinate work and observations on UV\,Cet-type stars to investigate the temporal distribution of flares
\cite{Gershberg1972Ap&SS..19...75G}. 
In that time the temporal resolution, precision of the observations, and the amount of available data increased by orders of magnitudes.
\kovari{The Big Data era following the advent of space-based photometric observations has significantly advanced our understanding of stellar flares. Thanks to space missions like the \emph{Kepler} and \emph{TESS} observatories we have now continuous datasets of several thousand flare stars. These extensive datasets have enabled the discovery of flares in previously \vidak{under-observed stellar populations, e.g., giant stars \cite{Olah2022A&A...668A.101O}, }
 and provided new insights into the frequency, energy distribution, and temporal behavior of flares. The new findings indicate correlations between a star's fundamental properties and the nature of its flare activity, suggesting that the underlying magnetic dynamo processes are heavily influenced by these stellar characteristics, necessitating a deeper exploration of these relationships.}

\kovari{Our review highlights that flare activity is not limited to solar-type stars but spans a wide variety of stellar types, some of which exhibit unique flare characteristics. 
This diversity alone calls for a more flexible interpretation of the solar-stellar paradigm in case of flares and  underscores
the need for tailored models for different stellar environments.
The review identifies several observations that challenge the traditional solar-based flare model. The standard flare model may not account for the varying conditions found in different stellar environments, such as differing magnetic field strengths, rotation rates, and convection zones. Observations of stars with complex and dynamic magnetic field topologies suggest that the standard model’s assumptions about the magnetic reconnection process may not fully capture the intricacies of flare mechanisms in other stars. The universal applicability of the solar paradigm is also questioned by the fact that we know stars with significantly higher flare frequencies and intensities than the Sun. These stars may exhibit different energy release mechanisms or flare triggers. Going further, flare behaviors observed in stars at different stages of their evolution, from pre-main sequence to red giants, highlight the need for models that can accommodate changes in internal structure and magnetic activity over time. Also, 
\vidak{activity phenomena less typical for the Sun, }
such as active longitudes \cite{Mandal2017}, starspots persisting over many rotations \cite{Olah1999}, or extreme rotation modulation, etc., suggest that additional or alternative physical processes may influence flare activity in such cases. Moreover, the relationship between flares and CMEs in other stars often differs from the solar context. We still have limited statistical data on stellar CMEs, mainly due to observational constraints, but it appears, for example, that some stars show \petra{flares without associated CMEs, or vice versa} \vidak{(as seen on the Sun as well)}, indicating that the standard model’s coupling of these events may not be universally applicable.}


\kovari{Observing flares in close binary systems is even more challenging due to the need to disentangle the contributions from both stars, requiring advanced techniques and coordinated observations across multiple wavelengths. However, there are further problems to be worked out in connection with flares in close binaries. The energy distribution of flares may differ from single stars due to combined magnetic energy reservoirs and the complex magnetic field dynamics of the binary system. Synchronous rotation due to strong tidal interactions can lead to enhanced and more stable magnetic activity, influencing the frequency and energy of flares. Moreover, flares can impact any surrounding circumbinary disks, influencing disk dynamics, heating, and potential planet formation processes. In certain close binaries, especially those involving a compact object (e.g., a white dwarf, neutron star, or black hole), the accretion of material from one star to another can trigger flares through interactions with the accreted material and the magnetic field.
Furthermore, the relationship between flares and CMEs in close binaries can be influenced by the interactions between the two stars' magnetospheres, potentially leading to either enhanced or suppressed CME activity.
Last but not least, in systems with close-in planets, magnetic interactions between the planet and the host star can also influence flare activity, adding another layer of complexity to the study of flares.
It is clear that only by addressing the challenges mentioned above would it be possible to develop more comprehensive and flexible models that accurately describe flare phenomena in such a wide range of stellar environments.}

We started our introduction in Sect.\,\ref{intro} with the idea that flares and CMEs are known to affect nearby planets. So it is no surprise that the scientific community of exoplanet researchers is looking forward with great interest and excitement to space missions that are suitable for spectroscopic measurements, thereby being able to detect changes in the atmospheres of exoplanets, which occur due to flares from the host star. In fact, JWST is already capable of this \citep{JWST2023NatAs...7.1317L}, and ESA's mission in the near future, Planetary Transit and Oscillations of Stars (\emph{PLATO}), which \petra{is} aimed to study terrestrial planets orbiting in the habitable zone of Sun-like stars, will also be able to do so. But looking a little further, we can also mention ESA's \emph{Ariel} here.
Therefore, with continued advancements in these new observational technologies and collaborative efforts within the scientific community, we are optimistic about the prospects of deepening our understanding of stellar flares, CMEs, and their broader implications.


\newpage
\authorcontributions{
Conceptualization, K.V. and Zs.K.; formal analysis, K.V., Zs.K., L.K., M.L., P.O., K.O. and B.S.; investigation, K.V., Zs.K., L.K., M.L., P.O., K.O. and B.S.; software, L.K., B.S., and K.V.; Writing---original draft preparation, K.V., Zs.K., K.O., M.L. and P.O.; writing---review and editing, L.K., B.S. and R.G.; visualization, L.K., S.B. and A.G.; supervision, K.V., Zs.K., K.O., M.L. and P.O.; project administration, K.V., Zs.K., L.K., M.L. and P.O.; funding acquisition, K.V., Zs.K., L.K., M.L. and P.O. All authors have read and agreed to the published version of the manuscript.
}

\funding{This research was funded by the Hungarian National Research, Development and Innovation Office grant KKP-143986. Authors acknowledge the financial support of the Austrian-Hungarian Action Foundation grants 98\"ou5, 101\"ou13, \petra{104\"ou2,} 112\"ou1. L.K. acknowledges the support of the Hungarian National Research, Development and Innovation Office grant PD-134784. 
K.V. is supported by the Bolyai J\'anos Research Scholarship
of the Hungarian Academy of Sciences.
\petra{This research was funded in whole, or in part, by the Austrian Science Fund (FWF) 10.55776/I5711. For the purpose of open access, the author has applied a CC BY public copyright license to any Author Accepted Manuscript version arising from this submission.} On behalf of the \textit{"Looking for stellar CMEs on different wavelengths"} project we are grateful for the possibility of using HUN-REN Cloud \cite{MTACloud} which helped us achieve the results published in this paper.
}

\institutionalreview{Not applicable.}

\conflictsofinterest{The authors declare no conflict of interest. The funders had no role in the design of the study; in the collection, analyses, or interpretation of data; in the writing of the manuscript; or in the decision to publish the results.} 



\begin{adjustwidth}{-\extralength}{0cm}

\reftitle{References}


\bibliography{SpaceWeather}

\begin{thebibliography}{999}

\bibitem[{Vida} et~al.(2017){Vida}, {K{\H{o}}v{\'a}ri}, {P{\'a}l}, {Ol{\'a}h},
  and {Kriskovics}]{Vida2017ApJ...841..124V}
{Vida}, K.; {K{\H{o}}v{\'a}ri}, {\mbox Zs}.; {P{\'a}l}, A.; {Ol{\'a}h}, K.;
  {Kriskovics}, L.
\newblock {Frequent Flaring in the TRAPPIST-1 System{\textemdash}Unsuited for
  Life?}
\newblock {\em \apj} {\bf 2017}, {\em 841},~124,
  \href{http://arxiv.org/abs/1703.10130}{{\normalfont
  [arXiv:astro-ph.SR/1703.10130]}}.
\newblock {\url{https://doi.org/10.3847/1538-4357/aa6f05}}.

\bibitem[{Tsurutani} et~al.(2003){Tsurutani}, {Gonzalez}, {Lakhina}, and
  {Alex}]{2003JGRA..108.1268T}
{Tsurutani}, B.T.; {Gonzalez}, W.D.; {Lakhina}, G.S.; {Alex}, S.
\newblock {The extreme magnetic storm of 1-2 September 1859}.
\newblock {\em Journal of Geophysical Research (Space Physics)} {\bf 2003},
  {\em 108},~1268.
\newblock {\url{https://doi.org/10.1029/2002JA009504}}.

\bibitem[{Carrington}(1859)]{Carrington1859MNRAS..20...13C}
{Carrington}, R.C.
\newblock {Description of a Singular Appearance seen in the Sun on September 1,
  1859}.
\newblock {\em \mnras} {\bf 1859}, {\em 20},~13--15.
\newblock {\url{https://doi.org/10.1093/mnras/20.1.13}}.

\bibitem[{Crosby} et~al.(1993){Crosby}, {Aschwanden}, and {Dennis}]{Crosby1993}
{Crosby}, N.B.; {Aschwanden}, M.J.; {Dennis}, B.R.
\newblock {Frequency distributions and correlations of solar X-ray flare
  parameters}.
\newblock {\em \solphys} {\bf 1993}, {\em 143},~275--299.
\newblock {\url{https://doi.org/10.1007/BF00646488}}.

\bibitem[{Shibata} et~al.(2013){Shibata}, {Isobe}, {Hillier}, {Choudhuri},
  {Maehara}, {Ishii}, {Shibayama}, {Notsu}, {Notsu}, {Nagao}, {Honda}, and
  {Nogami}]{Shibata2013}
{Shibata}, K.; {Isobe}, H.; {Hillier}, A.; {Choudhuri}, A.R.; {Maehara}, H.;
  {Ishii}, T.T.; {Shibayama}, T.; {Notsu}, S.; {Notsu}, Y.; {Nagao}, T.;
  et~al.
\newblock {Can Superflares Occur on Our Sun?}
\newblock {\em \pasj} {\bf 2013}, {\em 65},~49,
  \href{http://arxiv.org/abs/1212.1361}{{\normalfont
  [arXiv:astro-ph.SR/1212.1361]}}.
\newblock {\url{https://doi.org/10.1093/pasj/65.3.49}}.

\bibitem[{Cliver} and {Dietrich}(2013)]{Cliver2013JSWSC...3A..31C}
{Cliver}, E.W.; {Dietrich}, W.F.
\newblock {The 1859 space weather event revisited: limits of extreme activity}.
\newblock {\em Journal of Space Weather and Space Climate} {\bf 2013}, {\em
  3},~A31.
\newblock {\url{https://doi.org/10.1051/swsc/2013053}}.

\bibitem[{Kopp} and {Pneuman}(1976)]{1976SoPh...50...85K}
{Kopp}, R.A.; {Pneuman}, G.W.
\newblock {Magnetic reconnection in the corona and the loo?p prominence
  phenomenon.}
\newblock {\em \solphys} {\bf 1976}, {\em 50},~85--98.
\newblock {\url{https://doi.org/10.1007/BF00206193}}.

\bibitem[{Heyvaerts} et~al.(1977){Heyvaerts}, {Priest}, and
  {Rust}]{1977SoPh...53..255H}
{Heyvaerts}, J.; {Priest}, E.; {Rust}, D.M.
\newblock {An emerging flux model for solar flares.}
\newblock {\em \solphys} {\bf 1977}, {\em 53},~255--258.
\newblock {\url{https://doi.org/10.1007/BF02260230}}.

\bibitem[{Feynman} and {Martin}(1995)]{1995JGR...100.3355F}
{Feynman}, J.; {Martin}, S.F.
\newblock {The initiation of coronal mass ejections by newly emerging magnetic
  flux}.
\newblock {\em \jgr} {\bf 1995}, {\em 100},~3355--3368.
\newblock {\url{https://doi.org/10.1029/94JA02591}}.

\bibitem[{Tanaka}(1991)]{1991SoPh..136..133T}
{Tanaka}, K.
\newblock {Studies on a very flare-active {\ensuremath{\delta}} group: Peculiar
  {\ensuremath{\delta}} spot evolution and inferred subsurface magnetic rope
  structure}.
\newblock {\em \solphys} {\bf 1991}, {\em 136},~133--149.
\newblock {\url{https://doi.org/10.1007/BF00151700}}.

\bibitem[{Forbes}(2010)]{2010hssr.book..159F}
{Forbes}, T.
\newblock {Models of coronal mass ejections and flares}. In {\em Heliophysics:
  Space Storms and Radiation: Causes and Effects}; {Schrijver}, C.J.; {Siscoe},
  G.L., Eds.; Cambridge University Press,  2010; p. 159.

\bibitem[{Priest}(2014)]{2014masu.book.....P}
{Priest}, E.
\newblock {\em {Magnetohydrodynamics of the Sun}}; Cambridge University Press,
  2014.
\newblock {\url{https://doi.org/10.1017/CBO9781139020732}}.

\bibitem[{Ruan} et~al.(2020){Ruan}, {Xia}, and {Keppens}]{2020ApJ...896...97R}
{Ruan}, W.; {Xia}, C.; {Keppens}, R.
\newblock {A Fully Self-consistent Model for Solar Flares}.
\newblock {\em \apj} {\bf 2020}, {\em 896},~97,
  \href{http://arxiv.org/abs/2005.08578}{{\normalfont
  [arXiv:astro-ph.SR/2005.08578]}}.
\newblock {\url{https://doi.org/10.3847/1538-4357/ab93db}}.

\bibitem[{Balona}(2015)]{Balona2015MNRAS.447.2714B}
{Balona}, L.A.
\newblock {Flare stars across the H-R diagram}.
\newblock {\em \mnras} {\bf 2015}, {\em 447},~2714--2725.
\newblock {\url{https://doi.org/10.1093/mnras/stu2651}}.

\bibitem[{He} et~al.(2018){He}, {Wang}, {Zhang}, {Mehrabi}, {Yan}, and
  {Yun}]{He2018ApJS..236....7H}
{He}, H.; {Wang}, H.; {Zhang}, M.; {Mehrabi}, A.; {Yan}, Y.; {Yun}, D.
\newblock {Activity Analyses for Solar-type Stars Observed with Kepler. II.
  Magnetic Feature versus Flare Activity}.
\newblock {\em \apjs} {\bf 2018}, {\em 236},~7,
  \href{http://arxiv.org/abs/1705.09028}{{\normalfont
  [arXiv:astro-ph.SR/1705.09028]}}.
\newblock {\url{https://doi.org/10.3847/1538-4365/aab779}}.

\bibitem[{Hathaway}(2015)]{Hathaway2015LRSP...12....4H}
{Hathaway}, D.H.
\newblock {The Solar Cycle}.
\newblock {\em Living Reviews in Solar Physics} {\bf 2015}, {\em 12},~4,
  \href{http://arxiv.org/abs/1502.07020}{{\normalfont
  [arXiv:astro-ph.SR/1502.07020]}}.
\newblock {\url{https://doi.org/10.1007/lrsp-2015-4}}.

\bibitem[{Lehtinen} et~al.(2020){Lehtinen}, {Spada}, {K{\"a}pyl{\"a}},
  {Olspert}, and {K{\"a}pyl{\"a}}]{Lehtinen2020NatAs...4..658L}
{Lehtinen}, J.J.; {Spada}, F.; {K{\"a}pyl{\"a}}, M.J.; {Olspert}, N.;
  {K{\"a}pyl{\"a}}, P.J.
\newblock {Common dynamo scaling in slowly rotating young and evolved stars}.
\newblock {\em Nature Astronomy} {\bf 2020}, {\em 4},~658--662,
  \href{http://arxiv.org/abs/2003.08997}{{\normalfont
  [arXiv:astro-ph.SR/2003.08997]}}.
\newblock {\url{https://doi.org/10.1038/s41550-020-1039-x}}.

\bibitem[{Charbonneau} et~al.(2001){Charbonneau}, {McIntosh}, {Liu}, and
  {Bogdan}]{Charbonneau2001SoPh..203..321C}
{Charbonneau}, P.; {McIntosh}, S.W.; {Liu}, H.L.; {Bogdan}, T.J.
\newblock {Avalanche models for solar flares (Invited Review)}.
\newblock {\em \solphys} {\bf 2001}, {\em 203},~321--353.
\newblock {\url{https://doi.org/10.1023/A:1013301521745}}.

\bibitem[{K{\H{o}}v{\'a}ri} et~al.(2020){K{\H{o}}v{\'a}ri}, {Ol{\'a}h},
  {G{\"u}nther}, {Vida}, {Kriskovics}, {Seli}, {Bakos}, {Hartman}, {Csubry},
  and {Bhatti}]{Kovari2020A&A...641A..83K}
{K{\H{o}}v{\'a}ri}, {\mbox Zs}.; {Ol{\'a}h}, K.; {G{\"u}nther}, M.N.; {Vida},
  K.; {Kriskovics}, L.; {Seli}, B.; {Bakos}, G.{\'A}.; {Hartman}, J.D.;
  {Csubry}, Z.; {Bhatti}, W.
\newblock {Superflares on the late-type giant KIC 2852961. Scaling effect
  behind flaring at different energy levels}.
\newblock {\em \aap} {\bf 2020}, {\em 641},~A83,
  \href{http://arxiv.org/abs/2005.05397}{{\normalfont
  [arXiv:astro-ph.SR/2005.05397]}}.
\newblock {\url{https://doi.org/10.1051/0004-6361/202038397}}.

\bibitem[{Wang} et~al.(2010){Wang}, {Cao}, {Chen}, {Zhang}, {Yu}, {Zheng},
  {Shen}, {Zhang}, and {Wang}]{Wang2010}
{Wang}, Y.; {Cao}, H.; {Chen}, J.; {Zhang}, T.; {Yu}, S.; {Zheng}, H.; {Shen},
  C.; {Zhang}, J.; {Wang}, S.
\newblock {Solar Limb Prominence Catcher and Tracker (SLIPCAT): An Automated
  System and its Preliminary Statistical Results}.
\newblock {\em \apj} {\bf 2010}, {\em 717},~973--986,
  \href{http://arxiv.org/abs/1004.4553}{{\normalfont
  [arXiv:astro-ph.SR/1004.4553]}}.
\newblock {\url{https://doi.org/10.1088/0004-637X/717/2/973}}.

\bibitem[{Martens} and {Kuin}(1989)]{Martens1989}
{Martens}, P.C.H.; {Kuin}, N.P.M.
\newblock {A Circuit Model for Filament Eruptions and Two-Ribbon Flares}.
\newblock {\em \solphys} {\bf 1989}, {\em 122},~263--302.
\newblock {\url{https://doi.org/10.1007/BF00912996}}.

\bibitem[{Yashiro} and {Gopalswamy}(2009)]{Yashiro2009}
{Yashiro}, S.; {Gopalswamy}, N.
\newblock {Statistical relationship between solar flares and coronal mass
  ejections}.
\newblock In Proceedings of the Universal Heliophysical Processes;
  {Gopalswamy}, N.; {Webb}, D.F., Eds.,  3 2009, Vol. 257, pp. 233--243.
\newblock {\url{https://doi.org/10.1017/S1743921309029342}}.

\bibitem[{Thalmann} et~al.(2015){Thalmann}, {Su}, {Temmer}, and
  {Veronig}]{Thalmann2015}
{Thalmann}, J.K.; {Su}, Y.; {Temmer}, M.; {Veronig}, A.M.
\newblock {The Confined X-class Flares of Solar Active Region 2192}.
\newblock {\em \apjl} {\bf 2015}, {\em 801},~L23,
  \href{http://arxiv.org/abs/1502.05157}{{\normalfont
  [arXiv:astro-ph.SR/1502.05157]}}.
\newblock {\url{https://doi.org/10.1088/2041-8205/801/2/L23}}.

\bibitem[{Li} et~al.(2020){Li}, {Hou}, {Yang}, {Zhang}, {Liu}, and
  {Veronig}]{Li2020}
{Li}, T.; {Hou}, Y.; {Yang}, S.; {Zhang}, J.; {Liu}, L.; {Veronig}, A.M.
\newblock {Magnetic Flux of Active Regions Determining the Eruptive Character
  of Large Solar Flares}.
\newblock {\em \apj} {\bf 2020}, {\em 900},~128,
  \href{http://arxiv.org/abs/2007.08127}{{\normalfont
  [arXiv:astro-ph.SR/2007.08127]}}.
\newblock {\url{https://doi.org/10.3847/1538-4357/aba6ef}}.

\bibitem[{Carley} et~al.(2020){Carley}, {Vilmer}, and {Vourlidas}]{Carley2020}
{Carley}, E.P.; {Vilmer}, N.; {Vourlidas}, A.
\newblock {Radio observations of coronal mass ejection initiation and
  development in the low solar corona}.
\newblock {\em Frontiers in Astronomy and Space Sciences} {\bf 2020}, {\em
  7},~79.
\newblock {\url{https://doi.org/10.3389/fspas.2020.551558}}.

\bibitem[{Dissauer} et~al.(2019){Dissauer}, {Veronig}, {Temmer}, and
  {Podladchikova}]{Dissauer2019}
{Dissauer}, K.; {Veronig}, A.M.; {Temmer}, M.; {Podladchikova}, T.
\newblock {Statistics of Coronal Dimmings Associated with Coronal Mass
  Ejections. II. Relationship between Coronal Dimmings and Their Associated
  CMEs}.
\newblock {\em \apj} {\bf 2019}, {\em 874},~123,
  \href{http://arxiv.org/abs/1810.01589}{{\normalfont
  [arXiv:astro-ph.SR/1810.01589]}}.
\newblock {\url{https://doi.org/10.3847/1538-4357/ab0962}}.

\bibitem[{Moschou} et~al.(2019){Moschou}, {Drake}, {Cohen},
  {Alvarado-G{\'o}mez}, {Garraffo}, and {Fraschetti}]{Moschou2019}
{Moschou}, S.P.; {Drake}, J.J.; {Cohen}, O.; {Alvarado-G{\'o}mez}, J.D.;
  {Garraffo}, C.; {Fraschetti}, F.
\newblock {The Stellar CME-Flare Relation: What Do Historic Observations
  Reveal?}
\newblock {\em \apj} {\bf 2019}, {\em 877},~105,
  \href{http://arxiv.org/abs/1904.09598}{{\normalfont
  [arXiv:astro-ph.SR/1904.09598]}}.
\newblock {\url{https://doi.org/10.3847/1538-4357/ab1b37}}.

\bibitem[{Houdebine} et~al.(1990){Houdebine}, {Foing}, and
  {Rodono}]{Houdebine1990}
{Houdebine}, E.R.; {Foing}, B.H.; {Rodono}, M.
\newblock {Dynamics of flares on late-type dMe stars. I - Flare mass ejections
  and stellar evolution}.
\newblock {\em \aap} {\bf 1990}, {\em 238},~249--255.

\bibitem[{Leitzinger} et~al.(2011){Leitzinger}, {Odert}, {Ribas}, {Hanslmeier},
  {Lammer}, {Khodachenko}, {Zaqarashvili}, and {Rucker}]{Leitzinger2011a}
{Leitzinger}, M.; {Odert}, P.; {Ribas}, I.; {Hanslmeier}, A.; {Lammer}, H.;
  {Khodachenko}, M.L.; {Zaqarashvili}, T.V.; {Rucker}, H.O.
\newblock {Search for indications of stellar mass ejections using FUV spectra}.
\newblock {\em \aap} {\bf 2011}, {\em 536},~A62.
\newblock {\url{https://doi.org/10.1051/0004-6361/201015985}}.

\bibitem[{Vida} et~al.(2016){Vida}, {Kriskovics}, {Ol{\'a}h}, {Leitzinger},
  {Odert}, {K{\H o}v{\'a}ri}, {Korhonen}, {Greimel}, {Robb}, {Cs{\'a}k}, and
  {Kov{\'a}cs}]{Vida2016}
{Vida}, K.; {Kriskovics}, L.; {Ol{\'a}h}, K.; {Leitzinger}, M.; {Odert}, P.;
  {K{\H o}v{\'a}ri}, Z.; {Korhonen}, H.; {Greimel}, R.; {Robb}, R.; {Cs{\'a}k},
  B.;  et~al.
\newblock {Investigating magnetic activity in very stable stellar magnetic
  fields. Long-term photometric and spectroscopic study of the fully convective
  M4 dwarf V374 Pegasi}.
\newblock {\em \aap} {\bf 2016}, {\em 590},~A11,
  \href{http://arxiv.org/abs/1603.00867}{{\normalfont
  [arXiv:astro-ph.SR/1603.00867]}}.
\newblock {\url{https://doi.org/10.1051/0004-6361/201527925}}.

\bibitem[{Argiroffi} et~al.(2019){Argiroffi}, {Reale}, {Drake}, {Ciaravella},
  {Testa}, {Bonito}, {Miceli}, {Orlando}, and {Peres}]{Argiroffi2019}
{Argiroffi}, C.; {Reale}, F.; {Drake}, J.J.; {Ciaravella}, A.; {Testa}, P.;
  {Bonito}, R.; {Miceli}, M.; {Orlando}, S.; {Peres}, G.
\newblock {A stellar flare-coronal mass ejection event revealed by X-ray plasma
  motions}.
\newblock {\em Nature Astronomy} {\bf 2019}, {\em 3},~742--748,
  \href{http://arxiv.org/abs/1905.11325}{{\normalfont
  [arXiv:astro-ph.SR/1905.11325]}}.
\newblock {\url{https://doi.org/10.1038/s41550-019-0781-4}}.

\bibitem[{Namekata} et~al.(2021){Namekata}, {Maehara}, {Honda}, {Notsu},
  {Okamoto}, {Takahashi}, {Takayama}, {Ohshima}, {Saito}, {Katoh}, {Tozuka},
  {Murata}, {Ogawa}, {Niwano}, {Adachi}, {Oeda}, {Shiraishi}, {Isogai}, {Seki},
  {Ishii}, {Ichimoto}, {Nogami}, and {Shibata}]{Namekata2021}
{Namekata}, K.; {Maehara}, H.; {Honda}, S.; {Notsu}, Y.; {Okamoto}, S.;
  {Takahashi}, J.; {Takayama}, M.; {Ohshima}, T.; {Saito}, T.; {Katoh}, N.;
  et~al.
\newblock {Probable detection of an eruptive filament from a superflare on a
  solar-type star}.
\newblock {\em Nature Astronomy} {\bf 2021},
  \href{http://arxiv.org/abs/2112.04808}{{\normalfont
  [arXiv:astro-ph.SR/2112.04808]}}.
\newblock {\url{https://doi.org/10.1038/s41550-021-01532-8}}.

\bibitem[{Leitzinger} and {Odert}(2022)]{Leitzinger2022c}
{Leitzinger}, M.; {Odert}, P.
\newblock {Stellar Coronal Mass Ejections}.
\newblock {\em Serbian Astronomical Journal} {\bf 2022}, {\em 205},~1--22,
  \href{http://arxiv.org/abs/2212.09079}{{\normalfont
  [arXiv:astro-ph.SR/2212.09079]}}.
\newblock {\url{https://doi.org/10.2298/SAJ2205001L}}.

\bibitem[{Osten}(2023)]{Osten2023}
{Osten}, R.A.
\newblock {Observations of Winds and CMEs of Low-Mass Stars}.
\newblock In Proceedings of the Winds of Stars and Exoplanets; {Vidotto}, A.A.;
  {Fossati}, L.; {Vink}, J.S., Eds.,  1 2023, Vol. 370, pp. 25--36.
\newblock {\url{https://doi.org/10.1017/S1743921322003714}}.

\bibitem[{Tian} et~al.(2023){Tian}, {Xu}, {Chen}, {Zhang}, {Lu}, {Chen},
  {Yang}, and {Wu}]{Tian2023}
{Tian}, H.; {Xu}, Y.; {Chen}, H.; {Zhang}, J.; {Lu}, H.; {Chen}, Y.; {Yang},
  Z.; {Wu}, Y.
\newblock {Observations and simulations of stellar coronal mass ejections}.
\newblock {\em SCIENTIA SINICA Technologica} {\bf 2023}, {\em 53},~2021--2038.
\newblock {\url{https://doi.org/10.1360/SST-2022-0212}}.

\bibitem[{Balona}(2012)]{Balona2012MNRAS.423.3420B}
{Balona}, L.A.
\newblock {Kepler observations of flaring in A-F type stars}.
\newblock {\em \mnras} {\bf 2012}, {\em 423},~3420--3429.
\newblock {\url{https://doi.org/10.1111/j.1365-2966.2012.21135.x}}.

\bibitem[{Pedersen} et~al.(2017){Pedersen}, {Antoci}, {Korhonen}, {White},
  {Jessen-Hansen}, {Lehtinen}, {Nikbakhsh}, and
  {Viuho}]{Pedersen2017MNRAS.466.3060P}
{Pedersen}, M.G.; {Antoci}, V.; {Korhonen}, H.; {White}, T.R.; {Jessen-Hansen},
  J.; {Lehtinen}, J.; {Nikbakhsh}, S.; {Viuho}, J.
\newblock {Do A-type stars flare?}
\newblock {\em \mnras} {\bf 2017}, {\em 466},~3060--3076,
  \href{http://arxiv.org/abs/1612.04575}{{\normalfont
  [arXiv:astro-ph.SR/1612.04575]}}.
\newblock {\url{https://doi.org/10.1093/mnras/stw3226}}.

\bibitem[{Davenport}(2016)]{Davenport2016ApJ...829...23D}
{Davenport}, J.R.A.
\newblock {The Kepler Catalog of Stellar Flares}.
\newblock {\em \apj} {\bf 2016}, {\em 829},~23,
  \href{http://arxiv.org/abs/1607.03494}{{\normalfont
  [arXiv:astro-ph.SR/1607.03494]}}.
\newblock {\url{https://doi.org/10.3847/0004-637X/829/1/23}}.

\bibitem[{Stelzer} et~al.(2016){Stelzer}, {Damasso}, {Scholz}, and
  {Matt}]{Stelzer2016MNRAS.463.1844S}
{Stelzer}, B.; {Damasso}, M.; {Scholz}, A.; {Matt}, S.P.
\newblock {A path towards understanding the rotation-activity relation of M
  dwarfs with K2 mission, X-ray and UV data}.
\newblock {\em \mnras} {\bf 2016}, {\em 463},~1844--1864,
  \href{http://arxiv.org/abs/1608.00772}{{\normalfont
  [arXiv:astro-ph.SR/1608.00772]}}.
\newblock {\url{https://doi.org/10.1093/mnras/stw1936}}.

\bibitem[{Vida} and {Roettenbacher}(2018)]{Vida2018A&A...616A.163V.FLATWRM}
{Vida}, K.; {Roettenbacher}, R.M.
\newblock {Finding flares in Kepler data using machine-learning tools}.
\newblock {\em \aap} {\bf 2018}, {\em 616},~A163,
  \href{http://arxiv.org/abs/1806.00334}{{\normalfont
  [arXiv:astro-ph.SR/1806.00334]}}.
\newblock {\url{https://doi.org/10.1051/0004-6361/201833194}}.

\bibitem[{Feinstein} et~al.(2020){Feinstein}, {Montet}, {Ansdell}, {Nord},
  {Bean}, {G{\"u}nther}, {Gully-Santiago}, and
  {Schlieder}]{Feinstein2020AJ....160..219F}
{Feinstein}, A.D.; {Montet}, B.T.; {Ansdell}, M.; {Nord}, B.; {Bean}, J.L.;
  {G{\"u}nther}, M.N.; {Gully-Santiago}, M.A.; {Schlieder}, J.E.
\newblock {Flare Statistics for Young Stars from a Convolutional Neural Network
  Analysis of TESS Data}.
\newblock {\em \aj} {\bf 2020}, {\em 160},~219,
  \href{http://arxiv.org/abs/2005.07710}{{\normalfont
  [arXiv:astro-ph.SR/2005.07710]}}.
\newblock {\url{https://doi.org/10.3847/1538-3881/abac0a}}.

\bibitem[{Vida} et~al.(2021){Vida}, {B{\'o}di}, {Szklen{\'a}r}, and
  {Seli}]{Vida2021A&A...652A.107V.FLATWRM2}
{Vida}, K.; {B{\'o}di}, A.; {Szklen{\'a}r}, T.; {Seli}, B.
\newblock {Finding flares in Kepler and TESS data with recurrent deep neural
  networks}.
\newblock {\em \aap} {\bf 2021}, {\em 652},~A107,
  \href{http://arxiv.org/abs/2105.11485}{{\normalfont
  [arXiv:astro-ph.SR/2105.11485]}}.
\newblock {\url{https://doi.org/10.1051/0004-6361/202141068}}.

\bibitem[{Kuerster} and {Schmitt}(1996)]{Kuerster1996A&A...311..211K}
{Kuerster}, M.; {Schmitt}, J.H.M.M.
\newblock {Forty days in the life of CF Tucanae (=HD 5303). The longest stellar
  X-ray flare observed with ROSAT.}
\newblock {\em \aap} {\bf 1996}, {\em 311},~211--229.

\bibitem[{Vida} et~al.(2009){Vida}, {Ol{\'a}h}, {K{\H{o}}v{\'a}ri}, {Korhonen},
  {Bartus}, {Hurta}, and {Posztob{\'a}nyi}]{Vida2009A&A...504.1021V}
{Vida}, K.; {Ol{\'a}h}, K.; {K{\H{o}}v{\'a}ri}, {\mbox Zs}.; {Korhonen}, H.;
  {Bartus}, J.; {Hurta}, Z.; {Posztob{\'a}nyi}, K.
\newblock {Photospheric and chromospheric activity in V405 Andromedae. An M
  dwarf binary with components on the two sides of the full convection limit}.
\newblock {\em \aap} {\bf 2009}, {\em 504},~1021--1029,
  \href{http://arxiv.org/abs/0907.1011}{{\normalfont
  [arXiv:astro-ph.SR/0907.1011]}}.
\newblock {\url{https://doi.org/10.1051/0004-6361/200912326}}.

\bibitem[{Davenport} et~al.(2014){Davenport}, {Hawley}, {Hebb}, {Wisniewski},
  {Kowalski}, {Johnson}, {Malatesta}, {Peraza}, {Keil}, {Silverberg}, {Jansen},
  {Scheffler}, {Berdis}, {Larsen}, and {Hilton}]{Davenport2014ApJ...797..122D}
{Davenport}, J.R.A.; {Hawley}, S.L.; {Hebb}, L.; {Wisniewski}, J.P.;
  {Kowalski}, A.F.; {Johnson}, E.C.; {Malatesta}, M.; {Peraza}, J.; {Keil}, M.;
  {Silverberg}, S.M.;  et~al.
\newblock {Kepler Flares. II. The Temporal Morphology of White-light Flares on
  GJ 1243}.
\newblock {\em \apj} {\bf 2014}, {\em 797},~122,
  \href{http://arxiv.org/abs/1411.3723}{{\normalfont
  [arXiv:astro-ph.SR/1411.3723]}}.
\newblock {\url{https://doi.org/10.1088/0004-637X/797/2/122}}.

\bibitem[{Mendoza} et~al.(2022){Mendoza}, {Davenport}, {Agol}, {Jackman}, and
  {Hawley}]{2022AJ....164...17M}
{Mendoza}, G.T.; {Davenport}, J.R.A.; {Agol}, E.; {Jackman}, J.A.G.; {Hawley},
  S.L.
\newblock {Llamaradas Estelares: Modeling the Morphology of White-light
  Flares}.
\newblock {\em \aj} {\bf 2022}, {\em 164},~17.
\newblock {\url{https://doi.org/10.3847/1538-3881/ac6fe6}}.

\bibitem[{Howard} and {MacGregor}(2022)]{2022ApJ...926..204H}
{Howard}, W.S.; {MacGregor}, M.A.
\newblock {No Such Thing as a Simple Flare: Substructure and Quasi-periodic
  Pulsations Observed in a Statistical Sample of 20 s Cadence TESS Flares}.
\newblock {\em \apj} {\bf 2022}, {\em 926},~204,
  \href{http://arxiv.org/abs/2110.13155}{{\normalfont
  [arXiv:astro-ph.SR/2110.13155]}}.
\newblock {\url{https://doi.org/10.3847/1538-4357/ac426e}}.

\bibitem[{Brown} et~al.(2011){Brown}, {Latham}, {Everett}, and
  {Esquerdo}]{KIC2011AJ....142..112B}
{Brown}, T.M.; {Latham}, D.W.; {Everett}, M.E.; {Esquerdo}, G.A.
\newblock {Kepler Input Catalog: Photometric Calibration and Stellar
  Classification}.
\newblock {\em \aj} {\bf 2011}, {\em 142},~112,
  \href{http://arxiv.org/abs/1102.0342}{{\normalfont
  [arXiv:astro-ph.SR/1102.0342]}}.
\newblock {\url{https://doi.org/10.1088/0004-6256/142/4/112}}.

\bibitem[{Yang} and {Liu}(2019)]{Yang2019ApJS..241...29Y}
{Yang}, H.; {Liu}, J.
\newblock {The Flare Catalog and the Flare Activity in the Kepler Mission}.
\newblock {\em \apjs} {\bf 2019}, {\em 241},~29,
  \href{http://arxiv.org/abs/1903.01056}{{\normalfont
  [arXiv:astro-ph.SR/1903.01056]}}.
\newblock {\url{https://doi.org/10.3847/1538-4365/ab0d28}}.

\bibitem[{Shibayama} et~al.(2013){Shibayama}, {Maehara}, {Notsu}, {Notsu},
  {Nagao}, {Honda}, {Ishii}, {Nogami}, and
  {Shibata}]{Shibayama2013ApJS..209....5S}
{Shibayama}, T.; {Maehara}, H.; {Notsu}, S.; {Notsu}, Y.; {Nagao}, T.; {Honda},
  S.; {Ishii}, T.T.; {Nogami}, D.; {Shibata}, K.
\newblock {Superflares on Solar-type Stars Observed with Kepler. I. Statistical
  Properties of Superflares}.
\newblock {\em \apjs} {\bf 2013}, {\em 209},~5,
  \href{http://arxiv.org/abs/1308.1480}{{\normalfont
  [arXiv:astro-ph.SR/1308.1480]}}.
\newblock {\url{https://doi.org/10.1088/0067-0049/209/1/5}}.

\bibitem[{Ol{\'a}h} et~al.(2022){Ol{\'a}h}, {Seli}, {K{\H{o}}v{\'a}ri},
  {Kriskovics}, and {Vida}]{Olah2022A&A...668A.101O}
{Ol{\'a}h}, K.; {Seli}, B.; {K{\H{o}}v{\'a}ri}, {\mbox Zs}.; {Kriskovics}, L.;
  {Vida}, K.
\newblock {Characteristics of flares on giant stars}.
\newblock {\em \aap} {\bf 2022}, {\em 668},~A101,
  \href{http://arxiv.org/abs/2210.09710}{{\normalfont
  [arXiv:astro-ph.SR/2210.09710]}}.
\newblock {\url{https://doi.org/10.1051/0004-6361/202243789}}.

\bibitem[{Vida} et~al.(2019){Vida}, {Leitzinger}, {Kriskovics}, {Seli},
  {Odert}, {Kov{\'a}cs}, {Korhonen}, and {van
  Driel-Gesztelyi}]{Vida2019A&A...623A..49V}
{Vida}, K.; {Leitzinger}, M.; {Kriskovics}, L.; {Seli}, B.; {Odert}, P.;
  {Kov{\'a}cs}, O.E.; {Korhonen}, H.; {van Driel-Gesztelyi}, L.
\newblock {The quest for stellar coronal mass ejections in late-type stars. I.
  Investigating Balmer-line asymmetries of single stars in Virtual Observatory
  data}.
\newblock {\em \aap} {\bf 2019}, {\em 623},~A49,
  \href{http://arxiv.org/abs/1901.04229}{{\normalfont
  [arXiv:astro-ph.SR/1901.04229]}}.
\newblock {\url{https://doi.org/10.1051/0004-6361/201834264}}.

\bibitem[{Cliver} et~al.(2022){Cliver}, {Schrijver}, {Shibata}, and
  {Usoskin}]{Cliver2022LRSP...19....2C}
{Cliver}, E.W.; {Schrijver}, C.J.; {Shibata}, K.; {Usoskin}, I.G.
\newblock {Extreme solar events}.
\newblock {\em Living Reviews in Solar Physics} {\bf 2022}, {\em 19},~2,
  \href{http://arxiv.org/abs/2205.09265}{{\normalfont
  [arXiv:astro-ph.SR/2205.09265]}}.
\newblock {\url{https://doi.org/10.1007/s41116-022-00033-8}}.

\bibitem[{Hayakawa} et~al.(2017){Hayakawa}, {Iwahashi}, {Ebihara}, {Tamazawa},
  {Shibata}, {Knipp}, {Kawamura}, {Hattori}, {Mase}, {Nakanishi}, and
  {Isobe}]{Hayakawa2017ApJ...850L..31H}
{Hayakawa}, H.; {Iwahashi}, K.; {Ebihara}, Y.; {Tamazawa}, H.; {Shibata}, K.;
  {Knipp}, D.J.; {Kawamura}, A.D.; {Hattori}, K.; {Mase}, K.; {Nakanishi}, I.;
  et~al.
\newblock {Long-lasting Extreme Magnetic Storm Activities in 1770 Found in
  Historical Documents}.
\newblock {\em \apjl} {\bf 2017}, {\em 850},~L31,
  \href{http://arxiv.org/abs/1711.00690}{{\normalfont
  [arXiv:astro-ph.SR/1711.00690]}}.
\newblock {\url{https://doi.org/10.3847/2041-8213/aa9661}}.

\bibitem[{Willis} and {Stephenson}(2001)]{Willis2001AnGeo..19..289W}
{Willis}, D.M.; {Stephenson}, F.R.
\newblock {Solar and auroral evidence for an intense recurrent geomagnetic
  storm during December in AD 1128}.
\newblock {\em Annales Geophysicae} {\bf 2001}, {\em 19},~289--302.
\newblock {\url{https://doi.org/10.5194/angeo-19-289-2001}}.

\bibitem[{Miyake} et~al.(2012){Miyake}, {Nagaya}, {Masuda}, and
  {Nakamura}]{Miyake2012Natur.486..240M}
{Miyake}, F.; {Nagaya}, K.; {Masuda}, K.; {Nakamura}, T.
\newblock {A signature of cosmic-ray increase in AD 774-775 from tree rings in
  Japan}.
\newblock {\em \nat} {\bf 2012}, {\em 486},~240--242.
\newblock {\url{https://doi.org/10.1038/nature11123}}.

\bibitem[{Miyake} et~al.(2013){Miyake}, {Masuda}, and
  {Nakamura}]{Miyake2013JGRA..118.7483M}
{Miyake}, F.; {Masuda}, K.; {Nakamura}, T.
\newblock {Lengths of Schwabe cycles in the seventh and eighth centuries
  indicated by precise measurement of carbon-14 content in tree rings}.
\newblock {\em Journal of Geophysical Research (Space Physics)} {\bf 2013},
  {\em 118},~7483--7487.
\newblock {\url{https://doi.org/10.1002/2012JA018320}}.

\bibitem[{Usoskin}(2013)]{Usoskin2013LRSP...10....1U}
{Usoskin}, I.G.
\newblock {A History of Solar Activity over Millennia}.
\newblock {\em Living Reviews in Solar Physics} {\bf 2013}, {\em 10},~1.
\newblock {\url{https://doi.org/10.12942/lrsp-2013-1}}.

\bibitem[{Mekhaldi} et~al.(2015){Mekhaldi}, {Muscheler}, {Adolphi}, {Aldahan},
  {Beer}, {McConnell}, {Possnert}, {Sigl}, {Svensson}, {Synal}, {Welten}, and
  {Woodruff}]{Mekhaldi2015NatCo...6.8611M}
{Mekhaldi}, F.; {Muscheler}, R.; {Adolphi}, F.; {Aldahan}, A.; {Beer}, J.;
  {McConnell}, J.R.; {Possnert}, G.; {Sigl}, M.; {Svensson}, A.; {Synal}, H.A.;
   et~al.
\newblock {Multiradionuclide evidence for the solar origin of the cosmic-ray
  events of AD 774/5 and 993/4}.
\newblock {\em Nature Communications} {\bf 2015}, {\em 6},~8611.
\newblock {\url{https://doi.org/10.1038/ncomms9611}}.

\bibitem[{Maehara} et~al.(2012){Maehara}, {Shibayama}, {Notsu}, {Notsu},
  {Nagao}, {Kusaba}, {Honda}, {Nogami}, and
  {Shibata}]{Maehara2012Natur.485..478M}
{Maehara}, H.; {Shibayama}, T.; {Notsu}, S.; {Notsu}, Y.; {Nagao}, T.;
  {Kusaba}, S.; {Honda}, S.; {Nogami}, D.; {Shibata}, K.
\newblock {Superflares on solar-type stars}.
\newblock {\em \nat} {\bf 2012}, {\em 485},~478--481.
\newblock {\url{https://doi.org/10.1038/nature11063}}.

\bibitem[{Okamoto} et~al.(2021){Okamoto}, {Notsu}, {Maehara}, {Namekata},
  {Honda}, {Ikuta}, {Nogami}, and {Shibata}]{Okamoto2021ApJ...906...72O}
{Okamoto}, S.; {Notsu}, Y.; {Maehara}, H.; {Namekata}, K.; {Honda}, S.;
  {Ikuta}, K.; {Nogami}, D.; {Shibata}, K.
\newblock {Statistical Properties of Superflares on Solar-type Stars: Results
  Using All of the Kepler Primary Mission Data}.
\newblock {\em \apj} {\bf 2021}, {\em 906},~72,
  \href{http://arxiv.org/abs/2011.02117}{{\normalfont
  [arXiv:astro-ph.SR/2011.02117]}}.
\newblock {\url{https://doi.org/10.3847/1538-4357/abc8f5}}.

\bibitem[{Karoff} et~al.(2016){Karoff}, {Knudsen}, {De Cat}, {Bonanno},
  {Fogtmann-Schulz}, {Fu}, {Frasca}, {Inceoglu}, {Olsen}, {Zhang}, {Hou},
  {Wang}, {Shi}, and {Zhang}]{Karoff2016}
{Karoff}, C.; {Knudsen}, M.F.; {De Cat}, P.; {Bonanno}, A.; {Fogtmann-Schulz},
  A.; {Fu}, J.; {Frasca}, A.; {Inceoglu}, F.; {Olsen}, J.; {Zhang}, Y.;  et~al.
\newblock {Observational evidence for enhanced magnetic activity of superflare
  stars}.
\newblock {\em Nature Communications} {\bf 2016}, {\em 7},~11058.
\newblock {\url{https://doi.org/10.1038/ncomms11058}}.

\bibitem[{Kriskovics} et~al.(2023){Kriskovics}, {K{\H{o}}v{\'a}ri}, {Seli},
  {Ol{\'a}h}, {Vida}, {Henry}, {Granzer}, and
  {G{\"o}rgei}]{Kriskovics2023A&A...674A.143K}
{Kriskovics}, L.; {K{\H{o}}v{\'a}ri}, {\mbox Zs}.; {Seli}, B.; {Ol{\'a}h}, K.;
  {Vida}, K.; {Henry}, G.W.; {Granzer}, T.; {G{\"o}rgei}, A.
\newblock {EI Eridani: A star under the influence. The effect of magnetic
  activity in the short and long term}.
\newblock {\em \aap} {\bf 2023}, {\em 674},~A143,
  \href{http://arxiv.org/abs/2304.13234}{{\normalfont
  [arXiv:astro-ph.SR/2304.13234]}}.
\newblock {\url{https://doi.org/10.1051/0004-6361/202245767}}.

\bibitem[{Gershberg}(1972)]{Gershberg1972Ap&SS..19...75G}
{Gershberg}, R.E.
\newblock {Some results of the cooperative photometric observations of the UV
  Cet-type flare stars in the years 1967 71}.
\newblock {\em \apss} {\bf 1972}, {\em 19},~75--92.
\newblock {\url{https://doi.org/10.1007/BF00643168}}.

\bibitem[{Lacy} et~al.(1976){Lacy}, {Moffett}, and
  {Evans}]{Lacy1976ApJS...30...85L}
{Lacy}, C.H.; {Moffett}, T.J.; {Evans}, D.S.
\newblock {UV Ceti stars: statistical analysis of observational data.}
\newblock {\em \apjs} {\bf 1976}, {\em 30},~85--96.
\newblock {\url{https://doi.org/10.1086/190358}}.

\bibitem[{Shakhovskaia}(1989)]{Shakhovskaia1989SoPh..121..375S}
{Shakhovskaia}, N.I.
\newblock {Stellar flare statistics {\textemdash} Physical consequences}.
\newblock {\em \solphys} {\bf 1989}, {\em 121},~375--386.
\newblock {\url{https://doi.org/10.1007/BF00161707}}.

\bibitem[{Schmidt} et~al.(2007){Schmidt}, {Cruz}, {Bongiorno}, {Liebert}, and
  {Reid}]{Schmidt2007AJ....133.2258S}
{Schmidt}, S.J.; {Cruz}, K.L.; {Bongiorno}, B.J.; {Liebert}, J.; {Reid}, I.N.
\newblock {Activity and Kinematics of Ultracool Dwarfs, Including an Amazing
  Flare Observation}.
\newblock {\em \aj} {\bf 2007}, {\em 133},~2258--2273,
  \href{http://arxiv.org/abs/astro-ph/0701055}{{\normalfont
  [arXiv:astro-ph/astro-ph/0701055]}}.
\newblock {\url{https://doi.org/10.1086/512158}}.

\bibitem[{Hawley} et~al.(2014){Hawley}, {Davenport}, {Kowalski}, {Wisniewski},
  {Hebb}, {Deitrick}, and {Hilton}]{Hawley2014ApJ...797..121H}
{Hawley}, S.L.; {Davenport}, J.R.A.; {Kowalski}, A.F.; {Wisniewski}, J.P.;
  {Hebb}, L.; {Deitrick}, R.; {Hilton}, E.J.
\newblock {Kepler Flares. I. Active and Inactive M Dwarfs}.
\newblock {\em \apj} {\bf 2014}, {\em 797},~121,
  \href{http://arxiv.org/abs/1410.7779}{{\normalfont
  [arXiv:astro-ph.SR/1410.7779]}}.
\newblock {\url{https://doi.org/10.1088/0004-637X/797/2/121}}.

\bibitem[{Zhang} et~al.(2020){Zhang}, {Long}, {Shi}, {Lu}, {Gao}, {Han},
  {Wang}, {Prabhakar}, and {Lamost Mrs
  Collaboration}]{Zhang2020MNRAS.495.1252Z}
{Zhang}, L.Y.; {Long}, L.; {Shi}, J.; {Lu}, H.P.; {Gao}, Q.; {Han}, X.L.;
  {Wang}, H.; {Prabhakar}, M.; {Lamost Mrs Collaboration}.
\newblock {Magnetic activity based on LAMOST medium-resolution spectra and the
  Kepler survey}.
\newblock {\em \mnras} {\bf 2020}, {\em 495},~1252--1270.
\newblock {\url{https://doi.org/10.1093/mnras/staa942}}.

\bibitem[{Zhang} et~al.(2023){Zhang}, {Su}, {Misra}, {Han}, {Meng}, {Pi}, and
  {Yang}]{Zhang2023ApJS..264...17Z}
{Zhang}, L.y.; {Su}, T.; {Misra}, P.; {Han}, X.L.; {Meng}, G.; {Pi}, Q.;
  {Yang}, J.
\newblock {Stellar Parameters and Spectroscopic Properties of TESS Objects
  Observed in the LAMOST Low- and Medium-resolution Spectral Survey}.
\newblock {\em \apjs} {\bf 2023}, {\em 264},~17.
\newblock {\url{https://doi.org/10.3847/1538-4365/ac9b28}}.

\bibitem[{G{\"u}nther} et~al.(2020){G{\"u}nther}, {Zhan}, {Seager}, {Rimmer},
  {Ranjan}, {Stassun}, {Oelkers}, {Daylan}, {Newton}, {Kristiansen}, {Olah},
  {Gillen}, {Rappaport}, {Ricker}, {Vanderspek}, {Latham}, {Winn}, {Jenkins},
  {Glidden}, {Fausnaugh}, {Levine}, {Dittmann}, {Quinn}, {Krishnamurthy}, and
  {Ting}]{Gunther2020AJ....159...60G}
{G{\"u}nther}, M.N.; {Zhan}, Z.; {Seager}, S.; {Rimmer}, P.B.; {Ranjan}, S.;
  {Stassun}, K.G.; {Oelkers}, R.J.; {Daylan}, T.; {Newton}, E.; {Kristiansen},
  M.H.;  et~al.
\newblock {Stellar Flares from the First TESS Data Release: Exploring a New
  Sample of M Dwarfs}.
\newblock {\em \aj} {\bf 2020}, {\em 159},~60,
  \href{http://arxiv.org/abs/1901.00443}{{\normalfont
  [arXiv:astro-ph.EP/1901.00443]}}.
\newblock {\url{https://doi.org/10.3847/1538-3881/ab5d3a}}.

\bibitem[{Gao} et~al.(2022){Gao}, {Liu}, {Yang}, and
  {Zhou}]{Gao2022AJ....164..213G}
{Gao}, D.Y.; {Liu}, H.G.; {Yang}, M.; {Zhou}, J.L.
\newblock {Correcting Stellar Flare Frequency Distributions Detected by TESS
  and Kepler}.
\newblock {\em \aj} {\bf 2022}, {\em 164},~213,
  \href{http://arxiv.org/abs/2301.07552}{{\normalfont
  [arXiv:astro-ph.SR/2301.07552]}}.
\newblock {\url{https://doi.org/10.3847/1538-3881/ac937e}}.

\bibitem[{Mullan} and {Paudel}(2018)]{2018ApJ...854...14M}
{Mullan}, D.J.; {Paudel}, R.R.
\newblock {Frequencies of Flare Occurrence: Interaction between Convection and
  Coronal Loops}.
\newblock {\em \apj} {\bf 2018}, {\em 854},~14,
  \href{http://arxiv.org/abs/1801.07708}{{\normalfont
  [arXiv:astro-ph.SR/1801.07708]}}.
\newblock {\url{https://doi.org/10.3847/1538-4357/aaa960}}.

\bibitem[{Pietras} et~al.(2022){Pietras}, {Falewicz}, {Siarkowski}, {Bicz}, and
  {Pre{\'s}}]{Pietras2022ApJ...935..143P}
{Pietras}, M.; {Falewicz}, R.; {Siarkowski}, M.; {Bicz}, K.; {Pre{\'s}}, P.
\newblock {Statistical Analysis of Stellar Flares from the First Three Years of
  TESS Observations}.
\newblock {\em \apj} {\bf 2022}, {\em 935},~143,
  \href{http://arxiv.org/abs/2207.11039}{{\normalfont
  [arXiv:astro-ph.SR/2207.11039]}}.
\newblock {\url{https://doi.org/10.3847/1538-4357/ac8352}}.

\bibitem[{Yang} et~al.(2023){Yang}, {Zhang}, {Meng}, {Han}, {Misra}, {Yang},
  and {Pi}]{YangZ2023}
{Yang}, Z.; {Zhang}, L.; {Meng}, G.; {Han}, X.L.; {Misra}, P.; {Yang}, J.;
  {Pi}, Q.
\newblock {Properties of flare events based on light curves from the TESS
  survey}.
\newblock {\em \aap} {\bf 2023}, {\em 669},~A15.
\newblock {\url{https://doi.org/10.1051/0004-6361/202142710}}.

\bibitem[{Hawley} and {Pettersen}(1991)]{Hawley1991ApJ...378..725H}
{Hawley}, S.L.; {Pettersen}, B.R.
\newblock {The Great Flare of 1985 April 12 on AD Leonis}.
\newblock {\em \apj} {\bf 1991}, {\em 378},~725.
\newblock {\url{https://doi.org/10.1086/170474}}.

\bibitem[{Dal} and {Evren}(2012)]{Dal2012NewA...17..399D}
{Dal}, H.A.; {Evren}, S.
\newblock {The statistical analyses of flares detected in B band photometry of
  UV Ceti type stars}.
\newblock {\em \na} {\bf 2012}, {\em 17},~399--410,
  \href{http://arxiv.org/abs/1206.3761}{{\normalfont
  [arXiv:astro-ph.SR/1206.3761]}}.
\newblock {\url{https://doi.org/10.1016/j.newast.2011.09.008}}.

\bibitem[{K\H{o}v{\'a}ri} et~al.(2007){K\H{o}v{\'a}ri}, {Vilardell}, {Ribas},
  {Vida}, {van Driel-Gesztelyi}, {Jordi}, and
  {Ol{\'a}h}]{Kovari2007AN....328..904K}
{K\H{o}v{\'a}ri}, {\mbox Zs}.; {Vilardell}, F.; {Ribas}, I.; {Vida}, K.; {van
  Driel-Gesztelyi}, L.; {Jordi}, C.; {Ol{\'a}h}, K.
\newblock {Optical flares from the faint mid-dM star 2MASS J00453912+4140395}.
\newblock {\em Astronomische Nachrichten} {\bf 2007}, {\em 328},~904--908,
  \href{http://arxiv.org/abs/0711.0742}{{\normalfont
  [arXiv:astro-ph/0711.0742]}}.
\newblock {\url{https://doi.org/10.1002/asna.200710756}}.

\bibitem[{Savanov}(2022)]{Savanov2022AstBu..77..431S}
{Savanov}, I.S.
\newblock {Activity of the Young Star TOI 837 with an Exoplanet}.
\newblock {\em Astrophysical Bulletin} {\bf 2022}, {\em 77},~431--436.
\newblock {\url{https://doi.org/10.1134/S1990341322040113}}.

\bibitem[{Roettenbacher} and {Vida}(2018)]{Roettenbacher2018ApJ...868....3R}
{Roettenbacher}, R.M.; {Vida}, K.
\newblock {The Connection between Starspots and Flares on Main-sequence Kepler
  Stars}.
\newblock {\em \apj} {\bf 2018}, {\em 868},~3,
  \href{http://arxiv.org/abs/1810.04762}{{\normalfont
  [arXiv:astro-ph.SR/1810.04762]}}.
\newblock {\url{https://doi.org/10.3847/1538-4357/aae77e}}.

\bibitem[{Ara{\'u}jo} and {Valio}(2021)]{Araujo2021ApJ...922L..23A}
{Ara{\'u}jo}, A.; {Valio}, A.
\newblock {Kepler-411 Star Activity: Connection between Starspots and
  Superflares}.
\newblock {\em \apjl} {\bf 2021}, {\em 922},~L23,
  \href{http://arxiv.org/abs/2111.05452}{{\normalfont
  [arXiv:astro-ph.SR/2111.05452]}}.
\newblock {\url{https://doi.org/10.3847/2041-8213/ac3767}}.

\bibitem[{Lin} et~al.(2023){Lin}, {Wang}, {Deng}, {Deng}, {Mei}, and
  {Zhang}]{Lin2023}
{Lin}, J.; {Wang}, F.; {Deng}, L.; {Deng}, H.; {Mei}, Y.; {Zhang}, X.
\newblock {Evolutionary Relationship between Sunspot Groups and Soft X-Ray
  Flares over Solar Cycles 21-25}.
\newblock {\em \apj} {\bf 2023}, {\em 958},~1.
\newblock {\url{https://doi.org/10.3847/1538-4357/ad0469}}.

\bibitem[{Howard} and {Law}(2021)]{Howard2021ApJ...920...42H}
{Howard}, W.S.; {Law}, N.M.
\newblock {EvryFlare. IV. Detection of Periodicity in Flare Occurrence from
  Cool Stars with TESS}.
\newblock {\em \apj} {\bf 2021}, {\em 920},~42,
  \href{http://arxiv.org/abs/2107.06293}{{\normalfont
  [arXiv:astro-ph.SR/2107.06293]}}.
\newblock {\url{https://doi.org/10.3847/1538-4357/ac142a}}.

\bibitem[{Martin} et~al.(2024){Martin}, {Sethi}, {Armitage}, {Gilbert},
  {Rodr{\'\i}guez Mart{\'\i}nez}, and {Gilbert}]{Martin2024MNRAS.528..963M}
{Martin}, D.V.; {Sethi}, R.; {Armitage}, T.; {Gilbert}, G.J.; {Rodr{\'\i}guez
  Mart{\'\i}nez}, R.; {Gilbert}, E.A.
\newblock {The benchmark M dwarf eclipsing binary CM Draconis with TESS: spots,
  flares, and ultra-precise parameters}.
\newblock {\em \mnras} {\bf 2024}, {\em 528},~963--975,
  \href{http://arxiv.org/abs/2301.10858}{{\normalfont
  [arXiv:astro-ph.SR/2301.10858]}}.
\newblock {\url{https://doi.org/10.1093/mnras/stae015}}.

\bibitem[{Mavridis} and {Avgoloupis}(1986)]{Mavridis1986A&A...154..171M}
{Mavridis}, L.N.; {Avgoloupis}, S.
\newblock {The flare star EV Lac. I. The activity cycle.}
\newblock {\em \aap} {\bf 1986}, {\em 154},~171--175.

\bibitem[{Alekseev} et~al.(2000){Alekseev}, {Chalenko}, and
  {Shakhovsko{\u{i}}}]{Alekseev2000ARep...44..689A}
{Alekseev}, I.Y.; {Chalenko}, V.E.; {Shakhovsko{\u{i}}}, D.N.
\newblock {Rapid UBVRI Photometry of the Active Flare Stars EV Lac and AD Leo}.
\newblock {\em Astronomy Reports} {\bf 2000}, {\em 44},~689--695.
\newblock {\url{https://doi.org/10.1134/1.1312965}}.

\bibitem[{Alekseev}(2005)]{Alekseev2005Ap.....48...20A}
{Alekseev}, I.Y.
\newblock {Spots, activity cycles, and differential rotation on cool stars}.
\newblock {\em Astrophysics} {\bf 2005}, {\em 48},~20--31.
\newblock {\url{https://doi.org/10.1007/s10511-005-0003-x}}.

\bibitem[{Akopian}(2010)]{Akopian2010Ap.....53..544A}
{Akopian}, A.A.
\newblock {Cyclic flaring activity of flare stars}.
\newblock {\em Astrophysics} {\bf 2010}, {\em 53},~544--553.
\newblock {\url{https://doi.org/10.1007/s10511-010-9146-5}}.

\bibitem[{Davenport} et~al.(2020){Davenport}, {Mendoza}, and
  {Hawley}]{Davenport2020AJ....160...36D}
{Davenport}, J.R.A.; {Mendoza}, G.T.; {Hawley}, S.L.
\newblock {10 Years of Stellar Activity for GJ 1243}.
\newblock {\em \aj} {\bf 2020}, {\em 160},~36,
  \href{http://arxiv.org/abs/2005.10281}{{\normalfont
  [arXiv:astro-ph.SR/2005.10281]}}.
\newblock {\url{https://doi.org/10.3847/1538-3881/ab9536}}.

\bibitem[{Candelaresi} et~al.(2014){Candelaresi}, {Hillier}, {Maehara},
  {Brandenburg}, and {Shibata}]{Candelaresi2014ApJ...792...67C}
{Candelaresi}, S.; {Hillier}, A.; {Maehara}, H.; {Brandenburg}, A.; {Shibata},
  K.
\newblock {Superflare Occurrence and Energies on G-, K-, and M-type Dwarfs}.
\newblock {\em \apj} {\bf 2014}, {\em 792},~67,
  \href{http://arxiv.org/abs/1405.1453}{{\normalfont
  [arXiv:astro-ph.SR/1405.1453]}}.
\newblock {\url{https://doi.org/10.1088/0004-637X/792/1/67}}.

\bibitem[{Raetz} et~al.(2020){Raetz}, {Stelzer}, {Damasso}, and
  {Scholz}]{Raetz2020A&A...637A..22R}
{Raetz}, S.; {Stelzer}, B.; {Damasso}, M.; {Scholz}, A.
\newblock {Rotation-activity relations and flares of M dwarfs with K2 long- and
  short-cadence data}.
\newblock {\em \aap} {\bf 2020}, {\em 637},~A22,
  \href{http://arxiv.org/abs/2003.11937}{{\normalfont
  [arXiv:astro-ph.SR/2003.11937]}}.
\newblock {\url{https://doi.org/10.1051/0004-6361/201937350}}.

\bibitem[{Gizis} et~al.(2017{\natexlab{a}}){Gizis}, {Paudel}, {Schmidt},
  {Williams}, and {Burgasser}]{Gizis2017ApJ...838...22G}
{Gizis}, J.E.; {Paudel}, R.R.; {Schmidt}, S.J.; {Williams}, P.K.G.;
  {Burgasser}, A.J.
\newblock {K2 Ultracool Dwarfs Survey. I. Photometry of an L Dwarf Superflare}.
\newblock {\em \apj} {\bf 2017}, {\em 838},~22,
  \href{http://arxiv.org/abs/1611.07080}{{\normalfont
  [arXiv:astro-ph.SR/1611.07080]}}.
\newblock {\url{https://doi.org/10.3847/1538-4357/aa6197}}.

\bibitem[{Gizis} et~al.(2017{\natexlab{b}}){Gizis}, {Paudel}, {Mullan},
  {Schmidt}, {Burgasser}, and {Williams}]{Gizis2017ApJ...845...33G}
{Gizis}, J.E.; {Paudel}, R.R.; {Mullan}, D.; {Schmidt}, S.J.; {Burgasser},
  A.J.; {Williams}, P.K.G.
\newblock {K2 Ultracool Dwarfs Survey. II. The White Light Flare Rate of Young
  Brown Dwarfs}.
\newblock {\em \apj} {\bf 2017}, {\em 845},~33,
  \href{http://arxiv.org/abs/1703.08745}{{\normalfont
  [arXiv:astro-ph.SR/1703.08745]}}.
\newblock {\url{https://doi.org/10.3847/1538-4357/aa7da0}}.

\bibitem[{Murray} et~al.(2022){Murray}, {Queloz}, {Gillon}, {Demory}, {Triaud},
  {de Wit}, {Burdanov}, {Chinchilla}, {Delrez}, {Dransfield}, {Ducrot},
  {Garcia}, {G{\'o}mez Maqueo Chew}, {G{\"u}nther}, {Jehin}, {McCormac},
  {Niraula}, {Pedersen}, {Pozuelos}, {Rackham}, {Schanche}, {Sebastian},
  {Thompson}, {Timmermans}, and {Wells}]{2022MNRAS.513.2615M}
{Murray}, C.A.; {Queloz}, D.; {Gillon}, M.; {Demory}, B.O.; {Triaud}, A.H.M.J.;
  {de Wit}, J.; {Burdanov}, A.; {Chinchilla}, P.; {Delrez}, L.; {Dransfield},
  G.;  et~al.
\newblock {A study of flares in the ultra-cool regime from SPECULOOS-South}.
\newblock {\em \mnras} {\bf 2022}, {\em 513},~2615--2634,
  \href{http://arxiv.org/abs/2204.10417}{{\normalfont
  [arXiv:astro-ph.SR/2204.10417]}}.
\newblock {\url{https://doi.org/10.1093/mnras/stac1078}}.

\bibitem[{Petrucci} et~al.(2024){Petrucci}, {G{\'o}mez Maqueo Chew},
  {Jofr{\'e}}, {Segura}, and {Ferrero}]{2024MNRAS.527.8290P}
{Petrucci}, R.P.; {G{\'o}mez Maqueo Chew}, Y.; {Jofr{\'e}}, E.; {Segura}, A.;
  {Ferrero}, L.V.
\newblock {Exploring the photometric variability of ultra-cool dwarfs with
  TESS}.
\newblock {\em \mnras} {\bf 2024}, {\em 527},~8290--8304,
  \href{http://arxiv.org/abs/2311.13591}{{\normalfont
  [arXiv:astro-ph.SR/2311.13591]}}.
\newblock {\url{https://doi.org/10.1093/mnras/stad3720}}.

\bibitem[{Paudel} et~al.(2018){Paudel}, {Gizis}, {Mullan}, {Schmidt},
  {Burgasser}, {Williams}, and {Berger}]{2018ApJ...858...55P}
{Paudel}, R.R.; {Gizis}, J.E.; {Mullan}, D.J.; {Schmidt}, S.J.; {Burgasser},
  A.J.; {Williams}, P.K.G.; {Berger}, E.
\newblock {K2 Ultracool Dwarfs Survey. III. White Light Flares Are Ubiquitous
  in M6-L0 Dwarfs}.
\newblock {\em \apj} {\bf 2018}, {\em 858},~55,
  \href{http://arxiv.org/abs/1803.07708}{{\normalfont
  [arXiv:astro-ph.SR/1803.07708]}}.
\newblock {\url{https://doi.org/10.3847/1538-4357/aab8fe}}.

\bibitem[{Glazier} et~al.(2020){Glazier}, {Howard}, {Corbett}, {Law},
  {Ratzloff}, {Fors}, and {del Ser}]{2020ApJ...900...27G}
{Glazier}, A.L.; {Howard}, W.S.; {Corbett}, H.; {Law}, N.M.; {Ratzloff}, J.K.;
  {Fors}, O.; {del Ser}, D.
\newblock {Evryscope and K2 Constraints on TRAPPIST-1 Superflare Occurrence and
  Planetary Habitability}.
\newblock {\em \apj} {\bf 2020}, {\em 900},~27,
  \href{http://arxiv.org/abs/2006.14712}{{\normalfont
  [arXiv:astro-ph.EP/2006.14712]}}.
\newblock {\url{https://doi.org/10.3847/1538-4357/aba4a6}}.

\bibitem[{Howard} et~al.(2023){Howard}, {Kowalski}, {Flagg}, {MacGregor},
  {Lim}, {Radica}, {Piaulet}, {Roy}, {Lafreni{\`e}re}, {Benneke}, {Brown},
  {Espinoza}, {Doyon}, {Coulombe}, {Johnstone}, {Cowan}, {Jayawardhana},
  {Turner}, and {Dang}]{2023ApJ...959...64H}
{Howard}, W.S.; {Kowalski}, A.F.; {Flagg}, L.; {MacGregor}, M.A.; {Lim}, O.;
  {Radica}, M.; {Piaulet}, C.; {Roy}, P.A.; {Lafreni{\`e}re}, D.; {Benneke},
  B.;  et~al.
\newblock {Characterizing the Near-infrared Spectra of Flares from TRAPPIST-1
  during JWST Transit Spectroscopy Observations}.
\newblock {\em \apj} {\bf 2023}, {\em 959},~64,
  \href{http://arxiv.org/abs/2310.03792}{{\normalfont
  [arXiv:astro-ph.EP/2310.03792]}}.
\newblock {\url{https://doi.org/10.3847/1538-4357/acfe75}}.

\bibitem[{Seli} et~al.(2021){Seli}, {Vida}, {Mo{\'o}r}, {P{\'a}l}, and
  {Ol{\'a}h}]{2021A&A...650A.138S}
{Seli}, B.; {Vida}, K.; {Mo{\'o}r}, A.; {P{\'a}l}, A.; {Ol{\'a}h}, K.
\newblock {Activity of TRAPPIST-1 analog stars observed with TESS}.
\newblock {\em \aap} {\bf 2021}, {\em 650},~A138,
  \href{http://arxiv.org/abs/2103.13540}{{\normalfont
  [arXiv:astro-ph.SR/2103.13540]}}.
\newblock {\url{https://doi.org/10.1051/0004-6361/202040098}}.

\bibitem[{Strassmeier} et~al.(2008){Strassmeier}, {Briguglio}, {Granzer},
  {Tosti}, {Divarano}, {Savanov}, {Bagaglia}, {Castellini}, {Mancini},
  {Nucciarelli}, {Straniero}, {Distefano}, {Messina}, and
  {Cutispoto}]{2008A&A...490..287S}
{Strassmeier}, K.G.; {Briguglio}, R.; {Granzer}, T.; {Tosti}, G.; {Divarano},
  I.; {Savanov}, I.; {Bagaglia}, M.; {Castellini}, S.; {Mancini}, A.;
  {Nucciarelli}, G.;  et~al.
\newblock {First time-series optical photometry from Antarctica. sIRAIT
  monitoring of the RS CVn binary V841 Centauri and the
  {\ensuremath{\delta}}-Scuti star V1034 Centauri}.
\newblock {\em \aap} {\bf 2008}, {\em 490},~287--295.
\newblock {\url{https://doi.org/10.1051/0004-6361:200810379}}.

\bibitem[{Ol{\'a}h} et~al.(2021){Ol{\'a}h}, {K{\H{o}}v{\'a}ri}, {G{\"u}nther},
  {Vida}, {Gaulme}, {Seli}, and {P{\'a}l}]{Olah2021A&A...647A..62O}
{Ol{\'a}h}, K.; {K{\H{o}}v{\'a}ri}, {\mbox Zs}.; {G{\"u}nther}, M.N.; {Vida},
  K.; {Gaulme}, P.; {Seli}, B.; {P{\'a}l}, A.
\newblock {Toward the true number of flaring giant stars in the Kepler field.
  Are their flaring specialities associated with their being giant stars?}
\newblock {\em \aap} {\bf 2021}, {\em 647},~A62,
  \href{http://arxiv.org/abs/2010.07623}{{\normalfont
  [arXiv:astro-ph.SR/2010.07623]}}.
\newblock {\url{https://doi.org/10.1051/0004-6361/202039674}}.

\bibitem[{Gaulme} et~al.(2014){Gaulme}, {Jackiewicz}, {Appourchaux}, and
  {Mosser}]{2014ApJ...785....5G}
{Gaulme}, P.; {Jackiewicz}, J.; {Appourchaux}, T.; {Mosser}, B.
\newblock {Surface Activity and Oscillation Amplitudes of Red Giants in
  Eclipsing Binaries}.
\newblock {\em \apj} {\bf 2014}, {\em 785},~5,
  \href{http://arxiv.org/abs/1402.3027}{{\normalfont
  [arXiv:astro-ph.SR/1402.3027]}}.
\newblock {\url{https://doi.org/10.1088/0004-637X/785/1/5}}.

\bibitem[{Maehara} et~al.(2015){Maehara}, {Shibayama}, {Notsu}, {Notsu},
  {Honda}, {Nogami}, and {Shibata}]{2015EP&S...67...59M}
{Maehara}, H.; {Shibayama}, T.; {Notsu}, Y.; {Notsu}, S.; {Honda}, S.;
  {Nogami}, D.; {Shibata}, K.
\newblock {Statistical properties of superflares on solar-type stars based on
  1-min cadence data}.
\newblock {\em Earth, Planets and Space} {\bf 2015}, {\em 67},~59,
  \href{http://arxiv.org/abs/1504.00074}{{\normalfont
  [arXiv:astro-ph.SR/1504.00074]}}.
\newblock {\url{https://doi.org/10.1186/s40623-015-0217-z}}.

\bibitem[{Gehan} et~al.(2024){Gehan}, {Godoy-Rivera}, and
  {Gaulme}]{Gehan2024arXiv240113549G}
{Gehan}, C.; {Godoy-Rivera}, D.; {Gaulme}, P.
\newblock {Magnetic activity of red giants: a near-UV and H$\alpha$ view, and
  the enhancing role of tidal interactions}.
\newblock {\em arXiv e-prints} {\bf 2024}, p. arXiv:2401.13549,
  \href{http://arxiv.org/abs/2401.13549}{{\normalfont
  [arXiv:astro-ph.SR/2401.13549]}}.
\newblock {\url{https://doi.org/10.48550/arXiv.2401.13549}}.

\bibitem[{Ol{\'a}h}(2006)]{Olah2006Ap&SS.304..145O}
{Ol{\'a}h}, K.
\newblock {Active Longitudes in Close Binaries}.
\newblock {\em \apss} {\bf 2006}, {\em 304},~145--148.
\newblock {\url{https://doi.org/10.1007/s10509-006-9096-x}}.

\bibitem[{Moss} and {Tuominen}(1997)]{Moss1997A&A...321..151M}
{Moss}, D.; {Tuominen}, I.
\newblock {Magnetic field generation in close binary systems.}
\newblock {\em \aap} {\bf 1997}, {\em 321},~151--158.

\bibitem[{Moss} et~al.(1995){Moss}, {Barker}, {Brandenburg}, and
  {Tuominen}]{Moss1995A&A...294..155M}
{Moss}, D.; {Barker}, D.M.; {Brandenburg}, A.; {Tuominen}, I.
\newblock {Nonaxisymmetric dynamo solutions and extended starspots on late-type
  stars.}
\newblock {\em \aap} {\bf 1995}, {\em 294},~155--164.

\bibitem[{K{\H{o}}v{\'a}ri} et~al.(2017){K{\H{o}}v{\'a}ri}, {Ol{\'a}h},
  {Kriskovics}, {Vida}, {Forg{\'a}cs-Dajka}, and
  {Strassmeier}]{Kovari2017AN....338..903K}
{K{\H{o}}v{\'a}ri}, {\mbox Zs}.; {Ol{\'a}h}, K.; {Kriskovics}, L.; {Vida}, K.;
  {Forg{\'a}cs-Dajka}, E.; {Strassmeier}, K.G.
\newblock {Rotation-differential rotation relationships for late-type single
  and binary stars from Doppler imaging}.
\newblock {\em Astronomische Nachrichten} {\bf 2017}, {\em 338},~903--909,
  \href{http://arxiv.org/abs/1709.09001}{{\normalfont
  [arXiv:astro-ph.SR/1709.09001]}}.
\newblock {\url{https://doi.org/10.1002/asna.201713400}}.

\bibitem[{Pi} et~al.(2019){Pi}, {Zhang}, {Bi}, {Han}, {Lu}, {Yue}, {Long}, and
  {Yan}]{Pi2019ApJ...877...75P}
{Pi}, Q.f.; {Zhang}, L.y.; {Bi}, S.l.; {Han}, X.L.; {Lu}, H.p.; {Yue}, Q.;
  {Long}, L.; {Yan}, Y.
\newblock {Magnetic Activity and Orbital Period Study for the Short-period RS
  CVn-type Eclipsing Binary DV Psc}.
\newblock {\em \apj} {\bf 2019}, {\em 877},~75.
\newblock {\url{https://doi.org/10.3847/1538-4357/ab19c3}}.

\bibitem[{Holzwarth} and
  {Sch{\"u}ssler}(2003{\natexlab{a}})]{Holzwarth2003A&A...405..291H}
{Holzwarth}, V.; {Sch{\"u}ssler}, M.
\newblock {Dynamics of magnetic flux tubes in close binary stars. I.
  Equilibrium and stability properties}.
\newblock {\em \aap} {\bf 2003}, {\em 405},~291--301,
  \href{http://arxiv.org/abs/astro-ph/0304496}{{\normalfont
  [arXiv:astro-ph/astro-ph/0304496]}}.
\newblock {\url{https://doi.org/10.1051/0004-6361:20030582}}.

\bibitem[{Holzwarth} and
  {Sch{\"u}ssler}(2003{\natexlab{b}})]{Holzwarth2003A&A...405..303H}
{Holzwarth}, V.; {Sch{\"u}ssler}, M.
\newblock {Dynamics of magnetic flux tubes in close binary stars. II. Nonlinear
  evolution and surface distributions}.
\newblock {\em \aap} {\bf 2003}, {\em 405},~303--311,
  \href{http://arxiv.org/abs/astro-ph/0304498}{{\normalfont
  [arXiv:astro-ph/astro-ph/0304498]}}.
\newblock {\url{https://doi.org/10.1051/0004-6361:20030584}}.

\bibitem[{K{\H{o}}v{\'a}ri} et~al.(2021){K{\H{o}}v{\'a}ri}, {Kriskovics},
  {Ol{\'a}h}, {Odert}, {Leitzinger}, {Seli}, {Vida}, {Borkovits}, and
  {Carroll}]{Kovari2021A&A...650A.158K}
{K{\H{o}}v{\'a}ri}, {\mbox Zs}.; {Kriskovics}, L.; {Ol{\'a}h}, K.; {Odert}, P.;
  {Leitzinger}, M.; {Seli}, B.; {Vida}, K.; {Borkovits}, T.; {Carroll}, T.
\newblock {A confined dynamo: Magnetic activity of the K-dwarf component in the
  pre-cataclysmic binary system V471 Tauri}.
\newblock {\em \aap} {\bf 2021}, {\em 650},~A158,
  \href{http://arxiv.org/abs/2103.02041}{{\normalfont
  [arXiv:astro-ph.SR/2103.02041]}}.
\newblock {\url{https://doi.org/10.1051/0004-6361/202140707}}.

\bibitem[{{\v{S}}melcer} et~al.(2017){{\v{S}}melcer}, {Wolf},
  {Ku{\v{c}}{\'a}kov{\'a}}, {B{\'\i}lek}, {Dubovsk{\'y}}, {Ho{\v{n}}kov{\'a}},
  and {Vra{\v{s}}til}]{Smelcer2017MNRAS.466.2542S}
{{\v{S}}melcer}, L.; {Wolf}, M.; {Ku{\v{c}}{\'a}kov{\'a}}, H.; {B{\'\i}lek},
  F.; {Dubovsk{\'y}}, P.; {Ho{\v{n}}kov{\'a}}, K.; {Vra{\v{s}}til}, J.
\newblock {Flare activity on low-mass eclipsing binary GJ 3236*}.
\newblock {\em \mnras} {\bf 2017}, {\em 466},~2542--2546.
\newblock {\url{https://doi.org/10.1093/mnras/stw3179}}.

\bibitem[{Tan} and {Cheng}(2013)]{TanCheng2013}
{Tan}, B.; {Cheng}, Z.
\newblock {The mid-term and long-term solar quasi-periodic cycles and the
  possible relationship with planetary motions}.
\newblock {\em \apss} {\bf 2013}, {\em 343},~511--521,
  \href{http://arxiv.org/abs/1210.1450}{{\normalfont
  [arXiv:astro-ph.SR/1210.1450]}}.
\newblock {\url{https://doi.org/10.1007/s10509-012-1272-6}}.

\bibitem[{Stefani} et~al.(2024){Stefani}, {Horstmann}, {Klevs},
  {Mamatsashvili}, and {Weier}]{Stefani2024}
{Stefani}, F.; {Horstmann}, G.M.; {Klevs}, M.; {Mamatsashvili}, G.; {Weier}, T.
\newblock {Rieger, Schwabe, Suess-de Vries: The Sunny Beats of Resonance}.
\newblock {\em \solphys} {\bf 2024}, {\em 299},~51,
  \href{http://arxiv.org/abs/2309.00666}{{\normalfont
  [arXiv:astro-ph.SR/2309.00666]}}.
\newblock {\url{https://doi.org/10.1007/s11207-024-02295-x}}.

\bibitem[{Cherkis} and {Lyutikov}(2021)]{Cherkis2021ApJ...923...13C}
{Cherkis}, S.A.; {Lyutikov}, M.
\newblock {Magnetic Topology in Coupled Binaries, Spin-orbital Resonances, and
  Flares}.
\newblock {\em \apj} {\bf 2021}, {\em 923},~13,
  \href{http://arxiv.org/abs/2107.09702}{{\normalfont
  [arXiv:astro-ph.HE/2107.09702]}}.
\newblock {\url{https://doi.org/10.3847/1538-4357/ac29b8}}.

\bibitem[{Simon} et~al.(1980){Simon}, {Linsky}, and
  {Schiffer}]{Simon1980ApJ...239..911S}
{Simon}, T.; {Linsky}, J.L.; {Schiffer}, F.~H., I.
\newblock {IUE spectra of a flare in the RS Canum Venaticorum-type system UX
  Arietis.}
\newblock {\em \apj} {\bf 1980}, {\em 239},~911--918.
\newblock {\url{https://doi.org/10.1086/158178}}.

\bibitem[{Ferreira} and
  {Mendoza-Brice{\~n}o}(2005)]{Ferreire2005A&A...433.1055F}
{Ferreira}, J.M.; {Mendoza-Brice{\~n}o}, C.A.
\newblock {Coronal mass transfer in interbinary loops}.
\newblock {\em \aap} {\bf 2005}, {\em 433},~1055--1061.
\newblock {\url{https://doi.org/10.1051/0004-6361:20041926}}.

\bibitem[{Decampli} and {Baliunas}(1979)]{Decampli1979ApJ...230..815D}
{Decampli}, W.M.; {Baliunas}, S.L.
\newblock {What tides and flares do to RS Canum Venaticorum binaries.}
\newblock {\em \apj} {\bf 1979}, {\em 230},~815--821.
\newblock {\url{https://doi.org/10.1086/157140}}.

\bibitem[{Hall} and {Kreiner}(1980)]{Hall1980AcA....30..387H}
{Hall}, D.S.; {Kreiner}, J.M.
\newblock {Period changes and mass loss rates in 34 RS CVn binaries.}
\newblock {\em \actaa} {\bf 1980}, {\em 30},~387--451.

\bibitem[{Mathieu} et~al.(1997){Mathieu}, {Stassun}, {Basri}, {Jensen},
  {Johns-Krull}, {Valenti}, and {Hartmann}]{Mathieu1997AJ....113.1841M}
{Mathieu}, R.D.; {Stassun}, K.; {Basri}, G.; {Jensen}, E.L.N.; {Johns-Krull},
  C.M.; {Valenti}, J.A.; {Hartmann}, L.W.
\newblock {The Classical T Tauri Spectroscopic Binary DQ Tau.I.Orbital Elements
  and Light Curves}.
\newblock {\em \aj} {\bf 1997}, {\em 113},~1841.
\newblock {\url{https://doi.org/10.1086/118395}}.

\bibitem[{Salter} et~al.(2010){Salter}, {K{\'o}sp{\'a}l}, {Getman},
  {Hogerheijde}, {van Kempen}, {Carpenter}, {Blake}, and
  {Wilner}]{Salter2010A&A...521A..32S}
{Salter}, D.M.; {K{\'o}sp{\'a}l}, {\'A}.; {Getman}, K.V.; {Hogerheijde}, M.R.;
  {van Kempen}, T.A.; {Carpenter}, J.M.; {Blake}, G.A.; {Wilner}, D.
\newblock {Recurring millimeter flares as evidence for star-star magnetic
  reconnection events in the DQ Tauri PMS binary system}.
\newblock {\em \aap} {\bf 2010}, {\em 521},~A32,
  \href{http://arxiv.org/abs/1008.0981}{{\normalfont
  [arXiv:astro-ph.SR/1008.0981]}}.
\newblock {\url{https://doi.org/10.1051/0004-6361/201015197}}.

\bibitem[{Walter} et~al.(1980){Walter}, {Cash}, {Charles}, and
  {Bowyer}]{Walter1980ApJ...236..212W}
{Walter}, F.M.; {Cash}, W.; {Charles}, P.A.; {Bowyer}, C.S.
\newblock {X-rays from RS CVn systems : a HEAO 1 survey and the development of
  a coronal model.}
\newblock {\em \apj} {\bf 1980}, {\em 236},~212--218.
\newblock {\url{https://doi.org/10.1086/157735}}.

\bibitem[{Pres} et~al.(1995){Pres}, {Siarkowski}, and
  {Sylwester}]{Pres1995MNRAS.275...43P}
{Pres}, P.; {Siarkowski}, M.; {Sylwester}, J.
\newblock {Soft X-ray imaging of the TY Pyx binary system - II. Modelling the
  interconnecting loop-like structure}.
\newblock {\em \mnras} {\bf 1995}, {\em 275},~43--55.
\newblock {\url{https://doi.org/10.1093/mnras/275.1.43}}.

\bibitem[{Siarkowski} et~al.(1996){Siarkowski}, {Pres}, {Drake}, {White}, and
  {Singh}]{Siarkowski1996ApJ...473..470S}
{Siarkowski}, M.; {Pres}, P.; {Drake}, S.A.; {White}, N.E.; {Singh}, K.P.
\newblock {Corona(e) of AR Lacertae. II. The Spatial Structure}.
\newblock {\em \apj} {\bf 1996}, {\em 473},~470.
\newblock {\url{https://doi.org/10.1086/178159}}.

\bibitem[{Singh} and {Pandey}(2022)]{Singh2022ApJ...934...20S}
{Singh}, G.; {Pandey}, J.C.
\newblock {An X-Ray Study of Coronally Connected Active Eclipsing Binaries}.
\newblock {\em \apj} {\bf 2022}, {\em 934},~20,
  \href{http://arxiv.org/abs/2207.04253}{{\normalfont
  [arXiv:astro-ph.SR/2207.04253]}}.
\newblock {\url{https://doi.org/10.3847/1538-4357/ac7716}}.

\bibitem[{Ilin} et~al.(2024){Ilin}, {Poppenh{\"a}ger}, {Chebly}, {Ili{\'c}},
  and {Alvarado-G{\'o}mez}]{Ilin2024MNRAS.527.3395I}
{Ilin}, E.; {Poppenh{\"a}ger}, K.; {Chebly}, J.; {Ili{\'c}}, N.;
  {Alvarado-G{\'o}mez}, J.D.
\newblock {Planetary perturbers: flaring star-planet interactions in Kepler and
  TESS}.
\newblock {\em \mnras} {\bf 2024}, {\em 527},~3395--3417,
  \href{http://arxiv.org/abs/2311.04316}{{\normalfont
  [arXiv:astro-ph.SR/2311.04316]}}.
\newblock {\url{https://doi.org/10.1093/mnras/stad3398}}.

\bibitem[{Shkolnik} et~al.(2008){Shkolnik}, {Bohlender}, {Walker}, and {Collier
  Cameron}]{Shkolnik2008ApJ...676..628S}
{Shkolnik}, E.; {Bohlender}, D.A.; {Walker}, G.A.H.; {Collier Cameron}, A.
\newblock {The On/Off Nature of Star-Planet Interactions}.
\newblock {\em \apj} {\bf 2008}, {\em 676},~628--638,
  \href{http://arxiv.org/abs/0712.0004}{{\normalfont
  [arXiv:astro-ph/0712.0004]}}.
\newblock {\url{https://doi.org/10.1086/527351}}.

\bibitem[{Lanza}(2009)]{Lanza2009A&A...505..339L}
{Lanza}, A.F.
\newblock {Stellar coronal magnetic fields and star-planet interaction}.
\newblock {\em \aap} {\bf 2009}, {\em 505},~339--350,
  \href{http://arxiv.org/abs/0906.1738}{{\normalfont
  [arXiv:astro-ph.SR/0906.1738]}}.
\newblock {\url{https://doi.org/10.1051/0004-6361/200912367}}.

\bibitem[{Cohen} et~al.(2011){Cohen}, {Kashyap}, {Drake}, {Sokolov},
  {Garraffo}, and {Gombosi}]{Cohen2011ApJ...733...67C}
{Cohen}, O.; {Kashyap}, V.L.; {Drake}, J.J.; {Sokolov}, I.V.; {Garraffo}, C.;
  {Gombosi}, T.I.
\newblock {The Dynamics of Stellar Coronae Harboring Hot Jupiters. I. A
  Time-dependent Magnetohydrodynamic Simulation of the Interplanetary
  Environment in the HD 189733 Planetary System}.
\newblock {\em \apj} {\bf 2011}, {\em 733},~67,
  \href{http://arxiv.org/abs/1101.4825}{{\normalfont
  [arXiv:astro-ph.SR/1101.4825]}}.
\newblock {\url{https://doi.org/10.1088/0004-637X/733/1/67}}.

\bibitem[{Den} and {Kornienko}(1993)]{Den1993}
{Den}, O.E.; {Kornienko}, G.I.
\newblock {Mass ejection during the flare of 12 March 1989 based on H{$\alpha$}
  filtergrams and spectrograms}.
\newblock {\em Astronomy Reports} {\bf 1993}, {\em 37},~76--82.

\bibitem[{Ding} et~al.(2003){Ding}, {Chen}, {Li}, and {Chen}]{Ding2003}
{Ding}, M.D.; {Chen}, Q.R.; {Li}, J.P.; {Chen}, P.F.
\newblock {H{$\alpha$} and Hard X-Ray Observations of a Two-Ribbon Flare
  Associated with a Filament Eruption}.
\newblock {\em \apj} {\bf 2003}, {\em 598},~683--688,
  \href{http://arxiv.org/abs/arXiv:astro-ph/0308085}{{\normalfont
  [arXiv:astro-ph/0308085]}}.
\newblock {\url{https://doi.org/10.1086/378877}}.

\bibitem[{Ichimoto} et~al.(2017){Ichimoto}, {Ishii}, {Otsuji}, {Kimura},
  {Nakatani}, {Kaneda}, {Nagata}, {UeNo}, {Hirose}, {Cabezas}, and
  {Morita}]{Ichimoto2017}
{Ichimoto}, K.; {Ishii}, T.T.; {Otsuji}, K.; {Kimura}, G.; {Nakatani}, Y.;
  {Kaneda}, N.; {Nagata}, S.; {UeNo}, S.; {Hirose}, K.; {Cabezas}, D.;  et~al.
\newblock {A New Solar Imaging System for Observing High-Speed Eruptions: Solar
  Dynamics Doppler Imager (SDDI)}.
\newblock {\em \solphys} {\bf 2017}, {\em 292},~63.
\newblock {\url{https://doi.org/10.1007/s11207-017-1082-7}}.

\bibitem[{Namekata} et~al.(2022){Namekata}, {Ichimoto}, {Ishii}, and
  {Shibata}]{Namekata2022b}
{Namekata}, K.; {Ichimoto}, K.; {Ishii}, T.T.; {Shibata}, K.
\newblock {Sun-as-a-star Analysis of H{\ensuremath{\alpha}} Spectra of a Solar
  Flare Observed by SMART/SDDI: Time Evolution of Red Asymmetry and Line
  Broadening}.
\newblock {\em \apj} {\bf 2022}, {\em 933},~209,
  \href{http://arxiv.org/abs/2206.01395}{{\normalfont
  [arXiv:astro-ph.SR/2206.01395]}}.
\newblock {\url{https://doi.org/10.3847/1538-4357/ac75cd}}.

\bibitem[{Otsu} et~al.(2022){Otsu}, {Asai}, {Ichimoto}, {Ishii}, and
  {Namekata}]{Otsu2022}
{Otsu}, T.; {Asai}, A.; {Ichimoto}, K.; {Ishii}, T.T.; {Namekata}, K.
\newblock {Sun-as-a-star Analyses of Various Solar Active Events Using
  H{\ensuremath{\alpha}} Spectral Images Taken by SMART/SDDI}.
\newblock {\em \apj} {\bf 2022}, {\em 939},~98,
  \href{http://arxiv.org/abs/2210.02819}{{\normalfont
  [arXiv:astro-ph.SR/2210.02819]}}.
\newblock {\url{https://doi.org/10.3847/1538-4357/ac9730}}.

\bibitem[{Otsu} and {Asai}(2024)]{Otsu2024}
{Otsu}, T.; {Asai}, A.
\newblock {Multiwavelength Sun-as-a-star Analysis of the M8.7 Flare on 2022
  October 2 Using H{\ensuremath{\alpha}} and EUV Spectra Taken by SMART/SDDI
  and SDO/EVE}.
\newblock {\em \apj} {\bf 2024}, {\em 964},~75,
  \href{http://arxiv.org/abs/2402.00589}{{\normalfont
  [arXiv:astro-ph.SR/2402.00589]}}.
\newblock {\url{https://doi.org/10.3847/1538-4357/ad24ec}}.

\bibitem[{Leitzinger} et~al.(2021){Leitzinger}, {Odert}, {Leka}, {Heinzel}, and
  {Dissauer}]{Leitzinger2021}
{Leitzinger}, M.; {Odert}, P.; {Leka}, K.D.; {Heinzel}, P.; {Dissauer}, K.
\newblock {Constraining stellar CMEs by solar observations}.
\newblock In Proceedings of the AGU Fall Meeting Abstracts,  12 2021, Vol.
  2021, pp. U43B--06.

\bibitem[{Guenther} and {Emerson}(1997)]{Guenther1997}
{Guenther}, E.W.; {Emerson}, J.P.
\newblock {Spectrophotometry of flares and short time scale variations in weak
  line, and classical T Tauri stars in Chamaeleon.}
\newblock {\em \aap} {\bf 1997}, {\em 321},~803--810.

\bibitem[{Gunn} et~al.(1994){Gunn}, {Doyle}, {Mathioudakis}, {Houdebine}, and
  {Avgoloupis}]{Gunn1994}
{Gunn}, A.G.; {Doyle}, J.G.; {Mathioudakis}, M.; {Houdebine}, E.R.;
  {Avgoloupis}, S.
\newblock {High-velocity evaporation during a flare on AT Microscopii}.
\newblock {\em \aap} {\bf 1994}, {\em 285},~489--496.

\bibitem[{{\c{S}}enavc{\i}} et~al.(2018){{\c{S}}enavc{\i}}, {Bahar}, {Montes},
  {Zola}, {Hussain}, {Frasca}, {I{\c{s}}{\i}k}, and
  {Y{\"o}r{\"u}ko{\v{g}}lu}]{Senavci2018}
{{\c{S}}enavc{\i}}, H.V.; {Bahar}, E.; {Montes}, D.; {Zola}, S.; {Hussain},
  G.A.J.; {Frasca}, A.; {I{\c{s}}{\i}k}, E.; {Y{\"o}r{\"u}ko{\v{g}}lu}, O.
\newblock {Star-spot distributions and chromospheric activity on the RS CVn
  type eclipsing binary SV Cam}.
\newblock {\em \mnras} {\bf 2018}, {\em 479},~875--889,
  \href{http://arxiv.org/abs/1806.01801}{{\normalfont
  [arXiv:astro-ph.SR/1806.01801]}}.
\newblock {\url{https://doi.org/10.1093/mnras/sty1469}}.

\bibitem[{Leitzinger} et~al.(2024){Leitzinger}, {Odert}, and
  {Greimel}]{Leitzinger2024}
{Leitzinger}, M.; {Odert}, P.; {Greimel}, R.
\newblock {Observations and detectability of young Suns’ flaring and CME
  activity in optical spectra}.
\newblock {\em \mnras} {\bf 2024}, {\em submitted}.

\bibitem[{Namekata} et~al.(2024){Namekata}, {Airapetian}, {Petit}, {Maehara},
  {Ikuta}, {Inoue}, {Notsu}, {Paudel}, {Arzoumanian}, {Avramova-Boncheva},
  {Gendreau}, {Jeffers}, {Marsden}, {Morin}, {Neiner}, {Vidotto}, and
  {Shibata}]{Namekata2024}
{Namekata}, K.; {Airapetian}, V.S.; {Petit}, P.; {Maehara}, H.; {Ikuta}, K.;
  {Inoue}, S.; {Notsu}, Y.; {Paudel}, R.R.; {Arzoumanian}, Z.;
  {Avramova-Boncheva}, A.A.;  et~al.
\newblock {Multiwavelength Campaign Observations of a Young Solar-type Star, EK
  Draconis. I. Discovery of Prominence Eruptions Associated with Superflares}.
\newblock {\em \apj} {\bf 2024}, {\em 961},~23,
  \href{http://arxiv.org/abs/2311.07380}{{\normalfont
  [arXiv:astro-ph.SR/2311.07380]}}.
\newblock {\url{https://doi.org/10.3847/1538-4357/ad0b7c}}.

\bibitem[{Leitzinger} et~al.(2022){Leitzinger}, {Odert}, and
  {Heinzel}]{Leitzinger2022}
{Leitzinger}, M.; {Odert}, P.; {Heinzel}, P.
\newblock {Modeling Balmer line signatures of stellar CMEs}.
\newblock {\em \mnras} {\bf 2022}, {\em 513},~6058--6073,
  \href{http://arxiv.org/abs/2205.03110}{{\normalfont
  [arXiv:astro-ph.SR/2205.03110]}}.
\newblock {\url{https://doi.org/10.1093/mnras/stac1284}}.

\bibitem[{Fuhrmeister} and {Schmitt}(2004)]{FuhrmeisterSchmitt2004}
{Fuhrmeister}, B.; {Schmitt}, J.H.M.M.
\newblock {Detection and high-resolution spectroscopy of a huge flare on the
  old M 9 dwarf DENIS 104814.7-395606.1}.
\newblock {\em \aap} {\bf 2004}, {\em 420},~1079--1085,
  \href{http://arxiv.org/abs/arXiv:astro-ph/0403617}{{\normalfont
  [arXiv:astro-ph/0403617]}}.
\newblock {\url{https://doi.org/10.1051/0004-6361:20035644}}.

\bibitem[{Honda} et~al.(2018){Honda}, {Notsu}, {Namekata}, {Notsu}, {Maehara},
  {Ikuta}, {Nogami}, and {Shibata}]{Honda2018}
{Honda}, S.; {Notsu}, Y.; {Namekata}, K.; {Notsu}, S.; {Maehara}, H.; {Ikuta},
  K.; {Nogami}, D.; {Shibata}, K.
\newblock {Time-resolved spectroscopic observations of an M-dwarf flare star EV
  Lacertae during a flare}.
\newblock {\em \pasj} {\bf 2018}, {\em 70},~62,
  \href{http://arxiv.org/abs/1804.03771}{{\normalfont
  [arXiv:astro-ph.SR/1804.03771]}}.
\newblock {\url{https://doi.org/10.1093/pasj/psy055}}.

\bibitem[{Muheki} et~al.(2020{\natexlab{a}}){Muheki}, {Guenther}, {Mutabazi},
  and {Jurua}]{Muheki2020a}
{Muheki}, P.; {Guenther}, E.W.; {Mutabazi}, T.; {Jurua}, E.
\newblock {High-resolution spectroscopy of flares and CMEs on AD Leonis}.
\newblock {\em \aap} {\bf 2020}, {\em 637},~A13,
  \href{http://arxiv.org/abs/2003.06163}{{\normalfont
  [arXiv:astro-ph.SR/2003.06163]}}.
\newblock {\url{https://doi.org/10.1051/0004-6361/201936904}}.

\bibitem[{Muheki} et~al.(2020{\natexlab{b}}){Muheki}, {Guenther}, {Mutabazi},
  and {Jurua}]{Muheki2020b}
{Muheki}, P.; {Guenther}, E.W.; {Mutabazi}, T.; {Jurua}, E.
\newblock {Properties of flares and CMEs on EV Lac: possible erupting
  filament}.
\newblock {\em \mnras} {\bf 2020}, {\em 499},~5047--5058,
  \href{http://arxiv.org/abs/2010.03336}{{\normalfont
  [arXiv:astro-ph.SR/2010.03336]}}.
\newblock {\url{https://doi.org/10.1093/mnras/staa3152}}.

\bibitem[{Maehara} et~al.(2021){Maehara}, {Notsu}, {Namekata}, {Honda},
  {Kowalski}, {Katoh}, {Ohshima}, {Iida}, {Oeda}, {Murata}, {Yamanaka},
  {Takagi}, {Sasada}, {Akitaya}, {Ikuta}, {Okamoto}, {Nogami}, and
  {Shibata}]{Maehara2021}
{Maehara}, H.; {Notsu}, Y.; {Namekata}, K.; {Honda}, S.; {Kowalski}, A.F.;
  {Katoh}, N.; {Ohshima}, T.; {Iida}, K.; {Oeda}, M.; {Murata}, K.L.;  et~al.
\newblock {Time-resolved spectroscopy and photometry of M dwarf flare star YZ
  Canis Minoris with OISTER and TESS: Blue asymmetry in the
  H{\ensuremath{\alpha}} line during the non-white light flare}.
\newblock {\em \pasj} {\bf 2021}, {\em 73},~44--65,
  \href{http://arxiv.org/abs/2009.14412}{{\normalfont
  [arXiv:astro-ph.SR/2009.14412]}}.
\newblock {\url{https://doi.org/10.1093/pasj/psaa098}}.

\bibitem[{Johnson} et~al.(2021){Johnson}, {Czesla}, {Fuhrmeister},
  {Sch{\"o}fer}, {Shan}, {Cardona Guill{\'e}n}, {Reiners}, {Jeffers},
  {Lalitha}, {Luque}, {Rodr{\'\i}guez}, {B{\'e}jar}, {Caballero}, {Tal-Or},
  {Zechmeister}, {Ribas}, {Amado}, {Quirrenbach}, {Cort{\'e}s-Contreras},
  {Dreizler}, {Fukui}, {L{\'o}pez-Gonz{\'a}lez}, {Hatzes}, {Henning},
  {Kaminski}, {K{\"u}rster}, {Lafarga}, {Montes}, {Morales}, {Murgas},
  {Narita}, {Pall{\'e}}, {Parviainen}, {Pedraz}, {Pollacco}, and
  {Sota}]{Johnson2021}
{Johnson}, E.N.; {Czesla}, S.; {Fuhrmeister}, B.; {Sch{\"o}fer}, P.; {Shan},
  Y.; {Cardona Guill{\'e}n}, C.; {Reiners}, A.; {Jeffers}, S.V.; {Lalitha}, S.;
  {Luque}, R.;  et~al.
\newblock {Simultaneous photometric and CARMENES spectroscopic monitoring of
  fast-rotating M dwarf GJ 3270. Discovery of a post-flare corotating feature}.
\newblock {\em \aap} {\bf 2021}, {\em 651},~A105,
  \href{http://arxiv.org/abs/2104.07080}{{\normalfont
  [arXiv:astro-ph.SR/2104.07080]}}.
\newblock {\url{https://doi.org/10.1051/0004-6361/202040159}}.

\bibitem[{Wang} et~al.(2021){Wang}, {Xin}, {Li}, {Li}, {Sun}, {Gao}, {Han},
  {Dai}, {Liang}, {Wang}, and {Wei}]{Wang2021}
{Wang}, J.; {Xin}, L.P.; {Li}, H.L.; {Li}, G.W.; {Sun}, S.S.; {Gao}, C.; {Han},
  X.H.; {Dai}, Z.G.; {Liang}, E.W.; {Wang}, X.Y.;  et~al.
\newblock {Detection of Flare-associated CME Candidates on Two M-dwarfs by GWAC
  and Fast, Time-resolved Spectroscopic Follow-ups}.
\newblock {\em \apj} {\bf 2021}, {\em 916},~92,
  \href{http://arxiv.org/abs/2106.04774}{{\normalfont
  [arXiv:astro-ph.SR/2106.04774]}}.
\newblock {\url{https://doi.org/10.3847/1538-4357/ac096f}}.

\bibitem[{Wang} et~al.(2022){Wang}, {Li}, {Xin}, {Li}, {Bai}, {Gao}, {Ren},
  {Song}, {Deng}, {Han}, {Dai}, {Liang}, {Wang}, and {Wei}]{Wang2022}
{Wang}, J.; {Li}, H.L.; {Xin}, L.P.; {Li}, G.W.; {Bai}, J.Y.; {Gao}, C.; {Ren},
  B.; {Song}, D.; {Deng}, J.S.; {Han}, X.H.;  et~al.
\newblock {Flaring-associated Complex Dynamics in Two M Dwarfs Revealed by
  Fast, Time-resolved Spectroscopy}.
\newblock {\em \apj} {\bf 2022}, {\em 934},~98,
  \href{http://arxiv.org/abs/2206.10412}{{\normalfont
  [arXiv:astro-ph.SR/2206.10412]}}.
\newblock {\url{https://doi.org/10.3847/1538-4357/ac7a35}}.

\bibitem[{L{\'o}pez-Santiago} et~al.(2003){L{\'o}pez-Santiago}, {Montes},
  {Fern{\'a}ndez-Figueroa}, and {Ramsey}]{LopezSantiago2003}
{L{\'o}pez-Santiago}, J.; {Montes}, D.; {Fern{\'a}ndez-Figueroa}, M.J.;
  {Ramsey}, L.W.
\newblock {Rotational modulation of the photospheric and chromospheric activity
  in the young, single K2-dwarf PW And}.
\newblock {\em \aap} {\bf 2003}, {\em 411},~489--502,
  \href{http://arxiv.org/abs/astro-ph/0309072}{{\normalfont
  [arXiv:astro-ph/astro-ph/0309072]}}.
\newblock {\url{https://doi.org/10.1051/0004-6361:20031377}}.

\bibitem[{Hill} et~al.(2017){Hill}, {Carmona}, {Donati}, {Hussain}, {Gregory},
  {Alencar}, {Bouvier}, and {Matysse Collaboration}]{Hill2017}
{Hill}, C.A.; {Carmona}, A.; {Donati}, J.F.; {Hussain}, G.A.J.; {Gregory},
  S.G.; {Alencar}, S.H.P.; {Bouvier}, J.; {Matysse Collaboration}.
\newblock {Magnetic activity and radial velocity filtering of young Suns: the
  weak-line T-Tauri stars Par 1379 and Par 2244}.
\newblock {\em \mnras} {\bf 2017}, {\em 472},~1716--1735,
  \href{http://arxiv.org/abs/1708.09693}{{\normalfont
  [arXiv:astro-ph.SR/1708.09693]}}.
\newblock {\url{https://doi.org/10.1093/mnras/stx2042}}.

\bibitem[{Leitzinger} et~al.(2014){Leitzinger}, {Odert}, {Greimel}, {Korhonen},
  {Guenther}, {Hanslmeier}, {Lammer}, and {Khodachenko}]{Leitzinger2014}
{Leitzinger}, M.; {Odert}, P.; {Greimel}, R.; {Korhonen}, H.; {Guenther}, E.W.;
  {Hanslmeier}, A.; {Lammer}, H.; {Khodachenko}, M.L.
\newblock {A search for flares and mass ejections on young late-type stars in
  the open cluster Blanco-1}.
\newblock {\em \mnras} {\bf 2014}, {\em 443},~898--910,
  \href{http://arxiv.org/abs/1406.2734}{{\normalfont
  [arXiv:astro-ph.SR/1406.2734]}}.
\newblock {\url{https://doi.org/10.1093/mnras/stu1161}}.

\bibitem[{Korhonen} et~al.(2017){Korhonen}, {Vida}, {Leitzinger}, {Odert}, and
  {Kov{\'a}cs}]{Korhonen2017}
{Korhonen}, H.; {Vida}, K.; {Leitzinger}, M.; {Odert}, P.; {Kov{\'a}cs}, O.E.
\newblock {Hunting for Stellar Coronal Mass Ejections}.
\newblock In Proceedings of the Living Around Active Stars; {Nandy}, D.;
  {Valio}, A.; {Petit}, P., Eds.,  10 2017, Vol. 328, pp. 198--203,
  \href{http://arxiv.org/abs/1612.06643}{{\normalfont
  [arXiv:astro-ph.SR/1612.06643]}}.
\newblock {\url{https://doi.org/10.1017/S1743921317003969}}.

\bibitem[{Vida} et~al.(2024){Vida}, {Seli}, {Roettenbacher}, {G\"orgei},
  {Kriskovics}, {K{\H{o}}v{\'a}ri}, and {Ol\'ah}]{Vida2024b}
{Vida}, K.; {Seli}, B.; {Roettenbacher}, R.; {G\"orgei}, A.; {Kriskovics}, L.;
  {K{\H{o}}v{\'a}ri}, Z.; {Ol\'ah}, K.
\newblock {Searching for stellar CMEs in the Praesepe and Pleiades clusters}.
\newblock In Proceedings of the {IAU Symposium 388: Solar \& stellar coronal
  mass ejections, submitted},  2024,
  \href{http://arxiv.org/abs/2407.11461}{{\normalfont
  [arXiv:astro-ph.SR/2407.11461]}}.

\bibitem[{Vida} et~al.(2019){Vida}, {Leitzinger}, {Kriskovics}, {Seli},
  {Odert}, {Kov{\'a}cs}, {Korhonen}, and {van Driel-Gesztelyi}]{Vida2019}
{Vida}, K.; {Leitzinger}, M.; {Kriskovics}, L.; {Seli}, B.; {Odert}, P.;
  {Kov{\'a}cs}, O.E.; {Korhonen}, H.; {van Driel-Gesztelyi}, L.
\newblock {The quest for stellar coronal mass ejections in late-type stars. I.
  Investigating Balmer-line asymmetries of single stars in Virtual Observatory
  data}.
\newblock {\em \aap} {\bf 2019}, {\em 623},~A49,
  \href{http://arxiv.org/abs/1901.04229}{{\normalfont
  [arXiv:astro-ph.SR/1901.04229]}}.
\newblock {\url{https://doi.org/10.1051/0004-6361/201834264}}.

\bibitem[{Leitzinger} et~al.(2020){Leitzinger}, {Odert}, {Greimel}, {Vida},
  {Kriskovics}, {Guenther}, {Korhonen}, {Koller}, {Hanslmeier},
  {K{\H{o}}v{\'a}ri}, and {Lammer}]{Leitzinger2020}
{Leitzinger}, M.; {Odert}, P.; {Greimel}, R.; {Vida}, K.; {Kriskovics}, L.;
  {Guenther}, E.W.; {Korhonen}, H.; {Koller}, F.; {Hanslmeier}, A.;
  {K{\H{o}}v{\'a}ri}, Z.;  et~al.
\newblock {A census of coronal mass ejections on solar-like stars}.
\newblock {\em \mnras} {\bf 2020}, {\em 493},~4570--4589,
  \href{http://arxiv.org/abs/2002.04430}{{\normalfont
  [arXiv:astro-ph.SR/2002.04430]}}.
\newblock {\url{https://doi.org/10.1093/mnras/staa504}}.

\bibitem[{Fuhrmeister} et~al.(2018){Fuhrmeister}, {Czesla}, {Schmitt},
  {Jeffers}, {Caballero}, {Zechmeister}, {Reiners}, {Ribas}, {Amado},
  {Quirrenbach}, {B{\'e}jar}, {Galad{\'\i}-Enr{\'\i}quez}, {Guenther},
  {K{\"u}rster}, {Montes}, and {Seifert}]{Fuhrmeister2018}
{Fuhrmeister}, B.; {Czesla}, S.; {Schmitt}, J.H.M.M.; {Jeffers}, S.V.;
  {Caballero}, J.A.; {Zechmeister}, M.; {Reiners}, A.; {Ribas}, I.; {Amado},
  P.J.; {Quirrenbach}, A.;  et~al.
\newblock {The CARMENES search for exoplanets around M dwarfs. Wing asymmetries
  of H{\ensuremath{\alpha}}, Na I D, and He I lines}.
\newblock {\em \aap} {\bf 2018}, {\em 615},~A14,
  \href{http://arxiv.org/abs/1801.10372}{{\normalfont
  [arXiv:astro-ph.SR/1801.10372]}}.
\newblock {\url{https://doi.org/10.1051/0004-6361/201732204}}.

\bibitem[{Koller} et~al.(2021){Koller}, {Leitzinger}, {Temmer}, {Odert},
  {Beck}, and {Veronig}]{Koller2021}
{Koller}, F.; {Leitzinger}, M.; {Temmer}, M.; {Odert}, P.; {Beck}, P.G.;
  {Veronig}, A.
\newblock {Search for flares and associated CMEs on late-type main-sequence
  stars in optical SDSS spectra}.
\newblock {\em \aap} {\bf 2021}, {\em 646},~A34,
  \href{http://arxiv.org/abs/2012.00786}{{\normalfont
  [arXiv:astro-ph.SR/2012.00786]}}.
\newblock {\url{https://doi.org/10.1051/0004-6361/202039003}}.

\bibitem[{Lu} et~al.(2022){Lu}, {Tian}, {Zhang}, {Karoff}, {Chen}, {Shi},
  {Hou}, {Chen}, {Xu}, {Wu}, {Cao}, and {Wang}]{Lu2022}
{Lu}, H.p.; {Tian}, H.; {Zhang}, L.y.; {Karoff}, C.; {Chen}, H.c.; {Shi}, J.r.;
  {Hou}, Z.y.; {Chen}, Y.j.; {Xu}, Y.; {Wu}, Y.c.;  et~al.
\newblock {Possible detection of coronal mass ejections on late-type
  main-sequence stars in LAMOST medium-resolution spectra}.
\newblock {\em \aap} {\bf 2022}, {\em 663},~A140,
  \href{http://arxiv.org/abs/2205.09972}{{\normalfont
  [arXiv:astro-ph.SR/2205.09972]}}.
\newblock {\url{https://doi.org/10.1051/0004-6361/202142909}}.

\bibitem[{Chen} et~al.(2022){Chen}, {Tian}, {Li}, {Wang}, {Lu}, {Xu}, {Hou},
  and {Wu}]{Chen2022}
{Chen}, H.; {Tian}, H.; {Li}, H.; {Wang}, J.; {Lu}, H.; {Xu}, Y.; {Hou}, Z.;
  {Wu}, Y.
\newblock {Detection of Flare-induced Plasma Flows in the Corona of EV Lac with
  X-Ray Spectroscopy}.
\newblock {\em \apj} {\bf 2022}, {\em 933},~92,
  \href{http://arxiv.org/abs/2205.14293}{{\normalfont
  [arXiv:astro-ph.SR/2205.14293]}}.
\newblock {\url{https://doi.org/10.3847/1538-4357/ac739b}}.

\bibitem[{Xu} et~al.(2022){Xu}, {Tian}, {Hou}, {Yang}, {Gao}, and
  {Bai}]{Xu2022}
{Xu}, Y.; {Tian}, H.; {Hou}, Z.; {Yang}, Z.; {Gao}, Y.; {Bai}, X.
\newblock {Sun-as-a-star Spectroscopic Observations of the Line-of-sight
  Velocity of a Solar Eruption on 2021 October 28}.
\newblock {\em \apj} {\bf 2022}, {\em 931},~76,
  \href{http://arxiv.org/abs/2204.11722}{{\normalfont
  [arXiv:astro-ph.SR/2204.11722]}}.
\newblock {\url{https://doi.org/10.3847/1538-4357/ac69d5}}.

\bibitem[{Lu} et~al.(2023){Lu}, {Tian}, {Chen}, {Xu}, {Hou}, {Bai}, {Tan},
  {Yang}, and {Ren}]{Lu2023}
{Lu}, H.p.; {Tian}, H.; {Chen}, H.c.; {Xu}, Y.; {Hou}, Z.y.; {Bai}, X.y.;
  {Tan}, G.y.; {Yang}, Z.h.; {Ren}, J.
\newblock {Full Velocities and Propagation Directions of Coronal Mass Ejections
  Inferred from Simultaneous Full-disk Imaging and Sun-as-a-star Spectroscopic
  Observations}.
\newblock {\em \apj} {\bf 2023}, {\em 953},~68,
  \href{http://arxiv.org/abs/2305.08765}{{\normalfont
  [arXiv:astro-ph.SR/2305.08765]}}.
\newblock {\url{https://doi.org/10.3847/1538-4357/acd6a1}}.

\bibitem[{Abdul-Aziz} et~al.(1995){Abdul-Aziz}, {Abranin}, {Alekseev},
  {Avgoloupis}, {Bazelyan}, {Beskin}, {Brazhenko}, {Chalenko}, {Cutispoto},
  {Fuensalida}, {Gershberg}, {Kidger}, {Leto}, {Malkov}, {Mavridis}, {Pagano},
  {Panferova}, {Rodono}, {Seiradakis}, {Sergeev}, {Spencer}, {Shakhovskaya},
  and {Shakhovskoy}]{AbdulAziz1995}
{Abdul-Aziz}, H.; {Abranin}, E.P.; {Alekseev}, I.Y.; {Avgoloupis}, S.;
  {Bazelyan}, L.L.; {Beskin}, G.M.; {Brazhenko}, A.I.; {Chalenko}, N.N.;
  {Cutispoto}, G.; {Fuensalida}, J.J.;  et~al.
\newblock {Coordinated observations of the red dwarf flare star EV Lacertae in
  1992.}
\newblock {\em \aaps} {\bf 1995}, {\em 114},~509.

\bibitem[{Abranin} et~al.(1998){Abranin}, {Alekseev}, {Avgoloupis}, {Bazelyan},
  {Berdyugina}, {Cutispoto}, {Gershberg}, {Larionov}, {Leto}, {Lisachenko},
  {Marino}, {Mavridis}, {Messina}, {Mel'Nik}, {Pagano}, {Pustil'Nik},
  {Rodon{\`o}}, {Roizman}, {Seiradakis}, {Sigal}, {Shakhovskaya},
  {Shakhovskoy}, and {Shcherbakov}]{Abranin1998}
{Abranin}, E.P.; {Alekseev}, I.Y.; {Avgoloupis}, S.; {Bazelyan}, L.L.;
  {Berdyugina}, S.V.; {Cutispoto}, G.; {Gershberg}, R.E.; {Larionov}, V.M.;
  {Leto}, G.; {Lisachenko}, V.N.;  et~al.
\newblock {Coordinated Observations of the Red Dwarf Flare Star EV LAC in 1994
  and 1995}.
\newblock {\em Astronomical and Astrophysical Transactions} {\bf 1998}, {\em
  17},~221--262.
\newblock {\url{https://doi.org/10.1080/10556799808232093}}.

\bibitem[{Leitzinger} et~al.(2009){Leitzinger}, {Odert}, {Hanslmeier},
  {Konovalenko}, {Vanko}, {Khodachenko}, {Lammer}, and
  {Rucker}]{Leitzinger2009}
{Leitzinger}, M.; {Odert}, P.; {Hanslmeier}, A.; {Konovalenko}, A.A.; {Vanko},
  M.; {Khodachenko}, M.L.; {Lammer}, H.; {Rucker}, H.O.
\newblock {Decametric observations of active M-dwarfs}.
\newblock In Proceedings of the American Institute of Physics Conference
  Series; {E.~Stempels}., Ed.,  2 2009, Vol. 1094, {\em American Institute of
  Physics Conference Series}, pp. 680--683.
\newblock {\url{https://doi.org/10.1063/1.3099205}}.

\bibitem[{Boiko} et~al.(2012){Boiko}, {Konovalenko}, {Koliadin}, and
  {Melnik}]{Boiko2012}
{Boiko}, A.I.; {Konovalenko}, A.A.; {Koliadin}, V.L.; {Melnik}, V.N.
\newblock {Search of the radio emission from flare stars at decameter
  wavelengths}.
\newblock {\em Advances in Astronomy and Space Physics} {\bf 2012}, {\em
  2},~121--124.

\bibitem[{Konovalenko} et~al.(2012){Konovalenko}, {Koliadin}, {Boiko}, {Zarka},
  {Griessmeier}, {Denis}, {Coffre}, {Rucker}, {Zaitsev}, {Litvinenko},
  {Melnik}, {Stanislavsky}, {Stepkin}, {Mukha}, {Brazhenko}, {Leitzinger},
  {Odret}, and {Scherf}]{Konovalenko2012}
{Konovalenko}, A.A.; {Koliadin}, V.L.; {Boiko}, A.I.; {Zarka}, P.;
  {Griessmeier}, J.M.; {Denis}, L.; {Coffre}, A.; {Rucker}, H.O.; {Zaitsev},
  V.V.; {Litvinenko}, G.V.;  et~al.
\newblock {Analysis of the flare stars radio bursts parameters at the decameter
  wavelengths}.
\newblock In Proceedings of the European Planetary Science Congress 2012,  9
  2012, pp. EPSC2012--902.

\bibitem[{Crosley} et~al.(2016){Crosley}, {Osten}, {Broderick}, {Corbel},
  {Eisl{\"o}ffel}, {Grie{\ss}meier}, {van Leeuwen}, {Rowlinson}, {Zarka}, and
  {Norman}]{Crosley2016}
{Crosley}, M.K.; {Osten}, R.A.; {Broderick}, J.W.; {Corbel}, S.;
  {Eisl{\"o}ffel}, J.; {Grie{\ss}meier}, J.M.; {van Leeuwen}, J.; {Rowlinson},
  A.; {Zarka}, P.; {Norman}, C.
\newblock {The Search for Signatures of Transient Mass Loss in Active Stars}.
\newblock {\em \apj} {\bf 2016}, {\em 830},~24,
  \href{http://arxiv.org/abs/1606.02334}{{\normalfont
  [arXiv:astro-ph.SR/1606.02334]}}.
\newblock {\url{https://doi.org/10.3847/0004-637X/830/1/24}}.

\bibitem[{Crosley} and {Osten}(2018{\natexlab{a}})]{Crosley2018a}
{Crosley}, M.K.; {Osten}, R.A.
\newblock {Low-frequency Radio Transients on the Active M-dwarf EQ Peg and the
  Search for Coronal Mass Ejections}.
\newblock {\em \apj} {\bf 2018}, {\em 862},~113.
\newblock {\url{https://doi.org/10.3847/1538-4357/aacf02}}.

\bibitem[{Crosley} and {Osten}(2018{\natexlab{b}})]{Crosley2018b}
{Crosley}, M.K.; {Osten}, R.A.
\newblock {Constraining Stellar Coronal Mass Ejections through Multi-wavelength
  Analysis of the Active M Dwarf EQ Peg}.
\newblock {\em \apj} {\bf 2018}, {\em 856},~39,
  \href{http://arxiv.org/abs/1802.03440}{{\normalfont
  [arXiv:astro-ph.SR/1802.03440]}}.
\newblock {\url{https://doi.org/10.3847/1538-4357/aaaec2}}.

\bibitem[{Villadsen} and {Hallinan}(2019)]{Villadsen2019}
{Villadsen}, J.; {Hallinan}, G.
\newblock {Ultra-wideband Detection of 22 Coherent Radio Bursts on M Dwarfs}.
\newblock {\em \apj} {\bf 2019}, {\em 871},~214,
  \href{http://arxiv.org/abs/1810.00855}{{\normalfont
  [arXiv:astro-ph.SR/1810.00855]}}.
\newblock {\url{https://doi.org/10.3847/1538-4357/aaf88e}}.

\bibitem[{Mullan} and {Paudel}(2019)]{Mullan2019}
{Mullan}, D.J.; {Paudel}, R.R.
\newblock {Origin of Radio-quiet Coronal Mass Ejections in Flare Stars}.
\newblock {\em \apj} {\bf 2019}, {\em 873},~1,
  \href{http://arxiv.org/abs/1902.00810}{{\normalfont
  [arXiv:astro-ph.SR/1902.00810]}}.
\newblock {\url{https://doi.org/10.3847/1538-4357/ab041b}}.

\bibitem[{Zic} et~al.(2020){Zic}, {Murphy}, {Lynch}, {Heald}, {Lenc}, {Kaplan},
  {Cairns}, {Coward}, {Gendre}, {Johnston}, {MacGregor}, {Price}, and
  {Wheatland}]{Zic2020}
{Zic}, A.; {Murphy}, T.; {Lynch}, C.; {Heald}, G.; {Lenc}, E.; {Kaplan}, D.L.;
  {Cairns}, I.H.; {Coward}, D.; {Gendre}, B.; {Johnston}, H.;  et~al.
\newblock {A Flare-type IV Burst Event from Proxima Centauri and Implications
  for Space Weather}.
\newblock {\em \apj} {\bf 2020}, {\em 905},~23,
  \href{http://arxiv.org/abs/2012.04642}{{\normalfont
  [arXiv:astro-ph.SR/2012.04642]}}.
\newblock {\url{https://doi.org/10.3847/1538-4357/abca90}}.

\bibitem[{Kahler} et~al.(1982){Kahler}, {Golub}, {Harnden}, {Liller}, {Seward},
  {Vaiana}, {Lovell}, {Davis}, {Spencer}, {Whitehouse}, {Feldman}, {Viner},
  {Leslie}, {Kahn}, {Mason}, {Davis}, {Crannell}, {Hobbs}, {Schneeberger},
  {Worden}, {Schommer}, {Vogt}, {Pettersen}, {Coleman}, {Karpen}, {Giampapa},
  {Hege}, {Pazzani}, {Rodono}, {Romeo}, and {Chugainov}]{Kahler1982}
{Kahler}, S.; {Golub}, L.; {Harnden}, F.R.; {Liller}, W.; {Seward}, F.;
  {Vaiana}, G.; {Lovell}, B.; {Davis}, R.J.; {Spencer}, R.E.; {Whitehouse},
  D.R.;  et~al.
\newblock {Coordinated X-ray, optical and radio observations of flaring
  activityon YZ Canis Minoris.}
\newblock {\em \apj} {\bf 1982}, {\em 252},~239--249.
\newblock {\url{https://doi.org/10.1086/159551}}.

\bibitem[{Bloot} et~al.(2024){Bloot}, {Callingham}, {Vedantham}, {Kavanagh},
  {Pope}, {Climent}, {Guirado}, {Pe{\~n}a-Mo{\~n}ino}, and
  {P{\'e}rez-Torres}]{Bloot2024}
{Bloot}, S.; {Callingham}, J.R.; {Vedantham}, H.K.; {Kavanagh}, R.D.; {Pope},
  B.J.S.; {Climent}, J.B.; {Guirado}, J.C.; {Pe{\~n}a-Mo{\~n}ino}, L.;
  {P{\'e}rez-Torres}, M.
\newblock {Phenomenology and periodicity of radio emission from the stellar
  system AU Microscopii}.
\newblock {\em \aap} {\bf 2024}, {\em 682},~A170,
  \href{http://arxiv.org/abs/2312.09071}{{\normalfont
  [arXiv:astro-ph.SR/2312.09071]}}.
\newblock {\url{https://doi.org/10.1051/0004-6361/202348065}}.

\bibitem[{Mohan} et~al.(2024){Mohan}, {Mondal}, {Wedemeyer}, and
  {Gopalswamy}]{Mohan2024}
{Mohan}, A.; {Mondal}, S.; {Wedemeyer}, S.; {Gopalswamy}, N.
\newblock {Energetic particle activity in AD Leo: Detection of a solar-like
  type-IV burst}.
\newblock {\em arXiv e-prints} {\bf 2024}, p. arXiv:2402.00185,
  \href{http://arxiv.org/abs/2402.00185}{{\normalfont
  [arXiv:astro-ph.SR/2402.00185]}}.
\newblock {\url{https://doi.org/10.48550/arXiv.2402.00185}}.

\bibitem[{Veronig} et~al.(2021){Veronig}, {Odert}, {Leitzinger}, {Dissauer},
  {Fleck}, and {Hudson}]{Veronig2021}
{Veronig}, A.M.; {Odert}, P.; {Leitzinger}, M.; {Dissauer}, K.; {Fleck}, N.C.;
  {Hudson}, H.S.
\newblock {Indications of stellar coronal mass ejections through coronal
  dimmings}.
\newblock {\em Nature Astronomy} {\bf 2021}, {\em 5},~697--706.
\newblock {\url{https://doi.org/10.1038/s41550-021-01345-9}}.

\bibitem[{Loyd} et~al.(2022){Loyd}, {Mason}, {Jin}, {Shkolnik}, {France},
  {Youngblood}, {Villadsen}, {Schneider}, {Schneider}, {Llama},
  {Ramiaramanantsoa}, and {Richey-Yowell}]{Loyd2022}
{Loyd}, R.O.P.; {Mason}, J.P.; {Jin}, M.; {Shkolnik}, E.L.; {France}, K.;
  {Youngblood}, A.; {Villadsen}, J.; {Schneider}, C.; {Schneider}, A.C.;
  {Llama}, J.;  et~al.
\newblock {Constraining the Physical Properties of Stellar Coronal Mass
  Ejections with Coronal Dimming: Application to Far-ultraviolet Data of
  {\ensuremath{\epsilon}} Eridani}.
\newblock {\em \apj} {\bf 2022}, {\em 936},~170,
  \href{http://arxiv.org/abs/2207.05115}{{\normalfont
  [arXiv:astro-ph.SR/2207.05115]}}.
\newblock {\url{https://doi.org/10.3847/1538-4357/ac80c1}}.

\bibitem[{Haisch} et~al.(1983){Haisch}, {Linsky}, {Bornmann}, {Stencel},
  {Antiochos}, {Golub}, and {Vaiana}]{Haisch1983}
{Haisch}, B.M.; {Linsky}, J.L.; {Bornmann}, P.L.; {Stencel}, R.E.; {Antiochos},
  S.K.; {Golub}, L.; {Vaiana}, G.S.
\newblock {Coordinated Einstein and IUE observations of a disparitions brusques
  type flare event and quiescent emission from Proxima Centauri.}
\newblock {\em \apj} {\bf 1983}, {\em 267},~280--290.
\newblock {\url{https://doi.org/10.1086/160866}}.

\bibitem[{Tsuboi} et~al.(1998){Tsuboi}, {Koyama}, {Murakami}, {Hayashi},
  {Skinner}, and {Ueno}]{Tsuboi1998}
{Tsuboi}, Y.; {Koyama}, K.; {Murakami}, H.; {Hayashi}, M.; {Skinner}, S.;
  {Ueno}, S.
\newblock {ASCA Detection of a Superhot 100 Million K X-Ray Flare on the
  Weak-Lined T Tauri Star V773 Tauri}.
\newblock {\em \apj} {\bf 1998}, {\em 503},~894--901.
\newblock {\url{https://doi.org/10.1086/306024}}.

\bibitem[{Briggs} and {Pye}(2003)]{Briggs2003}
{Briggs}, K.R.; {Pye}, J.P.
\newblock {XMM-Newton and the Pleiades - I. Bright coronal sources and the
  X-ray emission from intermediate-type stars}.
\newblock {\em \mnras} {\bf 2003}, {\em 345},~714--726,
  \href{http://arxiv.org/abs/astro-ph/0307095}{{\normalfont
  [arXiv:astro-ph/astro-ph/0307095]}}.
\newblock {\url{https://doi.org/10.1046/j.1365-8711.2003.06991.x}}.

\bibitem[{Ottmann} and {Schmitt}(1996)]{Ottmann1996}
{Ottmann}, R.; {Schmitt}, J.H.M.M.
\newblock {ROSAT observation of a giant X-ray flare on Algol: evidence for
  abundance variations?}
\newblock {\em \aap} {\bf 1996}, {\em 307},~813--823.

\bibitem[{Favata} and {Schmitt}(1999)]{Favata1999}
{Favata}, F.; {Schmitt}, J.H.M.M.
\newblock {Spectroscopic analysis of a super-hot giant flare observed on Algol
  by BeppoSAX on 30 August 1997}.
\newblock {\em \aap} {\bf 1999}, {\em 350},~900--916,
  \href{http://arxiv.org/abs/astro-ph/9909041}{{\normalfont
  [arXiv:astro-ph/astro-ph/9909041]}}.

\bibitem[{Franciosini} et~al.(2001){Franciosini}, {Pallavicini}, and
  {Tagliaferri}]{Franciosini2001}
{Franciosini}, E.; {Pallavicini}, R.; {Tagliaferri}, G.
\newblock {BeppoSAX observation of a large long-duration X-ray flare from UX
  Arietis}.
\newblock {\em \aap} {\bf 2001}, {\em 375},~196--204.
\newblock {\url{https://doi.org/10.1051/0004-6361:20010830}}.

\bibitem[{Pandey} and {Singh}(2012)]{Pandey2012}
{Pandey}, J.C.; {Singh}, K.P.
\newblock {A study of X-ray flares - II. RS CVn-type binaries}.
\newblock {\em \mnras} {\bf 2012}, {\em 419},~1219--1237,
  \href{http://arxiv.org/abs/1110.2008}{{\normalfont
  [arXiv:astro-ph.SR/1110.2008]}}.
\newblock {\url{https://doi.org/10.1111/j.1365-2966.2011.19776.x}}.

\bibitem[{Moschou} et~al.(2017){Moschou}, {Drake}, {Cohen}, {Alvarado-Gomez},
  and {Garraffo}]{Moschou2017}
{Moschou}, S.P.; {Drake}, J.J.; {Cohen}, O.; {Alvarado-Gomez}, J.D.;
  {Garraffo}, C.
\newblock {A Monster CME Obscuring a Demon Star Flare}.
\newblock {\em \apj} {\bf 2017}, {\em 850},~191,
  \href{http://arxiv.org/abs/1710.07361}{{\normalfont
  [arXiv:astro-ph.SR/1710.07361]}}.
\newblock {\url{https://doi.org/10.3847/1538-4357/aa9520}}.

\bibitem[{Bond} et~al.(2001){Bond}, {Mullan}, {O'Brien}, and {Sion}]{Bond2001}
{Bond}, H.E.; {Mullan}, D.J.; {O'Brien}, M.S.; {Sion}, E.M.
\newblock {Detection of Coronal Mass Ejections in V471 Tauri with the Hubble
  Space Telescope}.
\newblock {\em \apj} {\bf 2001}, {\em 560},~919--927,
  \href{http://arxiv.org/abs/arXiv:astro-ph/0106400}{{\normalfont
  [arXiv:astro-ph/0106400]}}.
\newblock {\url{https://doi.org/10.1086/322980}}.

\bibitem[{Aarnio} et~al.(2012){Aarnio}, {Matt}, and {Stassun}]{Aarnio2012}
{Aarnio}, A.N.; {Matt}, S.P.; {Stassun}, K.G.
\newblock {Mass Loss in Pre-main-sequence Stars via Coronal Mass Ejections and
  Implications for Angular Momentum Loss}.
\newblock {\em \apj} {\bf 2012}, {\em 760},~9,
  \href{http://arxiv.org/abs/1209.6410}{{\normalfont
  [arXiv:astro-ph.SR/1209.6410]}}.
\newblock {\url{https://doi.org/10.1088/0004-637X/760/1/9}}.

\bibitem[{Drake} et~al.(2013){Drake}, {Cohen}, {Yashiro}, and
  {Gopalswamy}]{Drake2013}
{Drake}, J.J.; {Cohen}, O.; {Yashiro}, S.; {Gopalswamy}, N.
\newblock {Implications of Mass and Energy Loss due to Coronal Mass Ejections
  on Magnetically Active Stars}.
\newblock {\em \apj} {\bf 2013}, {\em 764},~170,
  \href{http://arxiv.org/abs/1302.1136}{{\normalfont
  [arXiv:astro-ph.SR/1302.1136]}}.
\newblock {\url{https://doi.org/10.1088/0004-637X/764/2/170}}.

\bibitem[{Odert} et~al.(2017){Odert}, {Leitzinger}, {Hanslmeier}, and
  {Lammer}]{Odert2017}
{Odert}, P.; {Leitzinger}, M.; {Hanslmeier}, A.; {Lammer}, H.
\newblock {Stellar coronal mass ejections - I. Estimating occurrence
  frequencies and mass-loss rates}.
\newblock {\em \mnras} {\bf 2017}, {\em 472},~876--890,
  \href{http://arxiv.org/abs/1707.02165}{{\normalfont
  [arXiv:astro-ph.SR/1707.02165]}}.
\newblock {\url{https://doi.org/10.1093/mnras/stx1969}}.

\bibitem[{Wood} et~al.(2005){Wood}, {M{\"u}ller}, {Zank}, {Linsky}, and
  {Redfield}]{Wood2005}
{Wood}, B.E.; {M{\"u}ller}, H.R.; {Zank}, G.P.; {Linsky}, J.L.; {Redfield}, S.
\newblock {New Mass-Loss Measurements from Astrospheric Ly{$\alpha$}
  Absorption}.
\newblock {\em \apjl} {\bf 2005}, {\em 628},~L143--L146,
  \href{http://arxiv.org/abs/astro-ph/0506401}{{\normalfont
  [astro-ph/0506401]}}.
\newblock {\url{https://doi.org/10.1086/432716}}.

\bibitem[{Wood} et~al.(2021){Wood}, {M{\"u}ller}, {Redfield}, {Konow},
  {Vannier}, {Linsky}, {Youngblood}, {Vidotto}, {Jardine},
  {Alvarado-G{\'o}mez}, and {Drake}]{Wood2021}
{Wood}, B.E.; {M{\"u}ller}, H.R.; {Redfield}, S.; {Konow}, F.; {Vannier}, H.;
  {Linsky}, J.L.; {Youngblood}, A.; {Vidotto}, A.A.; {Jardine}, M.;
  {Alvarado-G{\'o}mez}, J.D.;  et~al.
\newblock {New Observational Constraints on the Winds of M dwarf Stars}.
\newblock {\em \apj} {\bf 2021}, {\em 915},~37,
  \href{http://arxiv.org/abs/2105.00019}{{\normalfont
  [arXiv:astro-ph.SR/2105.00019]}}.
\newblock {\url{https://doi.org/10.3847/1538-4357/abfda5}}.

\bibitem[{Osten} and {Wolk}(2015)]{Osten2015}
{Osten}, R.A.; {Wolk}, S.J.
\newblock {Connecting Flares and Transient Mass-loss Events in Magnetically
  Active Stars}.
\newblock {\em \apj} {\bf 2015}, {\em 809},~79,
  \href{http://arxiv.org/abs/1506.04994}{{\normalfont
  [arXiv:astro-ph.SR/1506.04994]}}.
\newblock {\url{https://doi.org/10.1088/0004-637X/809/1/79}}.

\bibitem[{Cranmer}(2017)]{Cranmer2017}
{Cranmer}, S.R.
\newblock {Mass-loss Rates from Coronal Mass Ejections: A Predictive
  Theoretical Model for Solar-type Stars}.
\newblock {\em \apj} {\bf 2017}, {\em 840},~114,
  \href{http://arxiv.org/abs/1704.06689}{{\normalfont
  [arXiv:astro-ph.SR/1704.06689]}}.
\newblock {\url{https://doi.org/10.3847/1538-4357/aa6f0e}}.

\bibitem[{Odert} et~al.(2020){Odert}, {Leitzinger}, {Guenther}, and
  {Heinzel}]{Odert2020}
{Odert}, P.; {Leitzinger}, M.; {Guenther}, E.W.; {Heinzel}, P.
\newblock {Stellar coronal mass ejections - II. Constraints from spectroscopic
  observations}.
\newblock {\em \mnras} {\bf 2020}, {\em 494},~3766--3783,
  \href{http://arxiv.org/abs/2004.04063}{{\normalfont
  [arXiv:astro-ph.SR/2004.04063]}}.
\newblock {\url{https://doi.org/10.1093/mnras/staa1021}}.

\bibitem[{Heinzel}(1995)]{Heinzel1995}
{Heinzel}, P.
\newblock {Multilevel NLTE radiative transfer in isolated atmospheric
  structures: implementation of the MALI-technique.}
\newblock {\em \aap} {\bf 1995}, {\em 299},~563.

\bibitem[{Heinzel} et~al.(1999){Heinzel}, {Mein}, and {Mein}]{Heinzel1999}
{Heinzel}, P.; {Mein}, N.; {Mein}, P.
\newblock {Cloud model with variable source function for solar
  H{\ensuremath{\alpha}} structures. II. Dynamical models}.
\newblock {\em \aap} {\bf 1999}, {\em 346},~322--328.

\bibitem[{Ikuta} and {Shibata}(2024)]{Ikuta2024}
{Ikuta}, K.; {Shibata}, K.
\newblock {Simple Model for Temporal Variations of H{\ensuremath{\alpha}}
  Spectrum by an Eruptive Filament from a Superflare on a Solar-type Star}.
\newblock {\em \apj} {\bf 2024}, {\em 963},~50,
  \href{http://arxiv.org/abs/2401.04279}{{\normalfont
  [arXiv:astro-ph.SR/2401.04279]}}.
\newblock {\url{https://doi.org/10.3847/1538-4357/ad1ce6}}.

\bibitem[{Namekata} et~al.(2021){Namekata}, {Maehara}, {Honda}, {Notsu},
  {Okamoto}, {Takahashi}, {Takayama}, {Ohshima}, {Saito}, {Katoh}, {Tozuka},
  {Murata}, {Ogawa}, {Niwano}, {Adachi}, {Oeda}, {Shiraishi}, {Isogai}, {Seki},
  {Ishii}, {Ichimoto}, {Nogami}, and {Shibata}]{Namekata2022}
{Namekata}, K.; {Maehara}, H.; {Honda}, S.; {Notsu}, Y.; {Okamoto}, S.;
  {Takahashi}, J.; {Takayama}, M.; {Ohshima}, T.; {Saito}, T.; {Katoh}, N.;
  et~al.
\newblock {Probable detection of an eruptive filament from a superflare on a
  solar-type star}.
\newblock {\em Nature Astronomy} {\bf 2021}, {\em 6},~241--248,
  \href{http://arxiv.org/abs/2112.04808}{{\normalfont
  [arXiv:astro-ph.SR/2112.04808]}}.
\newblock {\url{https://doi.org/10.1038/s41550-021-01532-8}}.

\bibitem[{Wilson} and {Raymond}(2022)]{Wilson2022}
{Wilson}, M.L.; {Raymond}, J.C.
\newblock {Solar Coronal Mass Ejections Plasma Diagnostics Expressed as
  Potential Stellar CME Signatures}.
\newblock {\em \aj} {\bf 2022}, {\em 164},~108,
  \href{http://arxiv.org/abs/2205.12985}{{\normalfont
  [arXiv:astro-ph.SR/2205.12985]}}.
\newblock {\url{https://doi.org/10.3847/1538-3881/ac80c4}}.

\bibitem[{Yang} et~al.(2024){Yang}, {Tian}, {Zhu}, {Xu}, {Chen}, and
  {Sun}]{Yang2024}
{Yang}, Z.; {Tian}, H.; {Zhu}, Y.; {Xu}, Y.; {Chen}, L.; {Sun}, Z.
\newblock {Is It Possible to Detect Coronal Mass Ejections on Solar-type Stars
  through Extreme-ultraviolet Spectral Observations?}
\newblock {\em \apj} {\bf 2024}, {\em 966},~24,
  \href{http://arxiv.org/abs/2402.12297}{{\normalfont
  [arXiv:astro-ph.SR/2402.12297]}}.
\newblock {\url{https://doi.org/10.3847/1538-4357/ad2a44}}.

\bibitem[{Cully} et~al.(1994){Cully}, {Fisher}, {Abbott}, and
  {Siegmund}]{Cully1994}
{Cully}, S.L.; {Fisher}, G.H.; {Abbott}, M.J.; {Siegmund}, O.H.W.
\newblock {A coronal mass ejection model for the 1992 July 15 flare on AU
  Microscopii observed by the extreme ultraviolet explorer}.
\newblock {\em \apj} {\bf 1994}, {\em 435},~449--+.
\newblock {\url{https://doi.org/10.1086/174827}}.

\bibitem[{Katsova} et~al.(1999){Katsova}, {Drake}, and {Livshits}]{Katsova1999}
{Katsova}, M.M.; {Drake}, J.J.; {Livshits}, M.A.
\newblock {New Insights into the Large 1992 July 15-17 Flare on AU Microscopii:
  The First Detection of Posteruptive Energy Release on a Red Dwarf Star}.
\newblock {\em \apj} {\bf 1999}, {\em 510},~986--998.
\newblock {\url{https://doi.org/10.1086/306587}}.

\bibitem[{Alvarado-G{\'o}mez} et~al.(2020){Alvarado-G{\'o}mez}, {Drake},
  {Fraschetti}, {Garraffo}, {Cohen}, {Vocks}, {Poppenh{\"a}ger}, {Moschou},
  {Yadav}, and {Manchester}]{Alvarado2020}
{Alvarado-G{\'o}mez}, J.D.; {Drake}, J.J.; {Fraschetti}, F.; {Garraffo}, C.;
  {Cohen}, O.; {Vocks}, C.; {Poppenh{\"a}ger}, K.; {Moschou}, S.P.; {Yadav},
  R.K.; {Manchester}, Ward~B., I.
\newblock {Tuning the Exospace Weather Radio for Stellar Coronal Mass
  Ejections}.
\newblock {\em \apj} {\bf 2020}, {\em 895},~47,
  \href{http://arxiv.org/abs/2004.05379}{{\normalfont
  [arXiv:astro-ph.SR/2004.05379]}}.
\newblock {\url{https://doi.org/10.3847/1538-4357/ab88a3}}.

\bibitem[{{\'O} Fionnag{\'a}in} et~al.(2022){{\'O} Fionnag{\'a}in}, {Kavanagh},
  {Vidotto}, {Jeffers}, {Petit}, {Marsden}, {Morin}, and
  {Golden}]{OFionnagain2022}
{{\'O} Fionnag{\'a}in}, D.; {Kavanagh}, R.D.; {Vidotto}, A.A.; {Jeffers}, S.V.;
  {Petit}, P.; {Marsden}, S.; {Morin}, J.; {Golden}, A.
\newblock {Coronal Mass Ejections and Type II Radio Emission Variability during
  a Magnetic Cycle on the Solar-type Star {\ensuremath{\epsilon}} Eridani}.
\newblock {\em \apj} {\bf 2022}, {\em 924},~115,
  \href{http://arxiv.org/abs/2111.02284}{{\normalfont
  [arXiv:astro-ph.SR/2111.02284]}}.
\newblock {\url{https://doi.org/10.3847/1538-4357/ac35de}}.

\bibitem[{Jin} et~al.(2020){Jin}, {Cheung}, {DeRosa}, {Nitta}, {Schrijver},
  {France}, {Kowalski}, {Mason}, and {Osten}]{Jin2020}
{Jin}, M.; {Cheung}, M.C.M.; {DeRosa}, M.L.; {Nitta}, N.V.; {Schrijver}, C.J.;
  {France}, K.; {Kowalski}, A.; {Mason}, J.P.; {Osten}, R.
\newblock {Coronal dimming as a proxy for stellar coronal mass ejections}.
\newblock In Proceedings of the Solar and Stellar Magnetic Fields: Origins and
  Manifestations; {Kosovichev}, A.; {Strassmeier}, S.; {Jardine}, M., Eds.,  1
  2020, Vol. 354, pp. 426--432,
  \href{http://arxiv.org/abs/2002.06249}{{\normalfont
  [arXiv:astro-ph.SR/2002.06249]}}.
\newblock {\url{https://doi.org/10.1017/S1743921320000575}}.

\bibitem[{Alvarado-G{\'o}mez} et~al.(2018){Alvarado-G{\'o}mez}, {Drake},
  {Cohen}, {Moschou}, and {Garraffo}]{Alvarado2018}
{Alvarado-G{\'o}mez}, J.D.; {Drake}, J.J.; {Cohen}, O.; {Moschou}, S.P.;
  {Garraffo}, C.
\newblock {Suppression of Coronal Mass Ejections in Active Stars by an
  Overlying Large-scale Magnetic Field: A Numerical Study}.
\newblock {\em \apj} {\bf 2018}, {\em 862},~93,
  \href{http://arxiv.org/abs/1806.02828}{{\normalfont
  [arXiv:astro-ph.SR/1806.02828]}}.
\newblock {\url{https://doi.org/10.3847/1538-4357/aacb7f}}.

\bibitem[{Alvarado-G{\'o}mez} et~al.(2019){Alvarado-G{\'o}mez}, {Drake},
  {Moschou}, {Garraffo}, {Cohen}, {NASA LWS Focus Science Team: Solar-Stellar
  Connection}, {Yadav}, and {Fraschetti}]{Alvarado2019}
{Alvarado-G{\'o}mez}, J.D.; {Drake}, J.J.; {Moschou}, S.P.; {Garraffo}, C.;
  {Cohen}, O.; {NASA LWS Focus Science Team: Solar-Stellar Connection}.;
  {Yadav}, R.K.; {Fraschetti}, F.
\newblock {Coronal Response to Magnetically Suppressed CME Events in M-dwarf
  Stars}.
\newblock {\em \apjl} {\bf 2019}, {\em 884},~L13,
  \href{http://arxiv.org/abs/1909.04092}{{\normalfont
  [arXiv:astro-ph.SR/1909.04092]}}.
\newblock {\url{https://doi.org/10.3847/2041-8213/ab44d0}}.

\bibitem[{Sun} et~al.(2022){Sun}, {T{\"o}r{\"o}k}, and {DeRosa}]{Sun2022}
{Sun}, X.; {T{\"o}r{\"o}k}, T.; {DeRosa}, M.L.
\newblock {Torus-stable zone above starspots}.
\newblock {\em \mnras} {\bf 2022}, {\em 509},~5075--5085,
  \href{http://arxiv.org/abs/2111.03665}{{\normalfont
  [arXiv:astro-ph.SR/2111.03665]}}.
\newblock {\url{https://doi.org/10.1093/mnras/stab3249}}.

\bibitem[{Lynch} et~al.(2019){Lynch}, {Airapetian}, {DeVore}, {Kazachenko},
  {L{\"u}ftinger}, {Kochukhov}, {Ros{\'e}n}, and {Abbett}]{Lynch2019}
{Lynch}, B.J.; {Airapetian}, V.S.; {DeVore}, C.R.; {Kazachenko}, M.D.;
  {L{\"u}ftinger}, T.; {Kochukhov}, O.; {Ros{\'e}n}, L.; {Abbett}, W.P.
\newblock {Modeling a Carrington-scale Stellar Superflare and Coronal Mass
  Ejection from $\kappa^1$ Cet}.
\newblock {\em \apj} {\bf 2019}, {\em 880},~97,
  \href{http://arxiv.org/abs/1906.03189}{{\normalfont
  [arXiv:astro-ph.SR/1906.03189]}}.
\newblock {\url{https://doi.org/10.3847/1538-4357/ab287e}}.

\bibitem[{Kay} et~al.(2016){Kay}, {Opher}, and {Kornbleuth}]{Kay2016}
{Kay}, C.; {Opher}, M.; {Kornbleuth}, M.
\newblock {Probability of CME Impact on Exoplanets Orbiting M Dwarfs and
  Solar-like Stars}.
\newblock {\em \apj} {\bf 2016}, {\em 826},~195,
  \href{http://arxiv.org/abs/1605.02683}{{\normalfont
  [arXiv:astro-ph.SR/1605.02683]}}.
\newblock {\url{https://doi.org/10.3847/0004-637X/826/2/195}}.

\bibitem[{Kay} et~al.(2019){Kay}, {Airapetian}, {L{\"u}ftinger}, and
  {Kochukhov}]{Kay2019}
{Kay}, C.; {Airapetian}, V.S.; {L{\"u}ftinger}, T.; {Kochukhov}, O.
\newblock {Frequency of Coronal Mass Ejection Impacts with Early Terrestrial
  Planets and Exoplanets around Active Solar-like Stars}.
\newblock {\em \apjl} {\bf 2019}, {\em 886},~L37,
  \href{http://arxiv.org/abs/1911.02701}{{\normalfont
  [arXiv:astro-ph.SR/1911.02701]}}.
\newblock {\url{https://doi.org/10.3847/2041-8213/ab551f}}.

\bibitem[{Menezes} et~al.(2023){Menezes}, {Valio}, {Netto}, {Ara{\'u}jo},
  {Kay}, and {Opher}]{Menezes2023}
{Menezes}, F.; {Valio}, A.; {Netto}, Y.; {Ara{\'u}jo}, A.; {Kay}, C.; {Opher},
  M.
\newblock {Trajectories of coronal mass ejection from solar-type stars}.
\newblock {\em \mnras} {\bf 2023}, {\em 522},~4392--4403,
  \href{http://arxiv.org/abs/2305.07159}{{\normalfont
  [arXiv:astro-ph.SR/2305.07159]}}.
\newblock {\url{https://doi.org/10.1093/mnras/stad1078}}.

\bibitem[{Mandal} et~al.(2017){Mandal}, {Chatterjee}, and
  {Banerjee}]{Mandal2017}
{Mandal}, S.; {Chatterjee}, S.; {Banerjee}, D.
\newblock {Solar Active Longitudes from Kodaikanal White-light Digitized Data}.
\newblock {\em \apj} {\bf 2017}, {\em 835},~62,
  \href{http://arxiv.org/abs/1611.07637}{{\normalfont
  [arXiv:astro-ph.SR/1611.07637]}}.
\newblock {\url{https://doi.org/10.3847/1538-4357/835/1/62}}.

\bibitem[{Ol{\'a}h} et~al.(1999){Ol{\'a}h}, {van Driel-Gesztelyi},
  {K{\H{o}}v{\'a}ri}, and {Bartus}]{Olah1999}
{Ol{\'a}h}, K.; {van Driel-Gesztelyi}, L.; {K{\H{o}}v{\'a}ri}, Z.; {Bartus}, J.
\newblock {Modelling the Sun as an active star. I. A diagnosis of photometric
  starspot models}.
\newblock {\em \aap} {\bf 1999}, {\em 344},~163--171.

\bibitem[{Lustig-Yaeger} et~al.(2023){Lustig-Yaeger}, {Fu}, {May}, {Ceballos},
  {Moran}, {Peacock}, {Stevenson}, {Kirk}, {L{\'o}pez-Morales}, {MacDonald},
  {Mayorga}, {Sing}, {Sotzen}, {Valenti}, {Redai}, {Alam}, {Batalha},
  {Bennett}, {Gonzalez-Quiles}, {Kruse}, {Lothringer}, {Rustamkulov}, and
  {Wakeford}]{JWST2023NatAs...7.1317L}
{Lustig-Yaeger}, J.; {Fu}, G.; {May}, E.M.; {Ceballos}, K.N.O.; {Moran}, S.E.;
  {Peacock}, S.; {Stevenson}, K.B.; {Kirk}, J.; {L{\'o}pez-Morales}, M.;
  {MacDonald}, R.J.;  et~al.
\newblock {A JWST transmission spectrum of the nearby Earth-sized exoplanet LHS
  475 b}.
\newblock {\em Nature Astronomy} {\bf 2023}, {\em 7},~1317--1328,
  \href{http://arxiv.org/abs/2301.04191}{{\normalfont
  [arXiv:astro-ph.EP/2301.04191]}}.
\newblock {\url{https://doi.org/10.1038/s41550-023-02064-z}}.

\bibitem[H{\'{e}}der et~al.(2022)H{\'{e}}der, Rig{\'{o}}, Medgyesi, Lovas,
  Tenczer, Török, Farkas, Em{\H{o}}di, Kadlecsik, Mez{\H{o}}, Pint{\'{e}}r,
  and Kacsuk]{MTACloud}
H{\'{e}}der, M.; Rig{\'{o}}, E.; Medgyesi, D.; Lovas, R.; Tenczer, S.; Török,
  F.; Farkas, A.; Em{\H{o}}di, M.; Kadlecsik, J.; Mez{\H{o}}, G.;  et~al.
\newblock The Past, Present and Future of the {ELKH} Cloud.
\newblock {\em Inform{\'{a}}ci{\'{o}}s T{\'{a}}rsadalom} {\bf 2022}, {\em
  22},~128.
\newblock {\url{https://doi.org/10.22503/inftars.xxii.2022.2.8}}.

\end{thebibliography}

%


\PublishersNote{}
\end{adjustwidth}
\end{document}